\documentclass[aps,prb,reprint,superscriptaddress]{revtex4-1}
\usepackage{amssymb}
\usepackage{amsmath}
\usepackage{graphicx}
\newcommand{\Dabel}[1]{\label{#1}}

\def\tTF{\text{TF}}
\def\tLDA{\text{LDA}}
\def\tKS{\text{KS}}
\def\teff{\text{eff}}
\def\tOF{\text{OF}}
\def\tmod{\text{Mod}}
\def\thomo{{\rm HOMO}}
\def\rr{{\bf r}}
\def\ro{(\rr)}

\begin{document}

\title{Quantum Hydrodynamic Theory for Plasmonics: Impact of the Electron Density Tail} 

\author{Cristian Cirac\`i}
\affiliation{Istituto Italiano di Tecnologia (IIT), Center for Biomolecular Nanotechnologies@UNILE, Via Barsanti, 73010 Arnesano, Italy.}
\email[]{cristian.ciraci@iit.it}

\author{Fabio Della Sala}
\affiliation{Istituto Nanoscienze-CNR, Euromediterranean Center for Nanomaterial Modelling and Technology, 
Via per Arnesano 16, 73100 Lecce, Italy.}
\affiliation{Istituto Italiano di Tecnologia (IIT), Center for Biomolecular Nanotechnologies@UNILE, Via Barsanti, 73010 Arnesano, Italy.}





\date{\today}

\begin{abstract}
 Multiscale plasmonic systems (\textit{e.g.} extended metallic nanostructures with sub-nanometer inter-distances) 
 play a key role in the development of next-generation nano-photonic devices. 
 An accurate modeling of the optical interactions in these systems requires an accurate description of both quantum effects and far-field properties. 
 Classical electromagnetism can only describe the latter, while 
 Time-Dependent Density Functional Theory (TD-DFT) can provide a full first-principles quantum treatment. 
 However, TD-DFT becomes computationally prohibitive for sizes that exceed few nanometers, which are instead very important for most applications.
In this article, we introduce a method based on the quantum hydrodynamic theory 
(QHT) that includes nonlocal contributions of the kinetic energy and the correct asymptotic description of the electron density.
 We show that our QHT method can predict both plasmon energy and spill-out effects in metal nanoparticles 
 in excellent agreement with TD-DFT predictions, thus allowing reliable and efficient calculations of both quantum and far-field properties
 in multiscale plasmonic systems.

\end{abstract}


\maketitle

\section{Introduction}
Plasmonic systems have received a renewed great deal of attention for their ability to localize electromagnetic radiation
 at visible frequencies well below the diffraction limit and enhance the local electric fields hundred times
 the incident radiation\cite{Gramotnev:2010ji,Schuller:2010fh,Maier:2007wq}.
These properties make plasmonic structures valuable candidates for enhancing
 nonlinear optical phenomena\cite{Aouani:2014iy},
 controlling surface reflectance properties\cite{Moreau:2012uba}, 
 enhancing the far-field coupling with nanometer-sized elements\cite{Akselrod:2014ek,Akselrod:2015cf,Rose:2014ko}, 
 such as quantum emitters, and studying fundamental phenomena\cite{Ciraci:2012fp,Savage:2012by,Hajisalem:2014cr}.
In particular, nanostructures supporting gap-plasmon modes\cite{Moreau:2013ei} constitute an important platform. 
Advances in nano-fabrication techniques\cite{Chen:2013hq,Chen:2014gn} have made it possible to 
 achieve separation between two metallic elements, \textit{i.e.} particles or nanowires, of only a fraction of a nanometer.
At such distances nonlocal or quantum effects becomes non-negligible.
It has been shown that the resonance of nanoparticle dimers or film-coupled nanospheres 
can be perturbed by such effects\cite{Ciraci:2012fp,Ciraci:2014jv,Hajisalem:2014cr}.
Form the theory stand point, it may be really challenging, if not impossible, to accurately describe at once the entire \textit{multiscale} physics involved in such systems.
On the one hand, one has a macroscopic electromagnetic system constituted by the whole plasmonic structure, on
 the other hand, as the gap closes it is crucial to take into account the quantum nature of the electrons
 in the metal\cite{Marinica:2012,Zuloaga:2009gm}.

A time-dependent density functional theory (TD-DFT) approach allows the 
exact calculation of plasmon, as well as single-particle, excitation energies in both finite and extended systems\cite{tddftbook}.
It can be implemented both in real-time propagation\cite{yabana96,octopus}, and frequency domain linear-response\cite{casi96}, and can serve as reference
for developing approximate schemes.
In the context of plasmonics, TD-DFT has been largely applied to nanosystems with an atomistic description\cite{morton11} 
or employing a jellium model\cite{brack93,ekardt85}.
Recently, TD-DFT has been applied to metallic wires with a diameter up to 20 nm\cite{Teperik:2013dd,Yan:2015gx}
and to metallic spheres (and sphere dimers) with around one thousand 
valence electrons\cite{Esteban:2012,barbry2015,Zhang:2014rubio,li2013,iida2014}.
 For larger systems TD-DFT rapidly becomes computationally prohibitive, as all single particle orbitals need to be computed.

An alternative approach is to use a simple linearized Thomas-Fermi hydrodynamic theory (TF-HT), also known simply as the hydrodynamic model, 
 which takes into account the nonlocal behavior of the electron response by including the electron pressure\cite{pitarke07,Raza:2011io,Ciraci:2013dz,Raza:2015ef}.
The introduction of an electron pressure term in the free-electron model accounts for the Pauli exclusion principle 
within the limit of the Thomas-Fermi (TF) theory\cite{parrbook}.
In contradistinction to the treatment of the electron response in classical electromagnetism, 
where induced charges are crushed into an 
infinitesimally thin layer at the surface of the metal, the induced electron density in the TF-HT approach
rather spreads out from the surface into the bulk region\cite{Ciraci:2013dz}. In fact, the TF-HT method 
is usually combined with
the assumption that the electrons cannot escape the metal boundaries (hard-wall boundary conditions).
The advantage of TF-HT with respect to full quantum methods is that it can be easily employed for structures 
of the order of several hundred nanometers in size.

The TF-HT model dates back to the '70s\cite{Heinrichs:1973hn,Eguiluz:1975be,Eguiluz:1976wk} and closed-form analytic solutions
exist for a homogeneous sphere \cite{ruppin75,ruppin78,fuchs81}.  
Nonetheless, its applicability in complex plasmonics has been limited by the absence of experimental confirmation or the validation by 
higher level theoretical methods, which are needed 
to verify  the assumptions and approximations used in constructing solutions.
In the last decade, however, the improvement of fabrication techniques and the proliferation of self-assembling colloidal
 plasmonic structures have provided a robust platform\cite{Anonymous:Vp6g18Xg,Hill:2010ez} for studying extremely sub-wavelength optical
 phenomena, thus reinvigorating the interest in TF-HT\cite{Mortensen:2014kc,Toscano:2013hr,Filter:2014,Ciraci:2013jt,FernandezDominguez:2012eg,Christensen:2014tm,Luo:2013jx,Wiener:2013gl,Wiener:2012fd}.
In particular, Cirac\`i \textit{et al.} applied the TF-HT method to plasmonic nanostructures consisting of film-coupled nanoparticles and found that the model 
can provide predictions that are both in qualitative and in quantitative agreement with experiments\cite{Ciraci:2012fp}.
More recently, TF-HT results have been compared with TD-DFT calculations \cite{Stella:2013by,Teperik:2013dd}, showing some limitations.
In fact, essential effects such as electron spill-out and 
quantum tunneling are completely neglected.
In order to include such effects, other methods based on effective descriptions 
have also been proposed \cite{lerme99,Esteban:2012,Luo:2013jx,Yan:2015gx,Zapata:2015fq}, as well as methods based on the real-time 
orbital-free TD-DFT\cite{domps98,xiang2014}.

To include spill-out effects in TF-HT the first step is to consider the spatial dependence of the electron density. This scheme traces back to the '70s both 
for finite systems (e.g. atoms) \cite{ball73,walecka76,mona74} and surfaces\cite{bennet70,Eguiluz:1975be,Schwartz82}, and has been recently reconsidered 
using equilibrium electron density from DFT calculations\cite{David:2014iw}. 
When spatial dependence of the electron density is included, however, one should consider nonlocal contributions, namely the von Weizs\"acker term, to the free-electron gas kinetic energy, 
in place of the simple TF kinetic energy.
This approach is usually named quantum hydrodynamic theory (QHT), and it 
has been widely used 
photoabsorption of atoms\cite{dreiz82},
metallic nanoparticles \cite{harbola00,harbola2008},
plasma physics\cite{plasmabook,manfredi05,shukla12,akbari15},
and two-dimensional magnetoplasmonics\cite{zaremba94,zaremba99}.
Very recently, the QHT method has been systematically investigated for surfaces\cite{Yan:2015ff} and 
 a self-consistent version of QHT has been presented and applied to plasmonic systems\cite{Toscano:2015iw,Li:2015io}.
However, the impact of the electronic ground-state density on the QHT optical response is yet unclear.

In this paper, we will first study the influence of ground-state electron density profile on the linear response of metallic nanospheres 
described by using the QHT method and compare our results with reference TD-DFT calculations.
We find that QHT can accurately  describe both the plasmon resonance and the spill-out effects only when it is combined with
the exact DFT ground-state density.

Secondly, we will show that by using an analytical model for the ground-state electronic density, it is possible to reproduce TD-DFT results 
and to include retardation effects simultaneously, thus
allowing the calculation of plasmonic systems exceeding hundred nanometers.

\section{Quantum Hydrodynamic Theory}
Within the hydrodynamic model, the many-body electron dynamic of an electronic system is described 
by two hydrodynamic quantities\cite{Tokatly:2000fy,Tokatly:1999id}: the electron density $n({\bf r},t)$ and the electron velocity field ${\bf v}({\bf r},t)$.
Under the influence of the electromagnetic fields ${\bf E}$ and ${\bf B}$ the electronic system can 
be described by the equation\cite{boardmanbook,Raza:2015ef}:
\begin{equation}
{m_e}\left( {\frac{\partial }{{\partial t}} + {\bf{v}} \cdot \nabla  + \gamma } \right){\bf{v}} 
=  - e\left( {{\bf{E}} + {\bf{v}} \times {\bf{B}}} \right) - \nabla {\frac{{\delta G[n]}}{{\delta n}}},
\Dabel{eq:nv}
\end{equation}
with $m_e$ and $e$, the electron mass and the electron charge (in absolute value) 
respectively, and $\gamma$ the phenomenological damping rate.
The energy functional $G[n]$ contains the sum of  the interacting kinetic energy ($T$) and the 
exchange-correlation (XC) potential energy ($U_{xc}$) of the electronic system.
In DFT the XC potential energy  is defined as $U_{xc}=E_{xc}-(T-T_s)$ \cite{parrbook,dreibook} where $E_{xc}$ is the XC energy and 
$T_s$ is the non-interacting kinetic energy: thus we have $G[n]=T[n]+U_{\text{XC}}[n]=T_s[n]+E_{\text{XC}}[n]$.
In this work we employ the following approximation for $G[n]$:
\begin{equation}
G[n]\approx G_\eta[n]=\left( T_s^{\tTF}[n]+\frac{1}{\eta} T_s^W[n]\right ) +E_{\text{XC}}^{\tLDA}[n],
\Dabel{eq:gdef}
 \end{equation}
 where $T_s^{\tTF}[n]$ is the kinetic energy functional in the TF approximation,
  $T_s^W[n]$ is the von Weizs\"acker kinetic energy functional \cite{parrbook}
and $E_{XC}^{\tLDA}[n]$ is the local density approximation (LDA) for the XC energy functional.
The expressions for these functionals can be easily found in the literature (see e.g. Ref. \onlinecite{Ho:2008jq});  here we report the expression 
for their potentials obtained by taking the functional derivative with respect to $n$:
\begin{subequations}
\Dabel{eq:potentials}
\begin{eqnarray}
\frac{\delta T_s^{\tTF}   }{\delta n} &=&  (E_h a_0^2)   \frac{5}{3}{c_{\tTF}}{n^{2/3}}   , \\
\frac{\delta T_s^{W}    }{\delta n} &=&  (E_h a_0^2)   \frac{1}{8}\left( {\frac{{\nabla n \cdot \nabla n}}{{{n^2}}} - 2\frac{{{\nabla ^2}n}}{n}} \right)  \\
\frac{\delta E_{XC}^{\tLDA}}{\delta n} &=&  (E_h) \left  ( - a_0 \frac{4}{3}{c_X}n^{1/3} +\mu_C[n] \right) =v_{xc}({\bf r})\quad
\end{eqnarray}
\end{subequations}
where $E_h=\frac{\hbar^2}{m_ea_0^2}$ is the Hartree energy, $a_0$ is the Bohr radius,  
$c_{\tTF}=\frac{3}{10} ( {3{\pi ^2}} )^{2/3}$, and
$c_X=\frac{3}{4} ( {\frac{3}{\pi } )^{1/3}}$.
Eqs. (\ref{eq:potentials}), as well as other formulas in this paper are in S.I. units; 
expressions in atomic units (a.u.) can also be easily obtained by considering that 
$E_h=a_0=m_e=\hbar=1$.
The term in Eq. (\ref{eq:potentials}c) is the XC potential
$v_{xc}({\bf r})$; the correlation potential $\mu_C[n]$ (in atomic units) is obtained from the
 Perdew-Zunger LDA parametrization \cite{Perdew:1981dv}:
 \begin{widetext}
\[
{\mu _C}[n]= \left\{ {\begin{array}{{l l}}
\ln (r_s)\left( {a + \frac{2}{3}c{r_s}} \right) + \left( {b - \frac{1}{3}a} \right) + \frac{1}{3}\left( {2d - c} \right){r_s}, &\quad {r_s} < 1\\
\frac{{\alpha + \left( {\frac{7}{6}\alpha {\beta _1}} \right)\sqrt {{r_s}}  + \left( {\frac{4}{3}\alpha {\beta _2}} \right){r_s}}}{{{{\left( {1 + {\beta _1}\sqrt {{r_s}}  + {\beta _2}{r_s}} \right)}^2}}}, &\quad {r_s} \ge 1
\end{array}} \right.
\]
 \end{widetext}
with $a_0 r_s = (\frac{3}{4\pi n})^{1/3}$ being the Wigner-Seitz radius. 
The coefficients are $a = 0.0311$, $b = -0.048$, $c = 0.002$, $d = -0.0116$, $\alpha = -0.1423$, $\beta_1 = 1.0529$ and $\beta_2 = 0.3334$.\cite{Perdew:1981dv}

The key parameter in Eq. (\ref{eq:gdef}) is $\eta$, which is in the range [1,$\infty$] (usually in the literature $\lambda=1/\eta$ is used).
While the TF approximation ($\eta=\infty$) is exact only in the bulk 
region where the electron density becomes uniform, the von Weizs\"acker term adds a correction that depends on the 
gradient (i.e. on the wavevector $k$ in the reciprocal space). 
In general choosing the parameter $\eta=9$ gives a good approximation for a slowly varying electron density ($k\ll1$), 
while taking $\eta=1$ gives exact results for large $k$. \cite{wangcarterof}
In this work we will consider both $\eta=1$ and $\eta=9$. The latter has been used in Ref. \onlinecite{Toscano:2015iw}.

Equation (\ref{eq:nv}) has to be coupled to Maxwell's equations.
By linearizing the system with the usual perturbation approach\cite{Yariv:1988vc}, 
taking into account the continuity equation, and the fact that 
$\partial {\bf{P}}/\partial t = {\bf{J}} =  - ne{\bf{v}}$, we obtain in the frequency 
domain the following system of equations:
\begin{subequations}
\Dabel{eq:sys}
\begin{eqnarray}
&&\nabla  \times \nabla  \times {\bf{E}} - \frac{{{\omega ^2}}}{{{c^2}}}{\bf{E}} 
= {\omega ^2}{\mu _0}{\bf{P}}, \\
&&\frac{{e{n_0}}}{{{m_e}}}
\nabla {\left( {\frac{{\delta G_\eta}}{{\delta n}}} \right)_1}
 + \left( {{\omega ^2} + i\gamma\omega  } \right){\bf{P}} 
=  - {\varepsilon _0}\omega _p^2{\bf{E}},
\end{eqnarray}
\end{subequations}
where  $\varepsilon_0$ is the vacuum permittivity, $c$ the speed of light, $n_0({\bf{r}})$ is the unperturbed (ground-state) electron density, 
and $\omega_p({\bf{r}})=\sqrt{e^2n_0({\bf{r}})/(m_e\varepsilon_0)}$ is the {\it spatially dependent} plasma frequency.
The first order terms for the potential can be calculated as:
\begin{equation}
 {\left( {\frac{{\delta G_\eta}}{{\delta n}}} \right)_1} 
={\int {{{\left. {\frac{{\delta G_\eta[n]}}{{\delta n({\bf{r}})\delta n({\bf{r'}})}}} \right|}_{n = {n_0}({\bf r})}}} 
{n_1}({\bf{r'}})d{\bf{r'}}} 
\Dabel{eq:fot}
\end{equation}
with $n_1=\frac{1}{e}\nabla\cdot{\bf P}$ being the electron density first order perturbation.
Clearly {\bf{E}}, {\bf{P}} and $n_1$ are complex quantities and depend on $\omega$. 
Using the expressions (\ref{eq:potentials}) and Eq. (\ref{eq:fot}) 
the first order terms for the potentials are:
\begin{widetext}
\begin{subequations}
\Dabel{eq:fop}
\begin{eqnarray}
{\left( \frac{\delta T_s^{\tTF}}{\delta n} \right)_1}& =&  (E_h a_0^2)   \frac{{10}}{9}c_{\tTF}{n_0}^{-1/3}{n_1} \\
{\left( \frac{\delta T_s^{W} }{\delta n} \right)_1}& = &  
 (E_h a_0^2)  \frac{1}{4}\left[ {\frac{{\nabla {n_0} \cdot \nabla {n_1}}}{{n_0^2}} + \frac{{{\nabla ^2}{n_0}}}{{n_0^2}}{n_1} - \frac{{{{\left| {\nabla {n_0}} \right|}^2}}}{{n_0^3}}{n_1} - \frac{{{\nabla ^2}{n_1}}}{{{n_0}}}} \right],\Dabel{eq:vW}\\
{\left( \frac{\delta E_{XC}^{LDA}}{\delta n} \right)_1}&=&  (E_h ) \left ( - a_0 \frac{4}{9}  {c_X}n_0^{-2/3}n_1+  a_0^3 \mu'_C[n_0] n_1 \right ),
\end{eqnarray}
\end{subequations}
with
\[ {\mu'_C}[n]  = 
\left\{ 
{\begin{array}{*{20}{l}}
{ - \frac{{4\pi }}{9}\left[ {a + \frac{1}{3}\left( {1 + 2\ln \left( {{r_s}} \right)} \right)c{r_s} + \frac{2}{3}d{r_s}} \right]r_s^3 ,}&{\quad {r_s} < 1}\\
{\frac{{\alpha \pi }}{{27}}\frac{{5{\beta _1}\sqrt {{r_s}}  
+ \left( {7\beta _1^2 + 8{\beta _2}} \right){r_s} + 21{\beta _1}{\beta _2}r_s^{3/2} 
+ 16\beta _2^2r_s^2}}{{{{\left( {1 + {\beta _1}\sqrt {{r_s}}  + {\beta _2}{r_s}} \right)}^3}}}r_s^3 ,}&{\quad {r_s} \ge 1}
\end{array}} 
\right.\]
\end{widetext}
In the following the system of equations (\ref{eq:sys}) will be named QHT$\eta$, 
i.e. QHT1 if $\eta=1$ or QHT9 if $\eta=9$.

It is useful to notice that Eq. (\ref{eq:sys}b) reduces to the TF-HT model if the XC 
and the von Weizs\"acker functionals are neglected, and assuming that 
the equilibrium electron density is uniform in space, $n_0({\bf r})\equiv n_0$. 
In this case, in fact, the first term on the right-hand side of Eq. (\ref{eq:sys}b) 
becomes $\beta^2 \nabla (\nabla \cdot {\bf P})$ 
with  ${\beta ^2} = \frac{{10}}{9}\frac{{{c_{\tTF}}}}{m_e}n_0^{2/3} =v_F^2/3$  
\cite{Ciraci:2013dz,Stella:2013by}.
As already pointed-out in the introduction, TF-HT will always be associated with hard-wall boundary 
conditions, \textit{i.e.}, ${\bf P}\cdot {\bf \hat n}=0$, at the metal boundaries, with ${\bf \hat n}$ being the unit vector normal to the surface.

 The equations QHT$\eta$ can be solved with a plane-wave excitation for a range of frequencies $\omega$; the solution 
 vectors ${\bf E}$ and ${\bf P}$
can then be used to compute the linear optical properties.
We have implemented a numerical solution of the system of Eqs. (\ref{eq:sys}) within 
a commercially available software based on the finite-element method, \textsc{Comsol} 
Multiphysics\cite{comsol} (see Appendix \ref{sec:ni}).
In particular, we have implemented the method using the \textit{2.5D technique}\cite{Ciraci:2013wi}, 
which allows to easily compute absorption spectra for spheres or, more generally, axis
symmetric structures of the order of few hundred nanometers in size.\\

\section{Jellium Nanospheres}
The results of the QHT approach will directly depend on the input ground state electronic 
density $n_0\ro$, which defines the system under consideration.
This is very different from classical plasmonics, where the system is defined 
by its local dielectric constant.

Ideal systems to test the QHT approach are represented by jellium nanospheres\cite{brack93}, where $N_e$ electrons are confined 
by the electrostatic potential generated by a uniformly charged sphere of radius $R=r_s N_e^{1/3}$  
with positive charge density $n^+=(r_s^3 4\pi/3)^{-1}$ inside, and zero outside; here $r_s$ is the Wigner-Seitz radius, ranging from 2 to 6 a.u. in real metals.
In this work we consider $r_s=4$ a.u., which represents sodium.
In order to exactly include all quantum effects, $n_0\ro$ 
should be the exact quantum-mechanical density of the system under consideration, obtained
for example from a full ground-state Kohn-Sham (KS) DFT calculation. 
We have developed an in-house code for the self-consistent solution of the KS equations for jellium nanospheres,
with the LDA XC functional. Calculations are performed with finite differences on a linear numerical grid up to $R$+50 bohr.
The final ground-state electron density can be written as:
\begin{equation}
n_0^{\tKS}\ro=\sum_{l=0}^{Lmax} \sum_{n=0}^{n_l} f \frac{2l+1}{4\pi} R_{nl}({\bf r})^2
\end{equation}
where $R_{nl}({\bf r})$ is the solution of the radial Schr\"odinger equation and $f=2$ (we consider only the
spin-restricted case).
The electronic configuration of a jellium nanosphere is characterized by $L_{max}$ and 
a sequence of shell-number $S=[n_0,n_1,\dots,n_{Lmax}]$ with  $n_0 \ge n_1 \ldots \ge n_{Lmax}$, 
i.e. there are $n_l$ occupied orbitals with angular
momentum $l$. The total number of electrons is then
 $N_e=\sum_{l=0}^{Lmax} n_l f (2l+1)$, which can be called shell-closing numbers.
The so called ``magic-number'' clusters not only have a shell-closing but also a (large) positive KS energy-gap\cite{ekardt97,rubio91,occup2000,vander93}.
 
We have implemented a code that computes {\it all} magic-number jellium nanospheres up to an 
arbitrary number of electrons. 
Starting from $S=[1]$, i.e. a system where there are only $N_e=2$ electrons in the lowest $1s$-shell, the program tries
to fill other shells in order to keep the KS energy-gap as large as possible.
In the first step the program thus compares the KS energy-gap between jellium 
nanospheres with $S=[2]$ and $S=[1,1]$, and obviously it finds that the latter is the next magic-number
jellium nanosphere. Then the same procedure is applied to $S=[1,1]$, comparing $S=[2,1]$ and $S=[1,1,1]$ and so on.
In this way the first jellium nanospheres obtained are $N_e=2,8,18,20,34,40,58,68,90,92,106,132..$, which 
are well established in the literature\cite{ekardt84}. However, jellium nanospheres with  $N_e=68,90,106,...$ are characterized 
by a negative KS energy-gap (i.e. there are occupation holes below the highest-occupied molecular orbitals):
this means that the so obtained electronic density is not a ground-state density 
and these clusters are not magic-number clusters. Moreover, we note that when $N_e$ is very large, the electronic 
configuration cannot be established using simple models\cite{ekardt84,vander93,occup2000}, due to the almost degeneracy of the high-lying KS orbitals.
In this work, we considered all  shell-closing magic-number jellium nanospheres up to $N_e=5032$ (see Table S1 in the Supplemental Material).


\begin{figure}[hbt]
        \centering
                \includegraphics[width=0.4\textwidth]{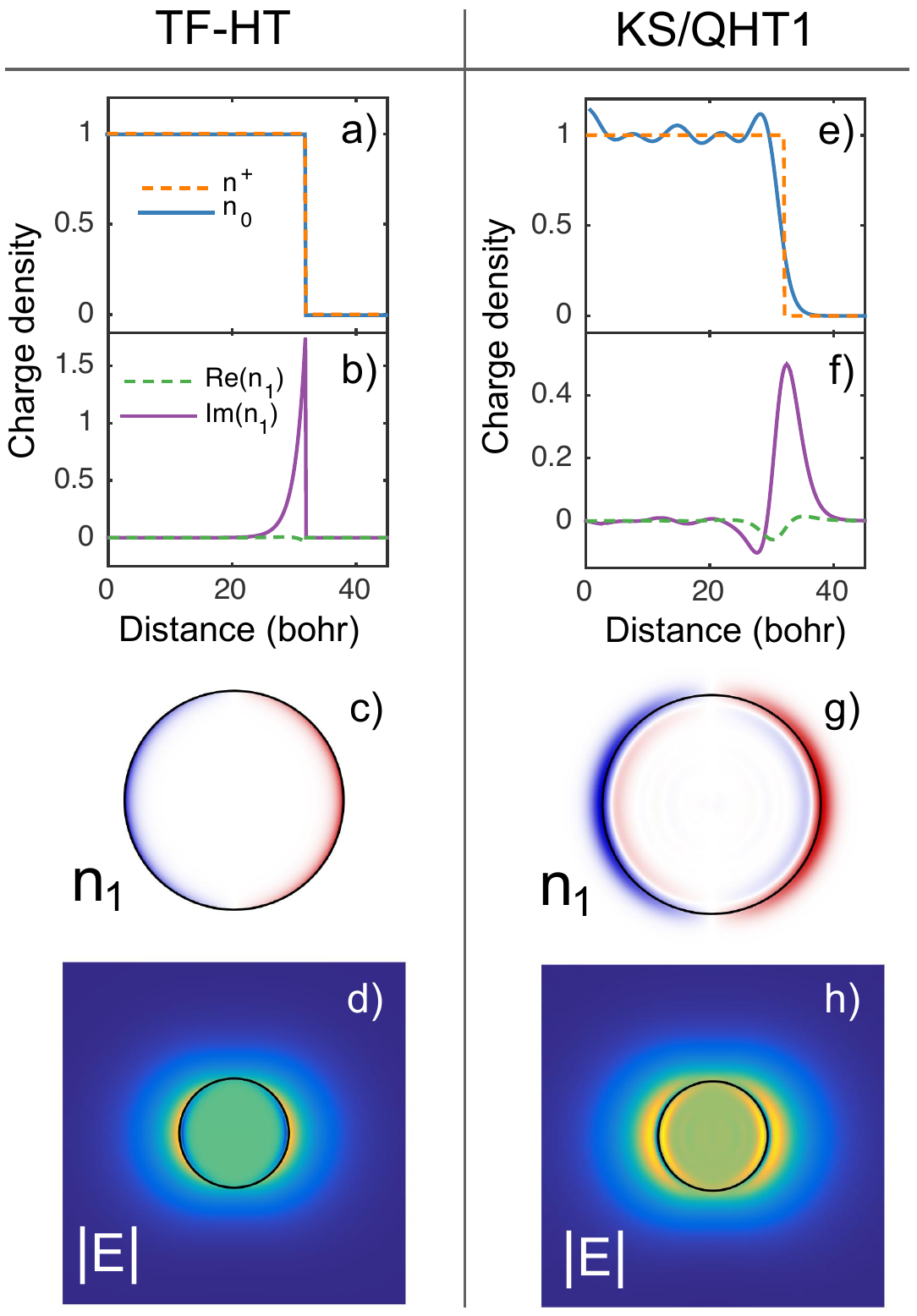}      
        \caption{Jellium nanosphere  ($r_s=4$) with $N_e=508$ electrons (R= 31.9 a.u.) as obtained from the TF-HT (panel a,b,c,d) and the KS/QHT1 (panels e,f,g,h) approaches. 
Ground-state density $n_0(r)$, panel a) and e);
Real and imaginary part of the induced charge density $n_1({\bf r})$ at the plasmon resonance, panel b) and f);
Imaginary part of $n_1({\bf r})$ at the cross-section plane, panel c) and g);
Norm of the induced electric field at the cross-section plane, panel d) and h).
}
         \Dabel{fig:plotjell}
\end{figure}

In Fig. \ref{fig:plotjell} we compare the results of  
the TF-HT approach for a jellium nanosphere with $N_e=508$ electrons with the solution of
the QHT equations with $\eta=1$ (QHT1) using the exact KS ground-state electronic density; this approach will be referred to as KS/QHT1.
While the TF-HT approach assumes that $n_0\ro=n^+$, see Fig. \ref{fig:plotjell}a,  the KS ground-state density 
spreads out from the jellium boundary, see Fig. \ref{fig:plotjell}e).
The resulting induced density $n_1\ro$ from the TF-HT model 
is confined inside the jellium boundary, see Fig. \ref{fig:plotjell}b) and c), whereas
there is a significant spill-out in the KS/QHT1 method, see   Fig. \ref{fig:plotjell} g) and h), as recently discussed in 
Ref. \onlinecite{Toscano:2015iw}.
This difference will lead to a different description of the electric field at the surface, which is the key quantity
for plasmonic applications, such as enhancement of the spontaneous 
emission rates\cite{Akselrod:2014ek}, sensing\cite{Chen:2014gn,King:2015gr}, and nonlinear optical effects\cite{Ciraci:2015tp,Argyropoulos:2014bw}.

In the following of the paper, we aim to verify if KS/QHT1 yields correct $n_1\ro$  and plasmon energies, and to investigate
 alternative paths to compute the ground-state density.
\section{Input Ground-state Densities}

The computation of the KS ground-state density is out-of-reach for
all but the smallest systems (computational cost scales as $O(N_e^3)$). 
An alternative, computationally cheaper but less accurate, is 
to use OF-DFT to compute the ground-state density ($n_0^{\tOF}\ro$).
In this case we have to solve the Euler equation \cite{parrbook}: 
%
\begin{equation}
 \frac{\delta T_s}{\delta n_0\ro} +v_{xc}\ro - e\phi\ro=
 \mu^{\tOF} 
\Dabel{eq:0th}
\end{equation}
where $\mu^{\tOF}$ is a constant representing  the chemical potential and $\phi\ro$ is the total (i.e. from both 
electrons and the bare positive background) electrostatic potential. 
The Euler equation (\ref{eq:0th}) can be recast into an eigenvalue equation for the square
root of the electron density \cite{levy84}; if the kinetic energy (KE) is approximated as $T_s^{\tTF}+(1/\eta_g) T_s^W$  it takes the form: 
\begin{eqnarray}
&& \left ( 
\frac{1}{\eta_g}\ \frac{\hbar^2 \nabla^2}{2m_e} + \frac{\delta T_s^{\tTF}   }{\delta n_0\ro} + v_{xc}\ro - e\phi\ro  
\right) \sqrt{n_0\ro} \nonumber \\
&&\;\;\;\;\;\;\;\;  =\mu^{\tOF} \sqrt{n_0\ro}
\Dabel{eq:eulersn}
\end{eqnarray}
which we solved as a self-consistent KS equation (see above), considering only the lowest eigenvalue (with angular momentum $l=0$).
The self-consistent calculation of $n_0^{\tOF}\ro$ can be, in principle, obtained for spherical nanoparticles of any size 
(computational cost is $O(N_e)$), even if we experienced very slow convergence, especially for $\eta=9$.

\def\decay{\kappa}

A third approach is to use a model expression that approximates the exact density. 
For a sphere, the approximated unperturbed electron density can be described by using the model\cite{snidersorbello83,brack89,harbola00,harbola2008}:
\begin{equation}
n_0^{Mod}({\bf r})= \frac{f_0}{1 + \exp{\left ( \kappa^{\tmod} (r - R) \right ) }}
\Dabel{eq:ana}
\end{equation}
where $r$ is the distance from the center of the sphere and $R$ is the radius of the nanosphere. 
The expression (\ref{eq:ana}) has to be normalized such that the total charge equals the total number of electron:
\begin{equation}
4\pi \int_0^{ + \infty } {n_0^{\tmod}(r){r^2}dr}  = \frac{4}{3}\pi {R^3}{n^+}=N_e.
\Dabel{eq:norm}
\end{equation}

This approach, if successful, is particularly useful to compute the spectral response of arbitrary big systems, since it provides the ground-state density without any computational cost.
We underline that Eq. (\ref{eq:ana}) is not employed for 
a variational calculation of the ground-state density\cite{snidersorbello83,brack89,harbola00,harbola2008}. 
Instead we will fix $\kappa_{\tmod}$, which describes 
the asymptotic decay of the electronic density and is the only parameter in Eq. (\ref{eq:ana}), as described in the next section.\\

\section{Asymptotic Analysis}
In the KS or OF approach, 
if  we assume that $v_s\ro=v_{xc}\ro-e\phi\ro$ goes exponentially to zero
(this is the case for a neutral system and using LDA for the XC functional), then the density asymptotically decays as\cite{parrbook}
\begin{equation}
n_0({\bf r})\rightarrow \frac{A}{r^2} \exp(-\kappa r) \; .
\Dabel{eq:n0asy}
\end{equation}
In the OF approach,  if the KE is approximated as $T_s^{\tTF}+(1/\eta_g) T_s^W$  we have:
\begin{equation}
\kappa^{\tOF}=\left(\frac{1}{a_0\sqrt{E_h}}\right) 2\sqrt{-2  \mu^{\tOF}} \sqrt{\eta_g} .
\Dabel{eq:alphadef}
\end{equation}
In the KS approach we have \cite{parrbook}
\begin{equation}
\kappa^{\tKS}=\left(\frac{1}{a_0\sqrt{E_h}}\right)2\sqrt{-2 \epsilon^{\thomo}}
\Dabel{eq:alphaksdef}
\end{equation}
and $\epsilon^{\thomo}$ is the eigenvalue of the highest occupied molecular orbital (HOMO). Note that $\epsilon^{\thomo}<0$ for stable electronic systems
and it coincides with the negative of the ionization potential only for the exact XC-functional\cite{perdewHOMO}. 

The values of $\mu^{\tOF1}$ (\textit{i.e.}, OF-DFT with $\eta=1$), $\mu^{\tOF9}$ (\textit{i.e.}, OF-DFT with $\eta=9$) and $\epsilon^{\thomo}$ for all the jellium nanospheres considered are reported in Fig. \ref{fig:plotmuhomo}.
It is found that  $|\mu^{\tOF9}|$ is only a factor 1.1-1.4 smaller than $\mu^{\tOF1}$.
Thus, unless $\eta_g=1$, we have that the density computed in OF-DFT
is decaying faster than the exact one, as numerically shown in Fig. \ref{fig:plotn0} for a jellium nanosphere with $N_e=338$ electrons.
This is consistent with the fact that the von Weizs\"acker KE approximation
is exact in the asymptotic region \cite{parrbook,dskato15}.

\begin{figure}[hbt]
        \centering
                \includegraphics[width=0.4\textwidth]{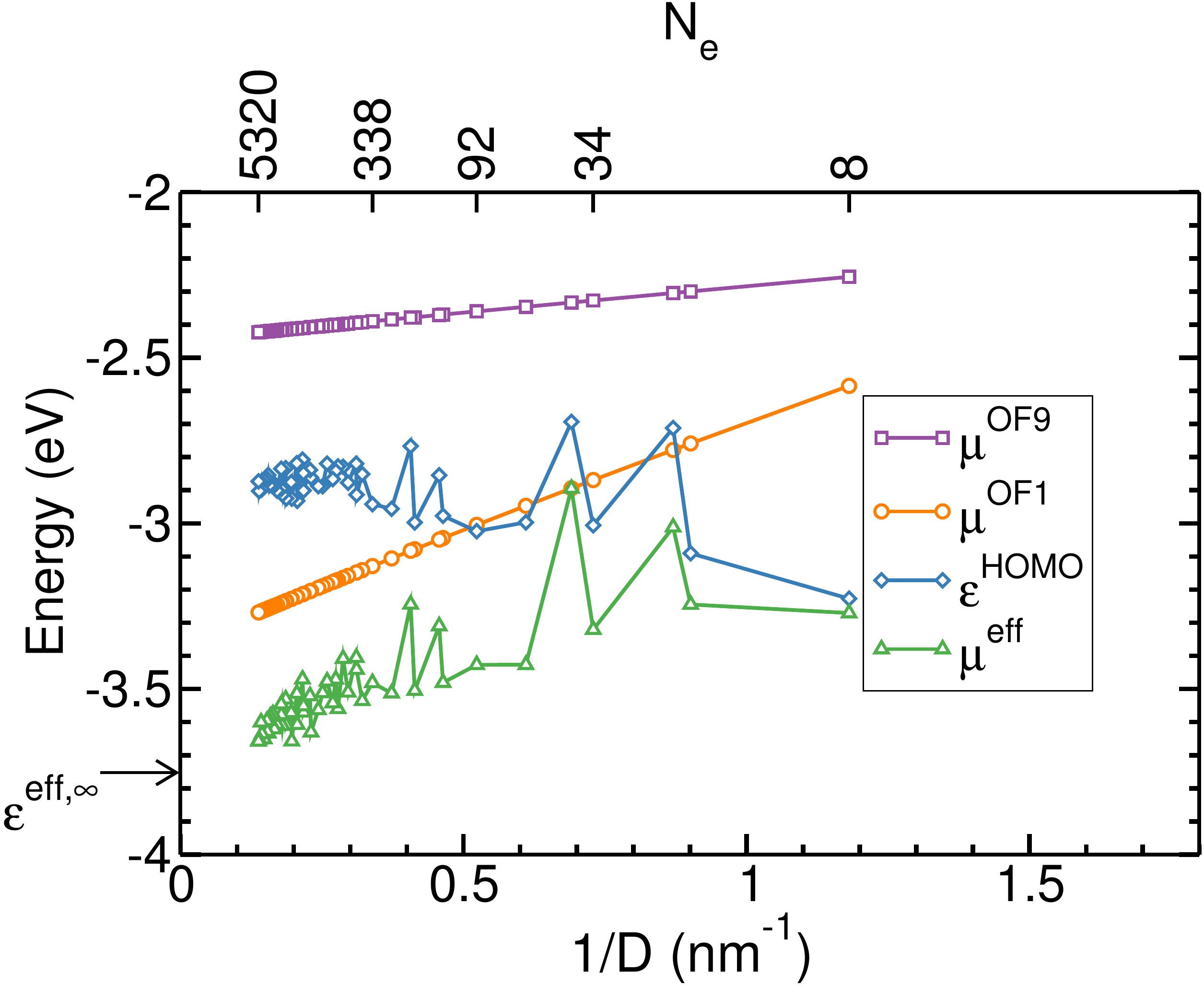}      
        \caption{Eigenvalues versus the inverse of the jellium nanosphere ($r_s=4$ a.u.) diameter;
chemical potential ($\mu^{\rm OF9}$) of orbital-free DFT calculations with $\eta=9$ (purple squares);
chemical potential ($\mu^{\rm OF1}$) of orbital-free DFT calculations with $\eta=1$ (orange circles);
HOMO eigenvalues ($\epsilon^{\thomo}$) of KS-DFT calculations (blue diamonds);
effective eigenvalue ($\mu^{\rm eff}$) from the KS-DFT electronic density decay  (green triangles), see text for details.
}
         \Dabel{fig:plotmuhomo}
\end{figure}
\begin{figure}[hbt]
        \centering
                \includegraphics[width=0.4\textwidth]{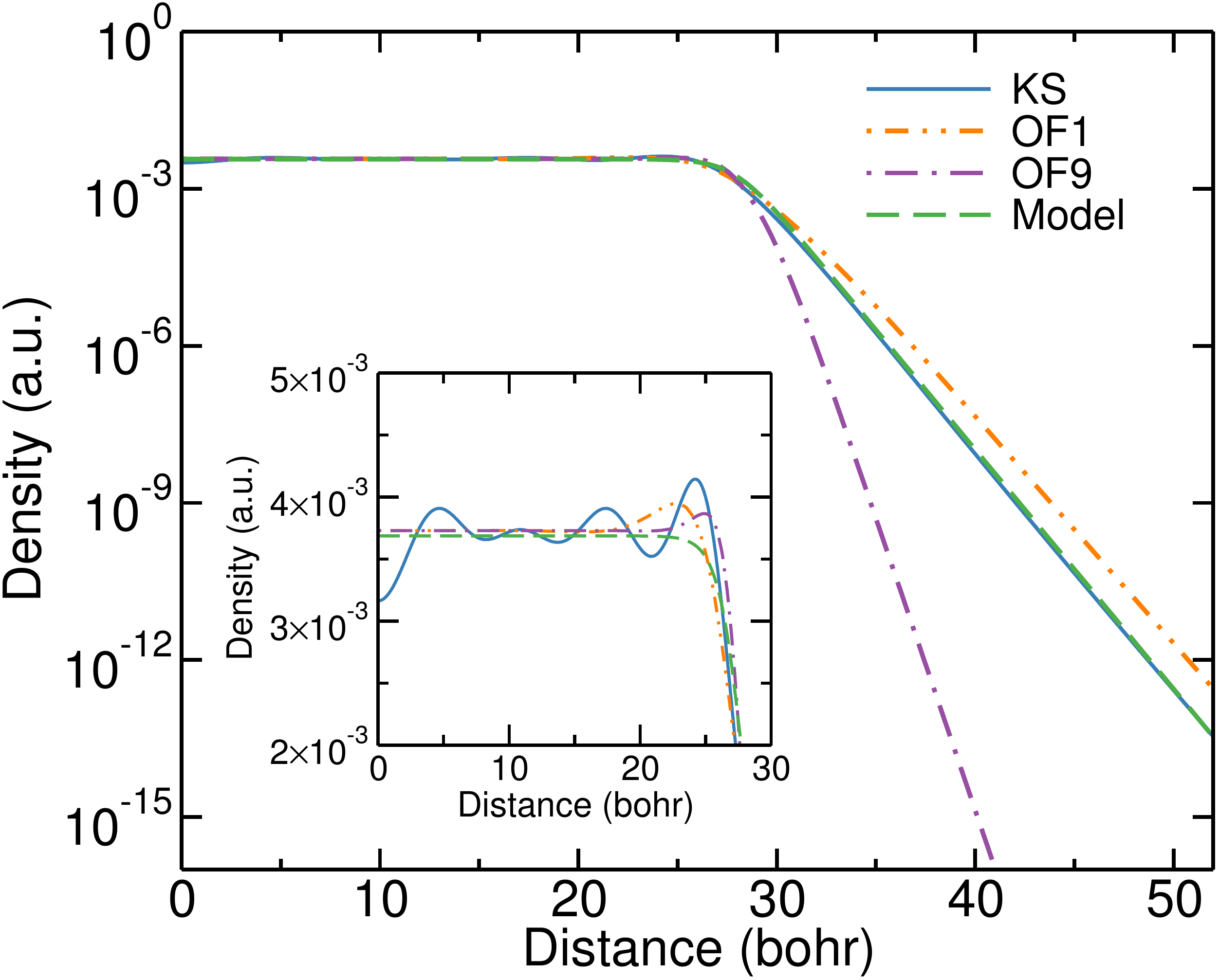}      
        \caption{Ground-state electronic density (in a log-scale)  for a jellium nanosphere ($r_s=4$) with 338 electrons (R=27.8 bohr)  computed
by KS-DFT, OF-DFT with $\eta_g=1$ (OF1) and with $\eta_g=9$ (OF9), and the model density. The inset shows the ground-state electronic density in a linear scale inside the
nanosphere.  
}
         \Dabel{fig:plotn0}
\end{figure}
We remark that Eq. (\ref{eq:alphaksdef}) is valid only in the asymptotic region, i.e. where the density is dominated
only by the HOMO. 
However, in the case of jellium nanospheres, there are several KS orbitals with energies very close
to the HOMO, so that the asymptotic limit will be reached only very far from the jellium boundary, in a region that is not 
relevant for total energies, nor for the optical properties (see Fig. \ref{fig:asymp}).
If in the ``near'' asymptotic region (i.e. within the simulation domain) we assume that the density decays as in Eq. (\ref{eq:n0asy}), 
with $\decay=\decay^\teff$ then we can define an effective energy: 
\begin{equation}
\mu^{\teff}=-(E_h a_0^2) \frac{(\decay^\teff)^2}{8}. 
\Dabel{eq:emod}
\end{equation}
The values of $\mu^{\teff}$ are also reported in   Fig. \ref{fig:plotmuhomo}, and they are clearly larger (in absolute value)
than the $\epsilon^{\thomo}$; the difference increases with the number of electrons, due to the increasing contribution of other (low-lying) orbitals.
For an infinite number of electrons, a linear extrapolation gives $\mu^{\teff,\infty}\approx-3.75 $ eV.
We then use this value to define 
\begin{equation}
\decay^{\tmod}=\left(\frac{1}{a_0\sqrt{E_h}}\right) \sqrt{-8\mu^{\teff,\infty}}\approx 1.05.
\end{equation} 
Figure \ref{fig:plotn0} shows that very good agreement is obtained in the asymptotic region, between the model and the KS density.

\begin{figure}
        \centering
                \includegraphics[width=0.4\textwidth]{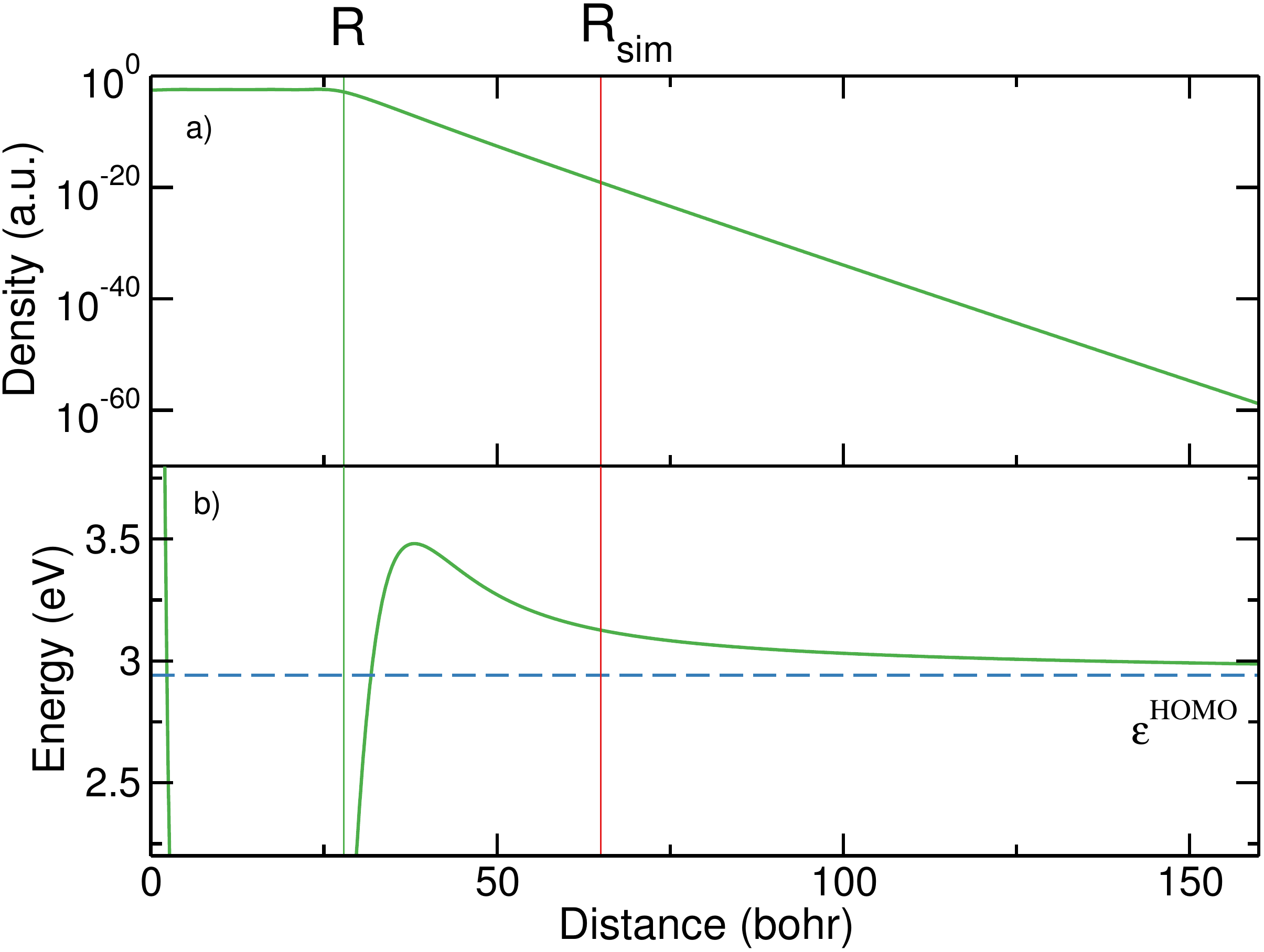}
        \caption{(a) KS Ground-state electronic density $n_0\ro$ for a jellium sphere 
with $N_e=338$ electrons ($R=27.86$ bohr, indicated by the solid-blue vertical line) ;
(b)  Plot of  
$-(E_h a_0^2) \frac{1}{8} \left (\frac{d \ln( r^2n_0(r))}{d r}\right)^2$, which represents the 'local' $\mu^{\teff}$ in 
the case of real density. Only in the far asymptotic region (e.g. for $r>150$ bohr)
 $\mu^{\teff}$ approaches $\epsilon^{\thomo}$ indicated by a horizontal
green dashed line. In the simulation domain, indicated by the vertical red-line, 
 $\epsilon^{\teff}$ is significantly larger than  $\epsilon^{\thomo}$.}
         \Dabel{fig:asymp}
\end{figure}
 
We now move to consider the asymptotic solution of the QHT$\eta$ equations for spherical systems, extending the work in 
Ref. \onlinecite{Yan:2015ff}, where only slabs have been considered, and the early one in Ref. \onlinecite{dreiz82}.
If we assume  the ground-state density decay in Eq. (\ref{eq:n0asy})
then we want to verify if Eq. (\ref{eq:sys}) has solutions of the type
\begin{equation}
n_1({\bf r})\rightarrow B \exp(-\beta r) \cos(\theta)
\Dabel{eq:n1dft}
\end{equation}
hereby limiting our investigation to dipolar excitations.
To proceed, we take the divergence of Eq. (\ref{eq:sys}b), and we use the quasistatic approximation (so that
$\varepsilon_0 \nabla\cdot{\bf E}=\nabla\cdot{\bf P}=e n_1$), obtaining:

\begin{equation}
\nabla\cdot \frac{{e{n_0}}}{{{m_e}}}\nabla {\left( {\frac{{\delta G}}{{\delta n}}} \right)_1} 
 + \left( {{\omega ^2}  } \right) e n_1 
=  - \frac{e^2}{m_e} \left( \frac{e}{\varepsilon_0} n_0 n_1 
+ \nabla n_0 \cdot {\bf E} \right ).
\Dabel{eq:asy1}
\end{equation}
In Eq. (\ref{eq:asy1}) we also assume no damping (i.e. $\gamma=0$) and no external field (i.e. we are considering only free oscillations).
The asymptotic solution of  Eq. (\ref{eq:asy1}) can be easily found considering
that the second term on the left-hand side is proportional to $n_1$: thus all terms which decay
exponentially faster than $n_1$ can be neglected. These are:
the TF and XC contributions in the first term on the left-hand side, which are proportional 
to $n_0^{2/3}n_1$ and $n_0^{1/3}n_1$, respectively (see Eq. (\ref{eq:potentials}a) and (\ref{eq:potentials}c)), 
and the first term to the right-hand side (proportional to $n_0n_1$).
The second term on the right-hand side requires special attention. Asymptotically it decays proportionally to 
 $(n_0 d)/r^3$, where $d$ is the dipole moment of $n_1$.
Thus Eq. (\ref{eq:asy1}) has an asymptotic solution only 
and only if $n_0$ decays faster than $n_1$, i.e.
if 
\begin{equation}
\beta <  \decay .
\Dabel{eq:bma}
\end{equation}
Using Eqs. (\ref{eq:n0asy}) and (\ref{eq:n1dft}) in Eq. (\ref{eq:asy1}), we obtain (after some algebra) 
that Eq. (\ref{eq:asy1}) is asymptotically satisfied if 
\begin{equation}
\left( \frac{E_h a_0^2}{m_e} \right) \frac{1}{\eta} \left(\frac{\decay^2 \beta^2}{4} +\frac{\beta^4}{4}
 -\frac{\decay\beta^3}{2}\right)= \omega^2  \;\; . 
\end{equation}
This equation has four solutions of the type:
\begin{equation}
\beta=\frac{\decay}{2} \pm \frac{\decay}{2} \sqrt{ 1\pm \frac{\omega}{\omega_c} } 
\Dabel{eq:allsol}
\end{equation}
where the critical energy is: 
\begin{equation}
\hbar\omega_c =
\hbar\frac{\kappa^2}{ 8} \sqrt{\left( \frac{E_h a_0^2 }{m_e}\right) \frac{1}{\eta}}=
|\mu| \frac{\eta_g}{\sqrt{\eta}}
\Dabel{eq:ceq}
\end{equation} 
where we used Eq. (\ref{eq:alphadef}). 
Note that it turned out that these solutions are identical to the slab case \cite{Yan:2015ff}.

The four solutions are shown in Fig. \ref{fig:plotbeta}.
\begin{figure}
        \centering
                \includegraphics[width=0.4\textwidth]{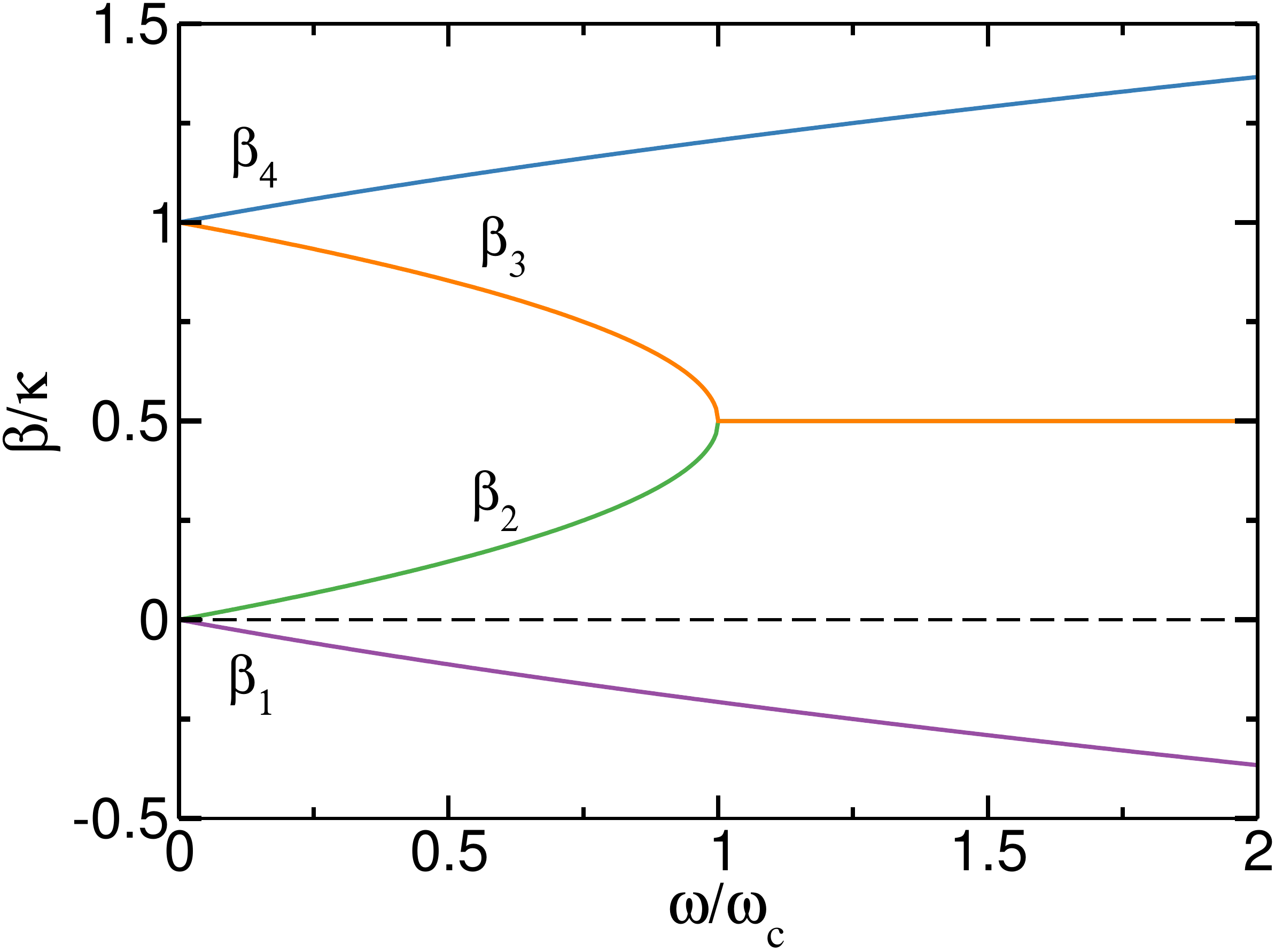}
        \caption{Graphical representation of the solutions in Eq. (\ref{eq:allsol}), see text for details.}
         \Dabel{fig:plotbeta}
\end{figure}
Solution $\beta_1$ is negative, i.e. it is asymptotically increasing, thus it is excluded
by the boundary conditions.
Solution $\beta_4$ is excluded by the condition in Eq. (\ref{eq:bma}).
Solution $\beta_2$ and $\beta_3$ are real only for $\omega<\omega_c$.
For $\omega>\omega_c$, $\beta_{2,3}$ are complex with the real part 
fixed to $\decay/2$.
The above results are consistent with the TD-DFT calculations of finite systems 
($\eta=\eta_g=1$), where $\hbar\omega_c=\epsilon^{\thomo}$ can be interpreted
as the ionization threshold \cite{casida98}.
In fact, in TD-DFT the computation of excitation energies higher than $\epsilon^{\thomo}$ (i.e. the plasmon peak, too) can be challenging because all the
continuum of virtual orbitals must be accurately described. 
In the same way the spectra calculated within QHT are well convergent up to energy $\hbar\omega_c$.

When the sphere is excited by photons with an energy larger than $\hbar\omega_c$,
we experienced a large dependence on the domain size. 
This is due to the fact that the induced charge density acquires a propagating characteristic 
typical of electrons in vacuum. 
The boundary condition we used (${\bf P}={\bf 0}$) is no longer valid since it produces 
an artificial scattering of the electrons at the simulation boundary and appropriate boundary conditions should be developed\cite{dreiz82,abc}. 
For the jellium nanospheres considered in this work we have that $\mu^{\teff}\approx$ 3.5 eV (see Fig. \ref{fig:plotmuhomo})
 which is a bit above the Mie energy $\hbar\omega_{Mie}=(E_h)\sqrt{1/r_s^3}=$3.4 eV.
Numerically we found that only the calculation of the main (first) plasmon peak is stable.\\

\section{Plasmon Resonance and Spill-out Effects.}
In Fig. \ref{fig:plot338} we plot the absorption cross-section, $\sigma$, normalized to the geometrical area $\sigma_0=\pi R^2$,   
for a Na jellium sphere ($r_s=4$ a.u.) with $N_e=338$ and thus $R=27.86$ a.u. ($D=2.94$ nm), using different approaches.

Figure \ref{fig:plot338}a) reports the  reference TD-DFT  results.
TD-DFT calculations (in the adiabatic LDA) have been performed using an in-house developed code, following the literature \cite{zangsoven80,ekardt85,bertsch90,prodan2002}. 
Details of our TD-DFT numerical implementation, which allows calculations for large nanospheres will be discussed elsewhere. 
In TD-DFT (where no retardation is included) the absorption cross-section can be computed as:
\begin{equation}
 \sigma(\omega)= \frac{\omega}{c \varepsilon_0} {\rm Im}\left [ \alpha_{zz}(\omega)\right ] \;\; , 
\end{equation}
where the frequency dependent polarizability is:
\begin{equation}
\alpha_{zz}(\omega)=-   e^2 \int {\rm d}{\bf r} {\rm d}{\bf r}' z \chi({\bf r},{\bf r}',\omega) z'  \;\; , 
\end{equation}
and  $\chi({\bf r},{\bf r}',\omega)={\delta n\ro}/{\delta (e V_{ext}({\bf r}'))}$ is the interacting density response function 
\cite{tddftbook}.

Panels b), c) and d) of  Fig. \ref{fig:plot338} report the QHT absorption cross section computed as: 
\begin{equation}
 \sigma(\omega) 
= \frac{1}{I_0}\frac{\omega}{2} \int {{\mathop{\rm Im}\nolimits} \left\{ {{\bf{E}} \cdot {{\bf{P}}^*}} \right\}dV}
\end{equation}
where $I_o$ is the energy flux of the incident plane-wave.



Panel \ref{fig:plot338}b) shows the spectrum obtained by applying the QHT method with $\eta=9$ (QHT9) to 
the OF9 density; this approach will be called self-consistent OF9/QHT9 and coincides with the approach 
of Toscano \textit{et al.}\cite{Toscano:2015iw}, with the only difference being the choice of the 
XC functional.
The energy position of the first peak ($\approx$ 3.2eV) is in very good agreement 
with the TD-DFT result ($\approx$ 3.15eV). Obviously TD-DFT results
are much broadened due to quantum size effects as $N_e$ is quite small.
However, the decay of $n_1$ is very different from the reference TD-DFT results. In fact
from Eq. (\ref{eq:ceq}) we obtain that   $\hbar \omega_c=3 |\mu^{{\tOF}9}|$. From Fig. \ref{fig:plotmuhomo} 
we see that $\mu^{{\tOF}9}\approx -2.4$ eV and thus the critial frequency is artificially moved 
to very high energy ($\hbar \omega_c\approx 7.2$ eV).
A good point of the OF9/QHT9 approach is that the computation of all the spectrum  (i.e. up to 7.2 eV) 
will be numerically stable. 
On the other hand, from Eqs. (\ref{eq:allsol}) and (\ref{eq:alphadef}) we see that the first solution will have a decay 
$\beta^{{\tOF}9/{\rm QHT}9}\approx 0.87\decay^{{\tOF}9}\approx 2.61 \decay^{\tKS}$, i.e. much more confined
than the reference TD-DFT results, as numerically shown in Fig. \ref{fig:tail}.
Recall that in TD-DFT the induced density will also decay as in Eq. (\ref{eq:n1dft}) with $\decay^{\tKS}/2 \le \beta \le \decay^\tKS$ as discussed in Ref. \onlinecite{Yan:2015gx}. 
The so-called spill-out effects in computational plasmonics, which indeed refer to the profile of induced density, are thus largely
underestimated in the  OF9/QHT9 approach. 
Thus the good accuracy of the OF9/QHT9 resonance energy seems originating from error cancellation between
between the too confined ground-state electron density (from OF9) and the approximated kinetic energy-kernel (QHT9, which is valid only for slowly varying density). 

In Fig. \ref{fig:plot338}c) we report the results from the KS/QHT1 approach, already introduced in Fig. \ref{fig:plotjell}.
In this case the resonance peak ($\approx$ 3.13 eV) is in even better agreement with the TD-DFT results.
Almost the same results are obtained by applying the QHT1 method  to the model ground-state density (Mod/QHT1), as shown in Fig. \ref{fig:plot338}d). 
More importantly Fig. \ref{fig:tail} shows that the induced density $n_1$ from KS/QHT1 has almost the same decay of the TD-DFT result, i.e.  the  
KS/QHT1 approach correctly describes the spill-out effects of the induced charge density.

It is useful to remark that, despite the fact the all the spectra presented in Fig. \ref{fig:plot338}b-d) result quite similar, QHT is very sensible to the density tail.
Using a model density with a larger (smaller) $\decay^{\tmod}$ yields a red-shifted (blue-shifted) plasmon peak. 
In a similar way, using the QHT1 method with the OF1 ground-state density 
yields a plasmon 
peak red-shifted by about 0.3 eV (data not reported). This is not surprising considering that the 
OF1 density is decaying much more slowly than the effective one, see Fig. \ref{fig:plotmuhomo}.
As mentioned at the end of Section V, spectral features appearing at energies higher than the ionization energy ($\hbar\omega_c$) are not stable and will be investigated elsewhere.  
We also point out that QHT with $\eta=9$ 
(which yields the exact dielectric response for small wavevectors\cite{akbari15})
 cannot be used in combination with a ground-state density with the exact asymptotic decay.
This seems surprising, but it can be easily justified by looking at Eq. (\ref{eq:ceq}).
If $\eta_g=1$ and $\eta=9$ we obtain $\hbar\omega_c=|\mu^\teff|/3\approx 1.16 eV$, i.e.
the critical frequency is three times smaller than the Mie frequency, so that
the whole absorption spectrum can be hardly computed.


\begin{figure}
        \centering
                \includegraphics[width=0.4\textwidth]{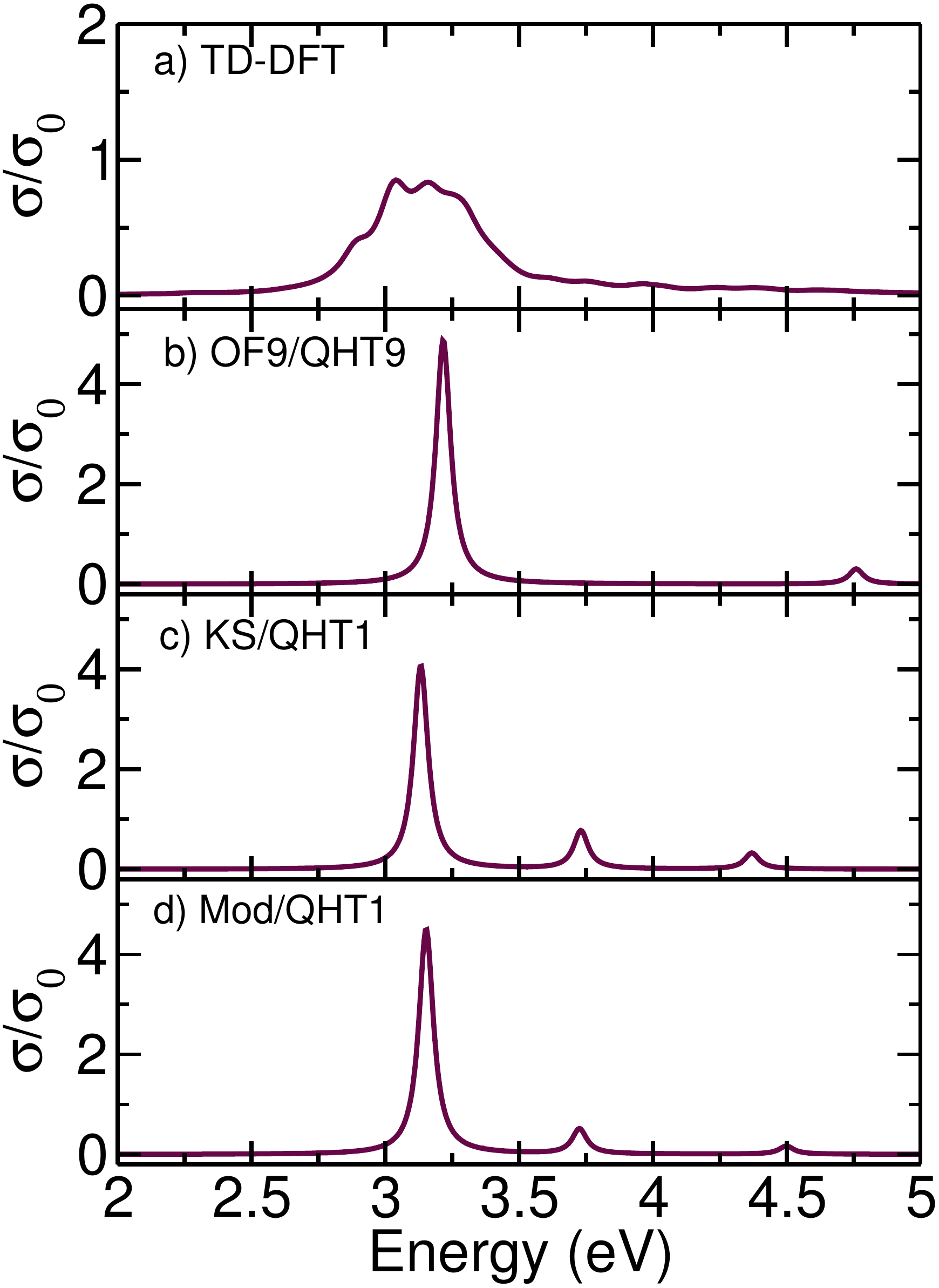}     
        \caption{Absorption cross section ($\sigma$) normalized to the geometrical area ($\sigma_0$) for a 
jellium nanosphere ($r_s=$ 4 a.u.)  with $N_e=338$ electrons as obtained form TD-DFT, OF9/QHT9, KS/QHT1 and Mod/QHT1. All the spectra have been obtained using an empirical broadening of 0.066 eV.}
         \Dabel{fig:plot338}
\end{figure}

 \begin{figure}
        \centering
                \includegraphics[width=0.4\textwidth]{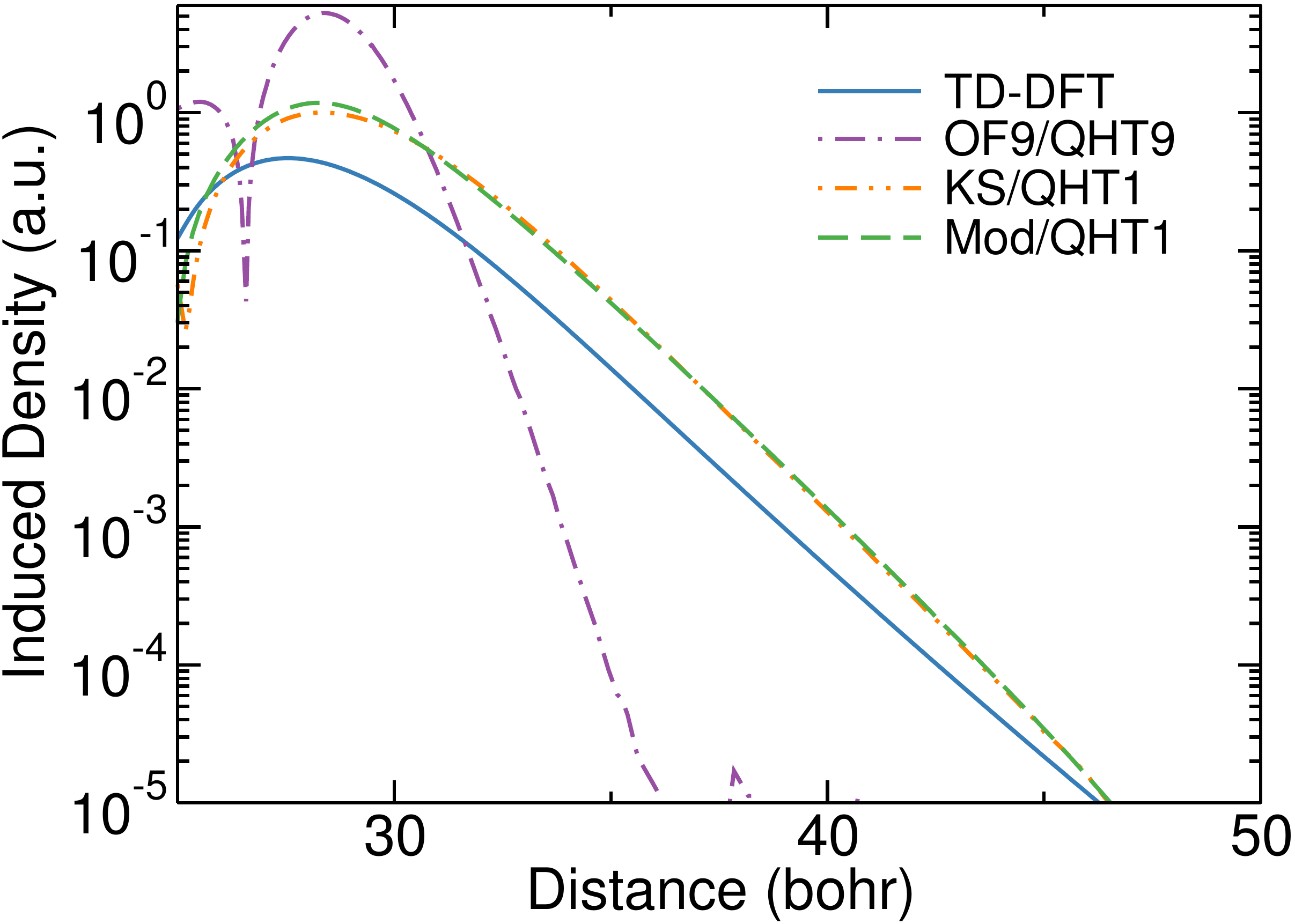}
        \caption{ Induced density (complex modulus of $n_1$) at the plasmon energy 
        for a jellium nanosphere ($r_s$=4 a.u.) with $N_e=338$ electrons ($R=27.86$ bohr), as computed
        from  TD-DFT, OF9/QHT9, KS/QHT1 and Mod/QHT1. 
}
         \Dabel{fig:tail}
\end{figure}


%
%
%



\begin{figure}[hbt]
        \centering
                \includegraphics[width=0.4\textwidth]{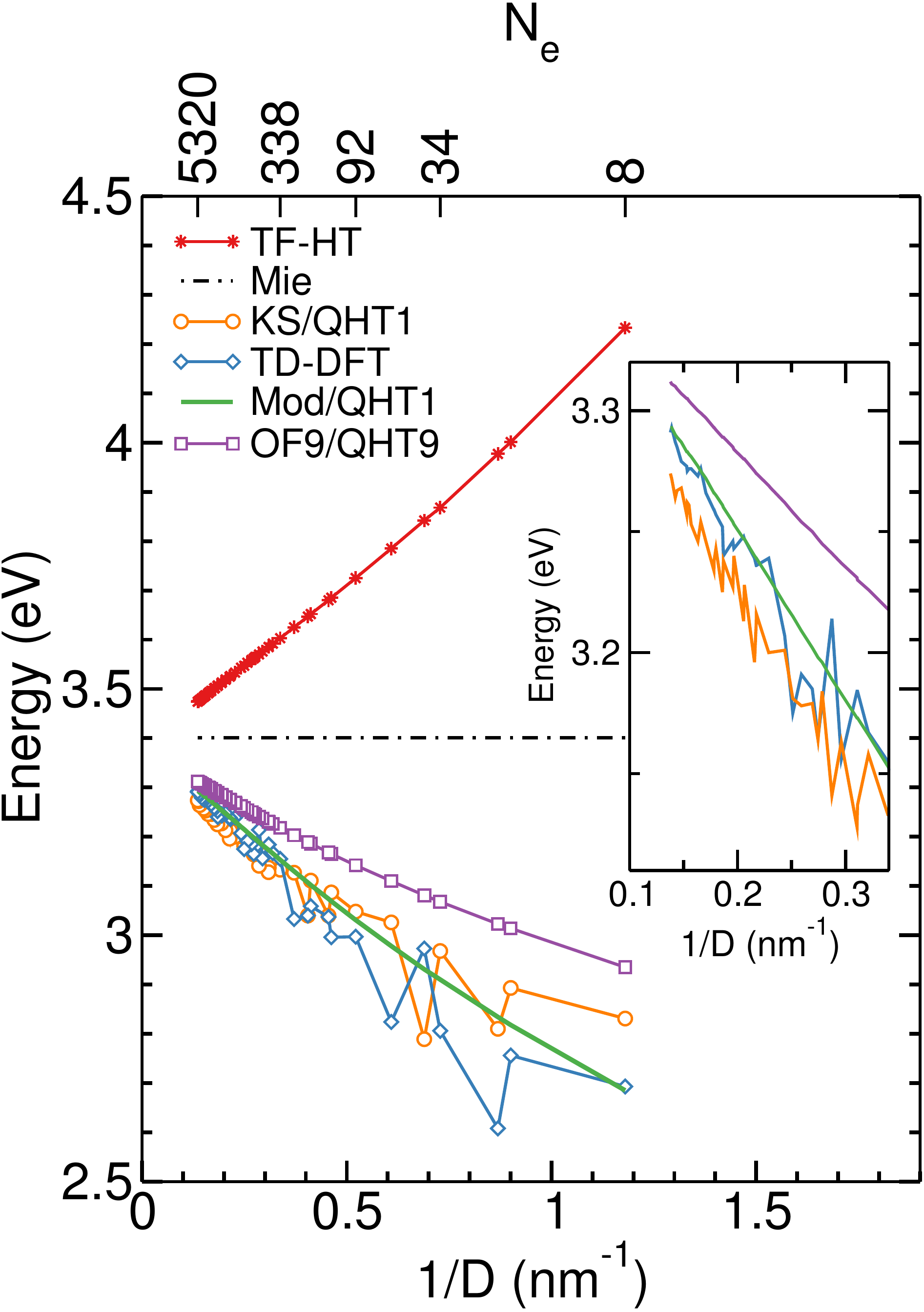}
        \caption{
         Plasmon resonance for jellium nanospheres ($r_s$=4 a.u.) as a function of the inverse of the sphere diameter, as computed by different approaches.
         The behavior for large particles is shown in the inset.
}
         \Dabel{fig:main}
\end{figure}




We now move on to describe the  shift of the plasmon resonance as a function of the particle size. This problem
 has been extensively studied in the literature,
both theoretically\cite{ruppin78,yanno93} and experimentally\cite{Reiners:1995gq,Parks:1989ui,Scholl:2012dsa}, and 
it represents a relevant benchmark for estimating the accuracy of QHT.
In Fig. \ref{fig:main} we report the energy position of the main resonance peak as a function of 
the inverse of the jellium nanosphere diameter, $D$.
The exact energy position of the main peak has been extracted from the computed spectra 
(with an empirical broadening of $\hbar\gamma=0.1$ eV and  using a spline interpolation).
These procedure is not unique for some of the smallest clusters, where there are many peaks with similar intensity (for large cluster there is always an unique main peak). 

The dot-dashed horizontal line represents the Mie plasmon energy  ($\hbar\omega_{Mie}=3.4$ eV). Note that for the
diameters considered in  Fig. \ref{fig:main}, retardation effects can be neglected.

The first thing to notice is the striking difference obtained using TF-HT. It predicts in fact a resonance shift toward 
higher energies (shorter wavelengths) as the particle radius gets smaller, as previously observed in other systems\cite{Raza:2011io,Toscano:2012fh,Raza:2013gn}.
For all the other cases
the peak resonances slide to lower energies (longer wavelengths) as the particle radius shrinks.
 While for noble metals like Au or Ag the plasmon resonance undergoes a blue shift as the radius $R$ decreases\cite{bore86,lerme98}, this is not the case for Na.
 The origin of the blue shift for noble metal nanoparticles
is due to size dependent changes of the optical interband transitions \cite{lieb93,apell2013}.

We now compare the QHT models investigated in this work, with respect to
the TD-DFT results, which can be considered as a reference.
As widely investigated in the literature for jellium nanospheres\cite{yanno93}, the TD-DFT main peak 
oscillates for small $N_e$, but it converges to $\hbar\omega_{Mie}$ for large $N_e$.

Results for the OF9/QHT9 approach are significantly blue-shifted with respect to TD-DFT (note that OF9/QHT9 predicts a red-shift with respect to TF-HT, as also found in Ref. \citenum{Toscano:2015iw}), and do not present quantum oscillations.  
In fact, it is well known\cite{PhysRev.136.B864} that orbital-free (OF1 or OF9) electronic density does not
show quantum (i.e. Friedel) oscillations inside the nanosphere (see inset of Fig. \ref{fig:plotn0}).
On the other hand when QHT1 is applied to the KS density, quantum-oscillations are clearly visible 
for small nanospheres, even if TD-DFT features are not fully reproduced. 
For $N_e\ge 338$, KS/QHT1 reproduces TD-DFT plasmon energies with great accuracy (with a maximum error of 20 
 meV, about a half of the error obtained with the OF9/QHT9 approach, see Table S1 in the Supplemental Material).

\begin{figure}
        \centering
               \includegraphics[width=0.45\textwidth]{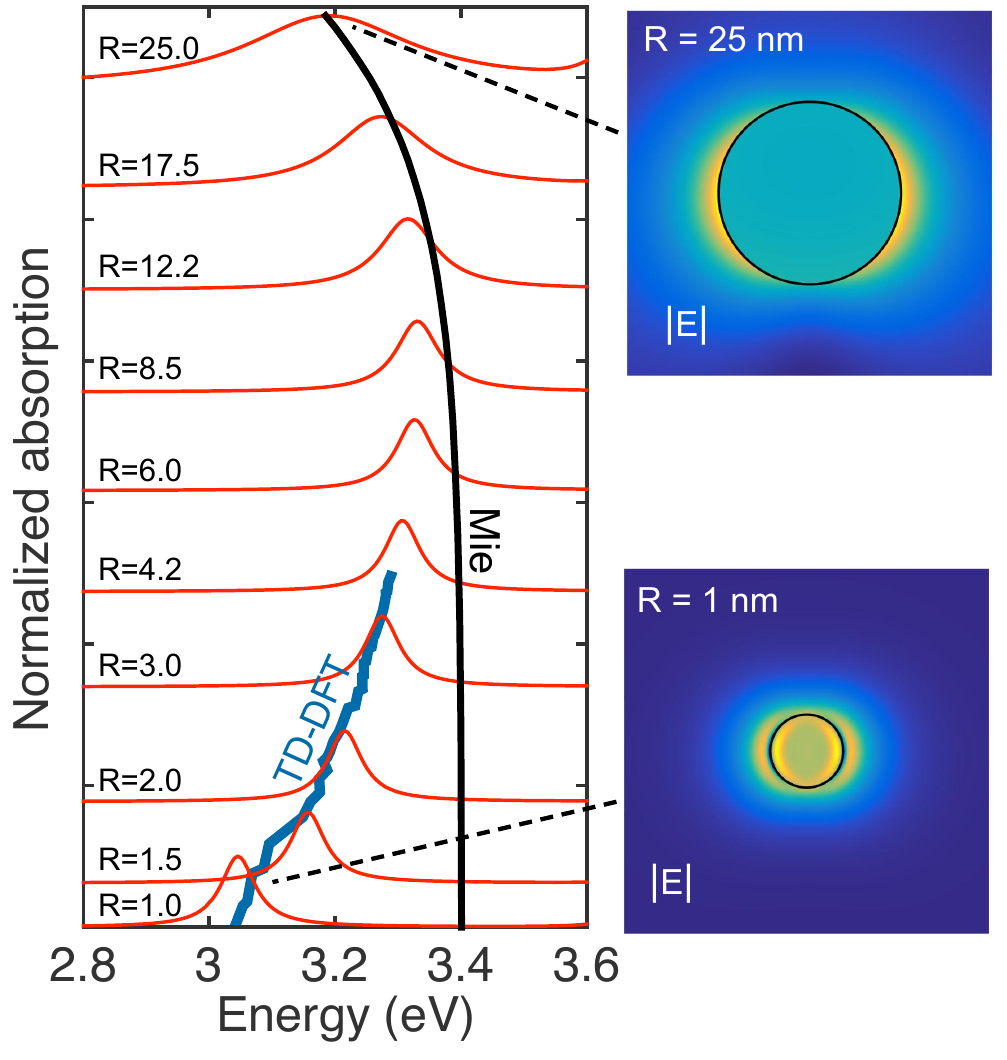}
         \caption{Mod/QHT1 spectra for jellium nanospheres ranging from $R=2$ to 25 nm (red curves). 
                  The black (blue) curve shows the Mie (TD-DFT) resonance trajectory.}
         \Dabel{fig:big}
\end{figure}

Finally, we analyze results for Mod/QHT1. Also in this case no quantum-oscillations 
are present, and for $N_e\ge 338$ TD-DFT results are reproduced almost exactly, 
with a maximum error of only 10 meV.
Thus using a simple model density it is possible to match the whole range of nanoparticle sizes. 
The comparison between TD-DFT and KS/QHT1 is important because both approaches use the same KS density (in the former 
additional information is used from the KS orbital and eigenvalues). The good accuracy in Fig. \ref{fig:main} means that for large
nanospheres the full TD-DFT linear response can be well approximated by the simpler QHT1 method.
The very good results obtained for Mod/QHT1 are even more important. In fact
it means that QHT 
can be use without the need of calculating the KS ground-state density, which is a bottleneck for large system (scales as $O(N_e^3)$).
Moreover, the simple model density employed here can be constructed at no cost for systems of any size, 
and can also be generalized to the non-spherical case.
The parameter $\decay^{\tmod}$, is clearly material-dependent (e.g. will depend on $r_s$) but can be parameterized once and for all. 

In Fig.~\ref{fig:big} we show the absorption spectrum for particles with a diameter going from 2 up to 50 nm.
The solid black line shows the trajectory of the Mie resonance as the particle size increases: for particles with $R>10$ nm
the Mie resonance peak undergoes a red-shift due to retardation effects.
 The red curves represent the spectrum calculated within the Mod/QHT1 method.
For small nanoparticles the peaks follow the TD-DFT trajectory, moving toward higher energies. As the particle size grows the plasmon energy tends toward the Mie trajectory, up to the big particle regime, where retardation effects become predominant (see Fig. \ref{fig:main}).
It is striking how Mod/QHT1 can describe the full range of effects going from the nonlocal/spill-out effects up to retardation 
effects. That is, the resonance shift due to microscopic and macroscopic effects are incorporated in a single model, which makes the 
potential of QHT with respect to DFT approach very clear.
Although this advantage was already outlined by the authors of Ref. \citenum{Toscano:2015iw}, we remark that their method was only qualitatively verified for nanowires.\\

\begin{figure}
        \centering
                \includegraphics[width=0.4\textwidth]{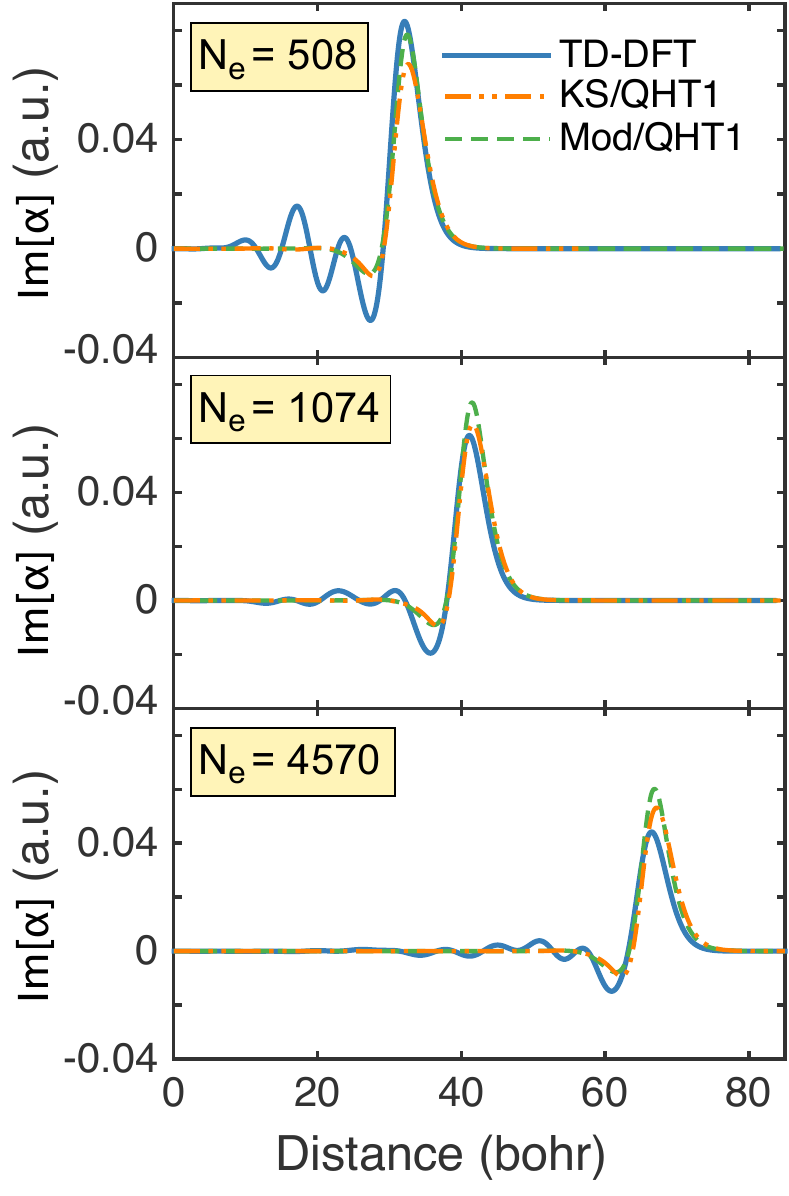}
        \caption{Induced polarization charge density (normalized by $R^3$) at the plasmonic resonance $\omega_0$ for different jellium nanospheres.}
         \Dabel{fig:rhoind}
\end{figure}

\section{The induced charge density}\Dabel{sec:rhoind}
So far we have seen that KS/QHT1 or Mod/QHT1  
reproduce with very good accuracy the reference TD-DFT results, both the energy position and the asymptotic decay of the 
induced density.
In this section we closely analyze the near-field properties. Accurate induced density translates into 
a good description of the local fields at the surface of the plasmonic system. Such knowledge is 
crucial for estimating the maximum field enhancements, and hence nonlinear optical efficiencies, and more in general light-matter interactions.

In particular we compare the QHT1 induced polarization charge density  to the full TD-DFT calculations.
The polarization charge density $\alpha_{zz}(r)$ of a 
sphere excited by the incident field ${\bf E}_{0}={\bf \hat z}E_0$ can be defined from:
\begin{equation}
n_1(r,\theta,\phi)= - e E_0 \frac{1}{r^2} \cos(\theta) \alpha_{zz}(r,\omega) 
\Dabel{eq:alphardef}
\end{equation}
so that
\begin{equation}
\alpha_{zz}(\omega) =\frac{4\pi}{3} \int_0^{ + \infty } {\alpha_{zz} (r,\omega)} r dr.
\Dabel{eq:alpharint}
\end{equation}

In Fig.~\ref{fig:rhoind} we plot the imaginary part of the induced polarization charge 
density for different particles sizes in correspondence with the plasmon resonance $\omega_0$.    
For the smaller particles big oscillations of the density can be seen in the case of the 
TD-DFT calculations that are not present if not in a very modest from in the case of QHT1. 
These oscillations are in fact due to a purely quantum size effect (Friedel oscillations). 
As the particle size increases however, these oscillations diminish.
The main induced peak, however, is very well reproduced by the QHT1 approach, both with the KS or the model ground-state density. 

\begin{figure}
        \centering
                \includegraphics[width=0.43\textwidth]{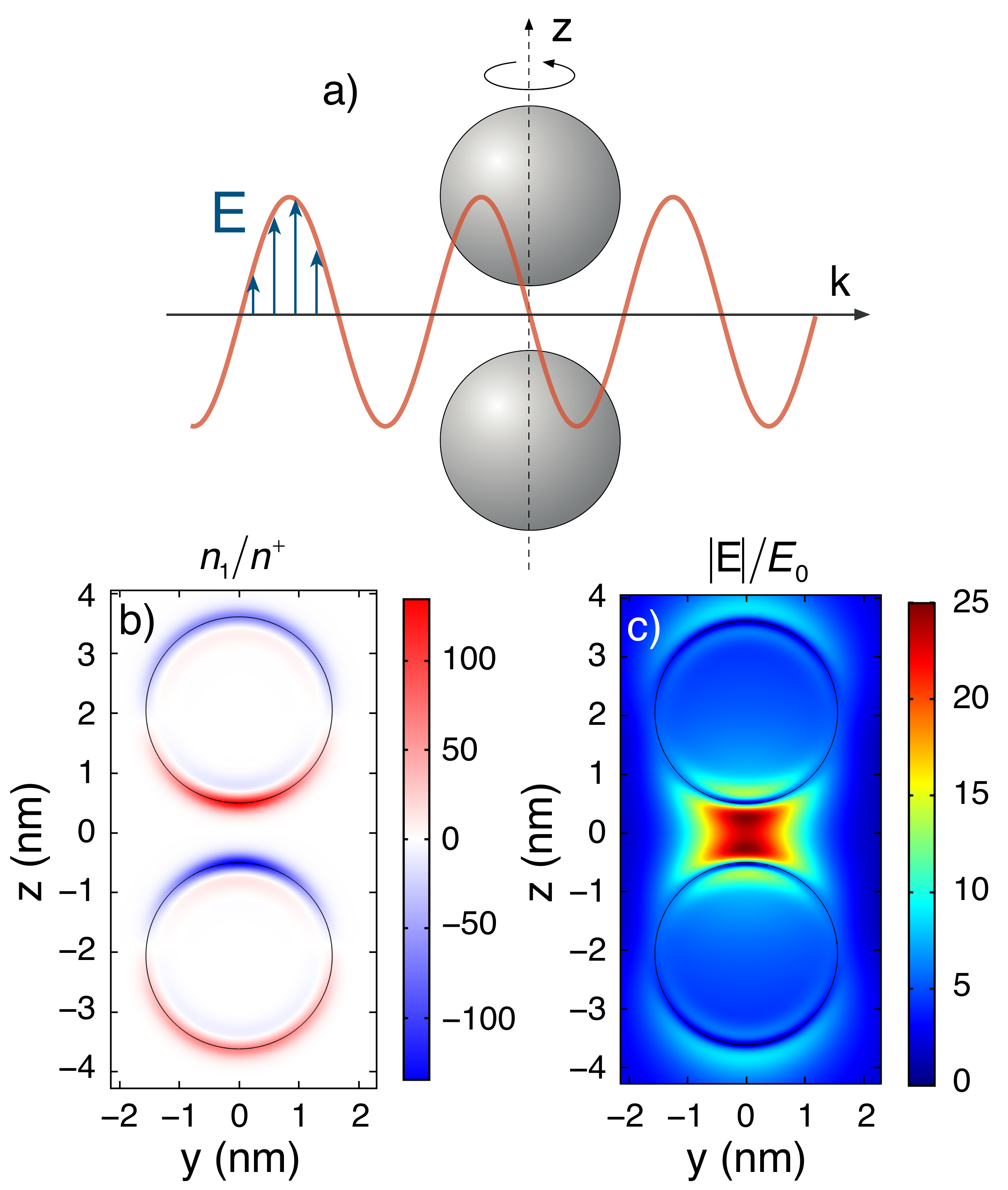}
        \caption{Dimer of Na spheres constituted by $N_e=398$ electrons each, and separated by a distance $g=1$ nm.
The dimer is excited by a plane wave oscillating with energy $\hbar\omega=-2.8$ eV, $\gamma=\gamma_0+v_F/R$ has been used with $\hbar\gamma_0=0.066$ eV. (a) Geometry and incident field. (b) Real part of the induced charge density normalized with respect to the bulk positive charge density. (c) Electric field norm distribution. }
         \Dabel{fig:dimer}
\end{figure}

\section{Application to the sphere dimer}
Up to this point we have considered spheres. In this section we are going to extend the applicability of QHT to axially symmetric structures.
Our 2.5D implementation (see Appendix \ref{sec:ni}) makes this task really easy and the only difference is in the excitation field.
Because the response of a spherical system is independent from the direction and polarization of the incident wave, in the previous calculations we have assumed for convenience a plane wave propagating along the $z$ axis so that we would need to solve our equations just for the cylindrical harmonic with azimuthal number $m=1$ (the case for $m=-1$ can be obtained by taking into account field parities).
For axis-symmetric structures, in general, it is not possible to arbitrarily choose the incident wave  and it becomes necessary to solve the problem for several azimuthal numbers.
For sub-wavelength structures, however, the number of cylindrical harmonics $m_{max}$ needed to accurately describe the problem remains very small ($m_{max}<3$).  

A relevant example of axially symmetric structure is the sphere dimer.
This structure has been extensively studied in the literature for its ability to strongly enhance local electric fields with respect to the incident radiation\cite{Romero:2006ip,FernandezDominguez:2012ih,FernandezDominguez:2010dq} and its potential to exhibit quantum effects\cite{Zuloaga:2009gm,Marinica:2012,barbry2015}.
Here, we consider a dimer of Na spheres constituted by $N_e=398$ electrons each, and separated by a distance $g=1$ nm.
The dimer is excited by a plane wave propagating orthogonally to the dimer axis whose electric field is polarized along $z$ (as depicted in Fig.~\ref{fig:dimer}a) and is oscillating with energy $\hbar\omega=-2.8$ eV. 
In Fig.~\ref{fig:dimer}b-c we plot the induced charge density and the electric field norm respectively for the case of Mod/QHT1 where the same equilibrium charge density as the single particle case has been used.
Although, our TD-DFT implementation can only be applied to spheres, Fig.~\ref{fig:dimer} and in particular Fig.~\ref{fig:dimer}c can be directly compared to results of Ref.\citenum{barbry2015} in which TD-DFT calculations for the same jellium Na dimer are reported.
It can be seen that the electric field distribution in the gap and in the vicinity of the jellium edge is accurately reproduced both qualitatively and quantitatively.

It is worth noting that results obtained with Mod/QHT1 are valid as long as the equilibrium electron density of a dimer can be approximated to that obtained by summing the densities of two single spheres.
For very small gaps ($g<0.4$ nm) this is not necessarily true and particular attention needs to be paid to the choice of the equilibrium density in the overlapping region.

\section{Conclusion}
In this work, we have investigated how different models based on QHT
can describe the plasmonic properties of spherical nanoparticles in comparison with reference TD-DFT results.

The main finding is twofold: 
\begin{itemize}
\item[i)] The accuracy of QHT strongly depends on the choice of the ground state density.
In particular the self-consistent approach with $\eta=9$ produces correct results
only for the plasmon resonance energy, whereas the induced density spill-out is largely underestimated. 
Using the exact KS electron equilibrium density within the QHT method with the full von Weizs\"acker kinetic energy, 
allows to predict the plasmon energy for Na jellium nanospheres within 
an error of about 20 meV in comparison with TD-DFT predictions.
\item[ii)] QHT yields similar high accuracy (with a maximum error of only 10 meV) if an 
analytical model density with the correct asymptotic behavior is used (the Mod/QHT1 approach).
This finding is of utmost importance because 
{\it it allows to circumvent the bottleneck given by the necessity of computing the exact KS
 ground state density}, allowing the QHT method to be directly applied to macroscopic systems that still require a precise 
microscopic description, such as gap-plasmon structures\cite{Savage:2012by,Hajisalem:2014cr,Ciraci:2012fp}.
\end{itemize}
By using a finite-element implementation based on the 2.5D technique we were able to investigate spherical nanoparticles under 
a plane-wave excitation and extend our calculations to big particles maintaining retardation effects.
We showed that our implementation can be used to study nanoparticle dimers and in general can be applied to arbitrary geometries that possess axial symmetry, such as cones, nanoparticle dimers, disks, or film-coupled nanoparticles. These systems are in fact quite frequent in experimental setups.
  
We believe that Mod/QHT1 is quite promising and can be further improved by adding extra terms\cite{Yan:2015ff} and more accurate 
kinetic energy functionals\cite{laricchia14}  in order to be  more reliable toward UV frequencies and for describing 
valence electrons in noble metals, for which a dynamical kinetic energy functional might be necessary. 
Although the QHT approach is not suitable to directly describe interband processes, these can be approximately taken into 
account by considering a local polarizability contribution \cite{lieb93,Toscano:2015iw}.
Moreover, QHT can be straightforwardly generalized to higher order terms so that nonlinear optical 
effects that are generated at the surface of a plasmonic system can be included in the calculations.



\appendix
\section{Numerical implementation of the QHT}\Dabel{sec:ni}

We solved the system of Eqs. (\ref{eq:sys}) using a commercially available software based on Finite-element method (FEM): \textsc{Comsol} Multiphysics\cite{comsol}.
The problem $L{\bf u} ={\bf 0}$, where $L$ is a linear differential operator and ${\bf u}$ the independent variable vector can be described by means of the weak formulation:
\begin{equation}
\int {{L_1}{\bf{u}} \cdot {L_2}{\bf{v}}} dV = 0,
\Dabel{eq:weak1}
\end{equation}
where ${\bf v}$ is a test function and the operators $L_1$ and $L_2$ are linear operators containing derivatives of order smaller than $L$.
In general, it is possible to go from $L$ to $L_1$ and $L_2$ simply by integrating by parts.
In FEM this step is necessary since one wants to keep the functions ${\bf u}_i$ approximating the
 solution ${\bf u}\simeq\sum_i \alpha_i{\bf u}_i$ as simple as possible.

In the case of the Eq. (\ref{eq:sys}b) we obtain integrating by parts and assuming the integral on the
 boundary to be equal zero the following weak expression:
 \begin{widetext}
\begin{equation}
 \int { -{\frac{e}{{{m_e}}}{{\left( {\frac{{\delta G}}{{\delta n}}} \right)}_1}
\left( {\nabla  \cdot {\bf{\tilde P}}} \right) + \frac{1}{{{n_0}}}\left[ {\left( {{\omega ^2} + i\gamma \omega } \right){\bf{P}} 
+ {\varepsilon _0}\omega _p^2{\bf{E}}} \right] \cdot {\bf{\tilde P}}} } dV = 0,
\Dabel{eq:weak}
\end{equation}
 \end{widetext}
where we distributed the derivatives to the test functions ${\bf{\tilde P}}$. This allows us to avoid calculating the gradient of the energy functional of Eq. (\ref{eq:fop}).
However, since the expression of the energy functional contains second order derivatives, we introduce 
the working variable ${\bf F}=\nabla n_1$ with $n_1=\frac{1}{e}\nabla\cdot{\bf P}$, so that $\nabla^2n_1=\nabla \cdot {\bf F}$,  
and our system of equations contains only first order derivatives.

In order to take advantage from the symmetry of the geometry, we implemented our equations assuming an azimuthal 
dependence of the form $e^{ - im\phi }$ with $m \in \mathbb{Z}$. That is, for a vector field ${\bf v}$, we have ${\bf v}(\rho,\phi,z)=\sum_{m \in \mathbb{Z}}{\bf v}^{(m)}(\rho,z)e^{ - im\phi }$. 
Maxwell's equation and the polarization equation are written assuming the following definitions:
\[\nabla  \cdot {{\bf{v}}^{(m)}} \equiv \left( {\frac{1}{\rho } +
 \frac{\partial }{{\partial \rho }}} \right)v_\rho ^{(m)} - \frac{{im}}{\rho }v_\phi ^{(m)} + \frac{{\partial v_z^{(m)}}}{{\partial z}},\]
\[\begin{array}{l}
\nabla  \times {{\bf{v}}^{(m)}} \equiv \hat \rho \left( { - \frac{{\partial v_\phi ^{(m)}}}{{\partial z}} - i\frac{m}{\rho }v_z^{(m)}} \right) + \\
\quad  + \hat \phi \left( {\frac{{\partial v_\rho ^{(m)}}}{{\partial z}} - \frac{{\partial v_z^{(m)}}}{{\partial \rho }}} \right) 
+ \hat z\left( {\frac{{v_\phi ^{(m)}}}{\rho } + \frac{{\partial v_\phi ^{(m)}}}{{\partial \rho }} + i\frac{m}{\rho }v_\rho ^{(m)}} \right)
\end{array}\]
Analogously, the test functions are assumed to have a dependence of the form $e^{ im\phi }$.
It is possible then to reduce the initially three-dimensional problem into $(2m_{max}+1)$ two-dimensional problems.
The system to solve in the unknown variables ${\bf E}$ (electric field), ${\bf P}$ (polarization field) and ${\bf F}$ (working variable), reads:
\begin{widetext}
\[\begin{array}{l}
2\pi \int {\left( {\nabla  \times {{\bf{E}}^{(m)}}} \right) \cdot \left( {\nabla  \times {{{\bf{\tilde E}}}^{(m)}}} \right) - \left( {k_0^2{{\bf{E}}^{(m)}} + {\mu _0}{\omega ^2}{{\bf{P}}^{(m)}}} \right) \cdot {{{\bf{\tilde E}}}^{(m)}}} \rho d\rho dz = 0,\\
2\pi \int { - \frac{e}{{{m_e}}}\left( {\frac{{\delta G}}{{\delta n}}} \right)_1^{(m)}\left( {\nabla  \cdot {{{\bf{\tilde P}}}^{(m)}}} \right) + \frac{1}{{{n_0}}}\left[ {\left( {{\omega ^2} + i\gamma \omega } \right){{\bf{P}}^{(m)}} + {\varepsilon _0}\omega _p^2\left( {{{\bf{E}}^{(m)}} + {\bf{E}}_{inc}^{(m)}} \right)} \right] \cdot {{{\bf{\tilde P}}}^{(m)}}} \rho d\rho dz = 0,\\
2\pi \int { - \left( {\nabla  \cdot {{\bf{P}}^{(m)}}} \right)\left( {\nabla  \cdot {{{\bf{\tilde F}}}^{(m)}}} \right) - e{{\bf{F}}^{(m)}} \cdot {{{\bf{\tilde F}}}^{(m)}}} \rho d\rho dz = 0.
\end{array}\]
\end{widetext}

Note that for the case of an incident plane wave propagating along the $z$ axis, one has to solve the problem just for $m=\pm 1$.
Moreover by taking into account field parities, the solution for $m=1$ can be related to the solution for $m=-1$, so that a single two-dimensional calculation becomes necessary\cite{Ciraci:2013jt,Ciraci:2013wi}.

Note that for the electromagnetic module \textsc{Comsol} uses by default \textit{curl} elements for the in-plane components and \textit{Lagrange} elements for the azimuthal component.
We found that using \textit{Lagrange} elements for each component provides much more stable solutions.
Since \textsc{Comsol} does not give the possibility to use different type of elements for the built-in physics (in our case electromagnetism) we had to re-implement the electromagnetic module ourselves by using the general weak form implementation.
Perfectly matched layers have been used in order to emulate an infinite domain and avoid unwanted reflections.


\begin{thebibliography}{121}%
\makeatletter
\providecommand \@ifxundefined [1]{%
 \@ifx{#1\undefined}
}%
\providecommand \@ifnum [1]{%
 \ifnum #1\expandafter \@firstoftwo
 \else \expandafter \@secondoftwo
 \fi
}%
\providecommand \@ifx [1]{%
 \ifx #1\expandafter \@firstoftwo
 \else \expandafter \@secondoftwo
 \fi
}%
\providecommand \natexlab [1]{#1}%
\providecommand \enquote  [1]{``#1''}%
\providecommand \bibnamefont  [1]{#1}%
\providecommand \bibfnamefont [1]{#1}%
\providecommand \citenamefont [1]{#1}%
\providecommand \href@noop [0]{\@secondoftwo}%
\providecommand \href [0]{\begingroup \@sanitize@url \@href}%
\providecommand \@href[1]{\@@startlink{#1}\@@href}%
\providecommand \@@href[1]{\endgroup#1\@@endlink}%
\providecommand \@sanitize@url [0]{\catcode `\\12\catcode `\$12\catcode
  `\&12\catcode `\#12\catcode `\^12\catcode `\_12\catcode `\%12\relax}%
\providecommand \@@startlink[1]{}%
\providecommand \@@endlink[0]{}%
\providecommand \url  [0]{\begingroup\@sanitize@url \@url }%
\providecommand \@url [1]{\endgroup\@href {#1}{\urlprefix }}%
\providecommand \urlprefix  [0]{URL }%
\providecommand \Eprint [0]{\href }%
\providecommand \doibase [0]{http://dx.doi.org/}%
\providecommand \selectlanguage [0]{\@gobble}%
\providecommand \bibinfo  [0]{\@secondoftwo}%
\providecommand \bibfield  [0]{\@secondoftwo}%
\providecommand \translation [1]{[#1]}%
\providecommand \BibitemOpen [0]{}%
\providecommand \bibitemStop [0]{}%
\providecommand \bibitemNoStop [0]{.\EOS\space}%
\providecommand \EOS [0]{\spacefactor3000\relax}%
\providecommand \BibitemShut  [1]{\csname bibitem#1\endcsname}%
\let\auto@bib@innerbib\@empty
\bibitem [{\citenamefont {Gramotnev}\ and\ \citenamefont
  {Bozhevolnyi}(2010)}]{Gramotnev:2010ji}%
  \BibitemOpen
  \bibfield  {author} {\bibinfo {author} {\bibfnamefont {D.~K.}\ \bibnamefont
  {Gramotnev}}\ and\ \bibinfo {author} {\bibfnamefont {S.~I.}\ \bibnamefont
  {Bozhevolnyi}},\ }\bibfield  {title} {\emph {\enquote {\bibinfo {title}
  {{Plasmonics beyond the diffraction limit}},}\ }}\href@noop {} {\bibfield
  {journal} {\bibinfo  {journal} {Nat. Photonics}\ }\textbf {\bibinfo {volume}
  {4}},\ \bibinfo {pages} {83} (\bibinfo {year} {2010})}\BibitemShut {NoStop}%
\bibitem [{\citenamefont {Schuller}\ \emph {et~al.}(2010)\citenamefont
  {Schuller}, \citenamefont {Barnard}, \citenamefont {Cai}, \citenamefont
  {Jun}, \citenamefont {White},\ and\ \citenamefont
  {Brongersma}}]{Schuller:2010fh}%
  \BibitemOpen
  \bibfield  {author} {\bibinfo {author} {\bibfnamefont {J.~A.}\ \bibnamefont
  {Schuller}}, \bibinfo {author} {\bibfnamefont {E.~S.}\ \bibnamefont
  {Barnard}}, \bibinfo {author} {\bibfnamefont {W.}~\bibnamefont {Cai}},
  \bibinfo {author} {\bibfnamefont {Y.~C.}\ \bibnamefont {Jun}}, \bibinfo
  {author} {\bibfnamefont {J.~S.}\ \bibnamefont {White}}, \ and\ \bibinfo
  {author} {\bibfnamefont {M.~L.}\ \bibnamefont {Brongersma}},\ }\bibfield
  {title} {\emph {\enquote {\bibinfo {title} {{Plasmonics for extreme light
  concentration and manipulation}},}\ }}\href@noop {} {\bibfield  {journal}
  {\bibinfo  {journal} {Nat. Mater.}\ }\textbf {\bibinfo {volume} {9}},\
  \bibinfo {pages} {193} (\bibinfo {year} {2010})}\BibitemShut {NoStop}%
\bibitem [{\citenamefont {Maier}(2007)}]{Maier:2007wq}%
  \BibitemOpen
  \bibfield  {author} {\bibinfo {author} {\bibfnamefont {S.~A.}\ \bibnamefont
  {Maier}},\ }\href@noop {} {\emph {\bibinfo {title} {{Plasmonics, Fundamentals
  and Applications}}}}\ (\bibinfo  {publisher} {Springer},\ \bibinfo {year}
  {2007})\BibitemShut {NoStop}%
\bibitem [{\citenamefont {Aouani}\ \emph {et~al.}(2014)\citenamefont {Aouani},
  \citenamefont {Rahmani}, \citenamefont {Navarro-C{\'\i}a},\ and\
  \citenamefont {Maier}}]{Aouani:2014iy}%
  \BibitemOpen
  \bibfield  {author} {\bibinfo {author} {\bibfnamefont {H.}~\bibnamefont
  {Aouani}}, \bibinfo {author} {\bibfnamefont {M.}~\bibnamefont {Rahmani}},
  \bibinfo {author} {\bibfnamefont {M.}~\bibnamefont {Navarro-C{\'\i}a}}, \
  and\ \bibinfo {author} {\bibfnamefont {S.~A.}\ \bibnamefont {Maier}},\
  }\bibfield  {title} {\emph {\enquote {\bibinfo {title}
  {{Third-harmonic-upconversion enhancement from a single semiconductor
  nanoparticle coupled to a plasmonic antenna}},}\ }}\href@noop {} {\bibfield
  {journal} {\bibinfo  {journal} {Nat. Nanotechnol.}\ }\textbf {\bibinfo
  {volume} {9}},\ \bibinfo {pages} {290} (\bibinfo {year} {2014})}\BibitemShut
  {NoStop}%
\bibitem [{\citenamefont {Moreau}\ \emph {et~al.}(2012)\citenamefont {Moreau},
  \citenamefont {Cirac{\`\i}}, \citenamefont {Mock}, \citenamefont {Hill},
  \citenamefont {Wang}, \citenamefont {Wiley}, \citenamefont {Chilkoti},\ and\
  \citenamefont {Smith}}]{Moreau:2012uba}%
  \BibitemOpen
  \bibfield  {author} {\bibinfo {author} {\bibfnamefont {A.}~\bibnamefont
  {Moreau}}, \bibinfo {author} {\bibfnamefont {C.}~\bibnamefont {Cirac{\`\i}}},
  \bibinfo {author} {\bibfnamefont {J.~J.}\ \bibnamefont {Mock}}, \bibinfo
  {author} {\bibfnamefont {R.~T.}\ \bibnamefont {Hill}}, \bibinfo {author}
  {\bibfnamefont {Q.}~\bibnamefont {Wang}}, \bibinfo {author} {\bibfnamefont
  {B.~J.}\ \bibnamefont {Wiley}}, \bibinfo {author} {\bibfnamefont
  {A.}~\bibnamefont {Chilkoti}}, \ and\ \bibinfo {author} {\bibfnamefont
  {D.~R.}\ \bibnamefont {Smith}},\ }\bibfield  {title} {\emph {\enquote
  {\bibinfo {title} {{Controlled-reflectance surfaces with film-coupled
  colloidal nanoantennas}},}\ }}\href@noop {} {\bibfield  {journal} {\bibinfo
  {journal} {Nature}\ }\textbf {\bibinfo {volume} {492}},\ \bibinfo {pages}
  {86} (\bibinfo {year} {2012})}\BibitemShut {NoStop}%
\bibitem [{\citenamefont {Akselrod}\ \emph {et~al.}(2014)\citenamefont
  {Akselrod}, \citenamefont {Argyropoulos}, \citenamefont {Hoang},
  \citenamefont {Cirac{\`\i}}, \citenamefont {Fang}, \citenamefont {Huang},
  \citenamefont {Smith},\ and\ \citenamefont {Mikkelsen}}]{Akselrod:2014ek}%
  \BibitemOpen
  \bibfield  {author} {\bibinfo {author} {\bibfnamefont {G.~M.}\ \bibnamefont
  {Akselrod}}, \bibinfo {author} {\bibfnamefont {C.}~\bibnamefont
  {Argyropoulos}}, \bibinfo {author} {\bibfnamefont {T.~B.}\ \bibnamefont
  {Hoang}}, \bibinfo {author} {\bibfnamefont {C.}~\bibnamefont {Cirac{\`\i}}},
  \bibinfo {author} {\bibfnamefont {C.}~\bibnamefont {Fang}}, \bibinfo {author}
  {\bibfnamefont {J.}~\bibnamefont {Huang}}, \bibinfo {author} {\bibfnamefont
  {D.~R.}\ \bibnamefont {Smith}}, \ and\ \bibinfo {author} {\bibfnamefont
  {M.~H.}\ \bibnamefont {Mikkelsen}},\ }\bibfield  {title} {\emph {\enquote
  {\bibinfo {title} {{Probing the mechanisms of large Purcell enhancement in
  plasmonic nanoantennas}},}\ }}\href@noop {} {\bibfield  {journal} {\bibinfo
  {journal} {Nat. Photonics}\ }\textbf {\bibinfo {volume} {8}},\ \bibinfo
  {pages} {835} (\bibinfo {year} {2014})}\BibitemShut {NoStop}%
\bibitem [{\citenamefont {Akselrod}\ \emph {et~al.}(2015)\citenamefont
  {Akselrod}, \citenamefont {Ming}, \citenamefont {Argyropoulos}, \citenamefont
  {Hoang}, \citenamefont {Lin}, \citenamefont {Ling}, \citenamefont {Smith},
  \citenamefont {Kong},\ and\ \citenamefont {Mikkelsen}}]{Akselrod:2015cf}%
  \BibitemOpen
  \bibfield  {author} {\bibinfo {author} {\bibfnamefont {G.~M.}\ \bibnamefont
  {Akselrod}}, \bibinfo {author} {\bibfnamefont {T.}~\bibnamefont {Ming}},
  \bibinfo {author} {\bibfnamefont {C.}~\bibnamefont {Argyropoulos}}, \bibinfo
  {author} {\bibfnamefont {T.~B.}\ \bibnamefont {Hoang}}, \bibinfo {author}
  {\bibfnamefont {Y.}~\bibnamefont {Lin}}, \bibinfo {author} {\bibfnamefont
  {X.}~\bibnamefont {Ling}}, \bibinfo {author} {\bibfnamefont {D.~R.}\
  \bibnamefont {Smith}}, \bibinfo {author} {\bibfnamefont {J.}~\bibnamefont
  {Kong}}, \ and\ \bibinfo {author} {\bibfnamefont {M.~H.}\ \bibnamefont
  {Mikkelsen}},\ }\bibfield  {title} {\emph {\enquote {\bibinfo {title}
  {{Leveraging Nanocavity Harmonics for Control of Optical Processes in 2D
  Semiconductors}},}\ }}\href@noop {} {\bibfield  {journal} {\bibinfo
  {journal} {Nano Lett.}\ }\textbf {\bibinfo {volume} {15}},\ \bibinfo {pages}
  {3578} (\bibinfo {year} {2015})}\BibitemShut {NoStop}%
\bibitem [{\citenamefont {Rose}\ \emph {et~al.}(2014)\citenamefont {Rose},
  \citenamefont {Hoang}, \citenamefont {McGuire}, \citenamefont {Mock},
  \citenamefont {Cirac{\`\i}}, \citenamefont {Smith},\ and\ \citenamefont
  {Mikkelsen}}]{Rose:2014ko}%
  \BibitemOpen
  \bibfield  {author} {\bibinfo {author} {\bibfnamefont {A.}~\bibnamefont
  {Rose}}, \bibinfo {author} {\bibfnamefont {T.~B.}\ \bibnamefont {Hoang}},
  \bibinfo {author} {\bibfnamefont {F.}~\bibnamefont {McGuire}}, \bibinfo
  {author} {\bibfnamefont {J.~J.}\ \bibnamefont {Mock}}, \bibinfo {author}
  {\bibfnamefont {C.}~\bibnamefont {Cirac{\`\i}}}, \bibinfo {author}
  {\bibfnamefont {D.~R.}\ \bibnamefont {Smith}}, \ and\ \bibinfo {author}
  {\bibfnamefont {M.~H.}\ \bibnamefont {Mikkelsen}},\ }\bibfield  {title}
  {\emph {\enquote {\bibinfo {title} {{Control of radiative processes using
  tunable plasmonic nanopatch antennas}},}\ }}\href@noop {} {\bibfield
  {journal} {\bibinfo  {journal} {Nano Lett.}\ }\textbf {\bibinfo {volume}
  {14}},\ \bibinfo {pages} {4797} (\bibinfo {year} {2014})}\BibitemShut
  {NoStop}%
\bibitem [{\citenamefont {Cirac{\`\i}}\ \emph {et~al.}(2012)\citenamefont
  {Cirac{\`\i}}, \citenamefont {Hill}, \citenamefont {Mock},\ and\
  \citenamefont {Urzhumov}}]{Ciraci:2012fp}%
  \BibitemOpen
  \bibfield  {author} {\bibinfo {author} {\bibfnamefont {C.}~\bibnamefont
  {Cirac{\`\i}}}, \bibinfo {author} {\bibfnamefont {R.}~\bibnamefont {Hill}},
  \bibinfo {author} {\bibfnamefont {J.~J.}\ \bibnamefont {Mock}}, \ and\
  \bibinfo {author} {\bibfnamefont {Y.~A.}\ \bibnamefont {Urzhumov}},\
  }\bibfield  {title} {\emph {\enquote {\bibinfo {title} {{Probing the ultimate
  limits of plasmonic enhancement}},}\ }}\href@noop {} {\bibfield  {journal}
  {\bibinfo  {journal} {Science}\ }\textbf {\bibinfo {volume} {337}},\ \bibinfo
  {pages} {1072} (\bibinfo {year} {2012})}\BibitemShut {NoStop}%
\bibitem [{\citenamefont {Savage}\ \emph {et~al.}(2012)\citenamefont {Savage},
  \citenamefont {Hawkeye}, \citenamefont {Esteban},\ and\ \citenamefont
  {Borisov}}]{Savage:2012by}%
  \BibitemOpen
  \bibfield  {author} {\bibinfo {author} {\bibfnamefont {K.~J.}\ \bibnamefont
  {Savage}}, \bibinfo {author} {\bibfnamefont {M.~M.}\ \bibnamefont {Hawkeye}},
  \bibinfo {author} {\bibfnamefont {R.}~\bibnamefont {Esteban}}, \ and\
  \bibinfo {author} {\bibfnamefont {A.~G.}\ \bibnamefont {Borisov}},\
  }\bibfield  {title} {\emph {\enquote {\bibinfo {title} {{Revealing the
  quantum regime in tunnelling plasmonics}},}\ }}\href@noop {} {\bibfield
  {journal} {\bibinfo  {journal} {Nature}\ }\textbf {\bibinfo {volume} {491}},\
  \bibinfo {pages} {574} (\bibinfo {year} {2012})}\BibitemShut {NoStop}%
\bibitem [{\citenamefont {Hajisalem}\ \emph {et~al.}(2014)\citenamefont
  {Hajisalem}, \citenamefont {Nezami},\ and\ \citenamefont
  {Gordon}}]{Hajisalem:2014cr}%
  \BibitemOpen
  \bibfield  {author} {\bibinfo {author} {\bibfnamefont {G.}~\bibnamefont
  {Hajisalem}}, \bibinfo {author} {\bibfnamefont {M.~S.}\ \bibnamefont
  {Nezami}}, \ and\ \bibinfo {author} {\bibfnamefont {R.}~\bibnamefont
  {Gordon}},\ }\bibfield  {title} {\emph {\enquote {\bibinfo {title} {{Probing
  the quantum tunneling limit of plasmonic enhancement by third harmonic
  generation}},}\ }}\href@noop {} {\bibfield  {journal} {\bibinfo  {journal}
  {Nano Lett.}\ }\textbf {\bibinfo {volume} {14}},\ \bibinfo {pages} {6651}
  (\bibinfo {year} {2014})}\BibitemShut {NoStop}%
\bibitem [{\citenamefont {Moreau}\ \emph {et~al.}(2013)\citenamefont {Moreau},
  \citenamefont {Cirac{\`\i}},\ and\ \citenamefont {Smith}}]{Moreau:2013ei}%
  \BibitemOpen
  \bibfield  {author} {\bibinfo {author} {\bibfnamefont {A.}~\bibnamefont
  {Moreau}}, \bibinfo {author} {\bibfnamefont {C.}~\bibnamefont {Cirac{\`\i}}},
  \ and\ \bibinfo {author} {\bibfnamefont {D.~R.}\ \bibnamefont {Smith}},\
  }\bibfield  {title} {\emph {\enquote {\bibinfo {title} {{Impact of nonlocal
  response on metallodielectric multilayers and optical patch antennas}},}\
  }}\href@noop {} {\bibfield  {journal} {\bibinfo  {journal} {Phys. Rev. B}\
  }\textbf {\bibinfo {volume} {87}},\ \bibinfo {pages} {045401} (\bibinfo
  {year} {2013})}\BibitemShut {NoStop}%
\bibitem [{\citenamefont {Chen}\ \emph {et~al.}(2013)\citenamefont {Chen},
  \citenamefont {Park}, \citenamefont {Pelton}, \citenamefont {Piao},
  \citenamefont {Lindquist}, \citenamefont {Im}, \citenamefont {Kim},
  \citenamefont {Ahn}, \citenamefont {Ahn}, \citenamefont {Park}, \citenamefont
  {Kim},\ and\ \citenamefont {Oh}}]{Chen:2013hq}%
  \BibitemOpen
  \bibfield  {author} {\bibinfo {author} {\bibfnamefont {X.}~\bibnamefont
  {Chen}}, \bibinfo {author} {\bibfnamefont {H.-R.}\ \bibnamefont {Park}},
  \bibinfo {author} {\bibfnamefont {M.}~\bibnamefont {Pelton}}, \bibinfo
  {author} {\bibfnamefont {X.}~\bibnamefont {Piao}}, \bibinfo {author}
  {\bibfnamefont {N.~C.}\ \bibnamefont {Lindquist}}, \bibinfo {author}
  {\bibfnamefont {H.}~\bibnamefont {Im}}, \bibinfo {author} {\bibfnamefont
  {Y.~J.}\ \bibnamefont {Kim}}, \bibinfo {author} {\bibfnamefont {J.~S.}\
  \bibnamefont {Ahn}}, \bibinfo {author} {\bibfnamefont {K.~J.}\ \bibnamefont
  {Ahn}}, \bibinfo {author} {\bibfnamefont {N.}~\bibnamefont {Park}}, \bibinfo
  {author} {\bibfnamefont {D.-S.}\ \bibnamefont {Kim}}, \ and\ \bibinfo
  {author} {\bibfnamefont {S.-H.}\ \bibnamefont {Oh}},\ }\bibfield  {title}
  {\emph {\enquote {\bibinfo {title} {{Atomic layer lithography of wafer-scale
  nanogap arrays for extreme confinement of electromagnetic waves}},}\
  }}\href@noop {} {\bibfield  {journal} {\bibinfo  {journal} {Nature Commun.}\
  }\textbf {\bibinfo {volume} {4}},\ \bibinfo {pages} {2361} (\bibinfo {year}
  {2013})}\BibitemShut {NoStop}%
\bibitem [{\citenamefont {Chen}\ \emph {et~al.}(2015)\citenamefont {Chen},
  \citenamefont {Cirac{\`\i}}, \citenamefont {Smith},\ and\ \citenamefont
  {Oh}}]{Chen:2014gn}%
  \BibitemOpen
  \bibfield  {author} {\bibinfo {author} {\bibfnamefont {X.}~\bibnamefont
  {Chen}}, \bibinfo {author} {\bibfnamefont {C.}~\bibnamefont {Cirac{\`\i}}},
  \bibinfo {author} {\bibfnamefont {D.~R.}\ \bibnamefont {Smith}}, \ and\
  \bibinfo {author} {\bibfnamefont {S.-H.}\ \bibnamefont {Oh}},\ }\bibfield
  {title} {\emph {\enquote {\bibinfo {title} {{Nanogap-enhanced Infrared
  Spectroscopy with Template-stripped Wafer-scale Arrays of Buried Plasmonic
  Cavities}},}\ }}\href@noop {} {\bibfield  {journal} {\bibinfo  {journal}
  {Nano Lett.}\ }\textbf {\bibinfo {volume} {15}},\ \bibinfo {pages} {107}
  (\bibinfo {year} {2015})}\BibitemShut {NoStop}%
\bibitem [{\citenamefont {Cirac{\`\i}}\ \emph {et~al.}(2014)\citenamefont
  {Cirac{\`\i}}, \citenamefont {Chen}, \citenamefont {Mock}, \citenamefont
  {McGuire}, \citenamefont {Liu}, \citenamefont {Oh},\ and\ \citenamefont
  {Smith}}]{Ciraci:2014jv}%
  \BibitemOpen
  \bibfield  {author} {\bibinfo {author} {\bibfnamefont {C.}~\bibnamefont
  {Cirac{\`\i}}}, \bibinfo {author} {\bibfnamefont {X.}~\bibnamefont {Chen}},
  \bibinfo {author} {\bibfnamefont {J.~J.}\ \bibnamefont {Mock}}, \bibinfo
  {author} {\bibfnamefont {F.}~\bibnamefont {McGuire}}, \bibinfo {author}
  {\bibfnamefont {X.}~\bibnamefont {Liu}}, \bibinfo {author} {\bibfnamefont
  {S.-H.}\ \bibnamefont {Oh}}, \ and\ \bibinfo {author} {\bibfnamefont {D.~R.}\
  \bibnamefont {Smith}},\ }\bibfield  {title} {\emph {\enquote {\bibinfo
  {title} {{Film-coupled nanoparticles by atomic layer deposition: Comparison
  with organic spacing layers}},}\ }}\href@noop {} {\bibfield  {journal}
  {\bibinfo  {journal} {Appl. Phys. Lett.}\ }\textbf {\bibinfo {volume}
  {104}},\ \bibinfo {pages} {023109} (\bibinfo {year} {2014})}\BibitemShut
  {NoStop}%
\bibitem [{\citenamefont {Marinica}\ \emph
  {et~al.}(2012{\natexlab{a}})\citenamefont {Marinica}, \citenamefont
  {Kazansky},\ and\ \citenamefont {Nordlander}}]{Marinica:2012}%
  \BibitemOpen
  \bibfield  {author} {\bibinfo {author} {\bibfnamefont {D.~C.}\ \bibnamefont
  {Marinica}}, \bibinfo {author} {\bibfnamefont {A.~K.}\ \bibnamefont
  {Kazansky}}, \ and\ \bibinfo {author} {\bibfnamefont {P.}~\bibnamefont
  {Nordlander}},\ }\bibfield  {title} {\emph {\enquote {\bibinfo {title}
  {{Quantum plasmonics: nonlinear effects in the field enhancement of a
  plasmonic nanoparticle dimer}},}\ }}\href@noop {} {\bibfield  {journal}
  {\bibinfo  {journal} {Nano Lett.}\ }\textbf {\bibinfo {volume} {12}},\
  \bibinfo {pages} {1333} (\bibinfo {year} {2012}{\natexlab{a}})}\BibitemShut
  {NoStop}%
\bibitem [{\citenamefont {Zuloaga}\ \emph {et~al.}(2009)\citenamefont
  {Zuloaga}, \citenamefont {Prodan},\ and\ \citenamefont
  {Nordlander}}]{Zuloaga:2009gm}%
  \BibitemOpen
  \bibfield  {author} {\bibinfo {author} {\bibfnamefont {J.}~\bibnamefont
  {Zuloaga}}, \bibinfo {author} {\bibfnamefont {E.}~\bibnamefont {Prodan}}, \
  and\ \bibinfo {author} {\bibfnamefont {P.}~\bibnamefont {Nordlander}},\
  }\bibfield  {title} {\emph {\enquote {\bibinfo {title} {{Quantum Description
  of the Plasmon Resonances of a Nanoparticle Dimer}},}\ }}\href@noop {}
  {\bibfield  {journal} {\bibinfo  {journal} {Nano Lett.}\ }\textbf {\bibinfo
  {volume} {9}},\ \bibinfo {pages} {887} (\bibinfo {year} {2009})}\BibitemShut
  {NoStop}%
\bibitem [{\citenamefont {Ullrich}(2011)}]{tddftbook}%
  \BibitemOpen
  \bibinfo {editor} {\bibfnamefont {C.~A.}\ \bibnamefont {Ullrich}},\ ed.,\
  \href@noop {} {\emph {\bibinfo {title} {Time-Dependent Density-Functional
  Theory: Concepts and Applications}}}\ (\bibinfo  {publisher} {Oxford
  University Press},\ \bibinfo {year} {2011})\BibitemShut {NoStop}%
\bibitem [{\citenamefont {Yabana}\ and\ \citenamefont
  {Bertsch}(1996)}]{yabana96}%
  \BibitemOpen
  \bibfield  {author} {\bibinfo {author} {\bibfnamefont {K.}~\bibnamefont
  {Yabana}}\ and\ \bibinfo {author} {\bibfnamefont {G.~F.}\ \bibnamefont
  {Bertsch}},\ }\bibfield  {title} {\emph {\enquote {\bibinfo {title}
  {Time-dependent local-density approximation in real time},}\ }}\href@noop {}
  {\bibfield  {journal} {\bibinfo  {journal} {Phys. Rev. B}\ }\textbf {\bibinfo
  {volume} {54}},\ \bibinfo {pages} {4484} (\bibinfo {year}
  {1996})}\BibitemShut {NoStop}%
\bibitem [{\citenamefont {Andrade}\ \emph {et~al.}(2012)\citenamefont
  {Andrade}, \citenamefont {Alberdi-Rodriguez}, \citenamefont {Strubbe},
  \citenamefont {Oliveira}, \citenamefont {Nogueira}, \citenamefont {Castro},
  \citenamefont {Muguerza}, \citenamefont {Arruabarrena}, \citenamefont
  {Louie}, \citenamefont {Aspuru-Guzik}, \citenamefont {Rubio},\ and\
  \citenamefont {Marques}}]{octopus}%
  \BibitemOpen
  \bibfield  {author} {\bibinfo {author} {\bibfnamefont {X.}~\bibnamefont
  {Andrade}}, \bibinfo {author} {\bibfnamefont {J.}~\bibnamefont
  {Alberdi-Rodriguez}}, \bibinfo {author} {\bibfnamefont {D.~A.}\ \bibnamefont
  {Strubbe}}, \bibinfo {author} {\bibfnamefont {M.~J.~T.}\ \bibnamefont
  {Oliveira}}, \bibinfo {author} {\bibfnamefont {F.}~\bibnamefont {Nogueira}},
  \bibinfo {author} {\bibfnamefont {A.}~\bibnamefont {Castro}}, \bibinfo
  {author} {\bibfnamefont {J.}~\bibnamefont {Muguerza}}, \bibinfo {author}
  {\bibfnamefont {A.}~\bibnamefont {Arruabarrena}}, \bibinfo {author}
  {\bibfnamefont {S.~G.}\ \bibnamefont {Louie}}, \bibinfo {author}
  {\bibfnamefont {A.}~\bibnamefont {Aspuru-Guzik}}, \bibinfo {author}
  {\bibfnamefont {A.}~\bibnamefont {Rubio}}, \ and\ \bibinfo {author}
  {\bibfnamefont {M.~A.~L.}\ \bibnamefont {Marques}},\ }\bibfield  {title}
  {\emph {\enquote {\bibinfo {title} {Time-dependent density-functional theory
  in massively parallel computer architectures: the octopus project},}\
  }}\href@noop {} {\bibfield  {journal} {\bibinfo  {journal} {J. Phys. Condens.
  Matter}\ }\textbf {\bibinfo {volume} {24}},\ \bibinfo {pages} {233202}
  (\bibinfo {year} {2012})}\BibitemShut {NoStop}%
\bibitem [{\citenamefont {Casida}(1996)}]{casi96}%
  \BibitemOpen
  \bibfield  {author} {\bibinfo {author} {\bibfnamefont {M.~E.}\ \bibnamefont
  {Casida}},\ }in\ \href@noop {} {\emph {\bibinfo {booktitle} {Recent
  developments and applications of modern density functional theory}}},\
  \bibinfo {editor} {edited by\ \bibinfo {editor} {\bibfnamefont {J.~M.}\
  \bibnamefont {Seminario}}}\ (\bibinfo  {publisher} {Elsevier},\ \bibinfo
  {address} {Amsterdam},\ \bibinfo {year} {1996})\ pp.\ \bibinfo {pages}
  {391--434}\BibitemShut {NoStop}%
\bibitem [{\citenamefont {Morton}\ \emph {et~al.}(2011)\citenamefont {Morton},
  \citenamefont {Silverstein},\ and\ \citenamefont {Jensen}}]{morton11}%
  \BibitemOpen
  \bibfield  {author} {\bibinfo {author} {\bibfnamefont {S.~M.}\ \bibnamefont
  {Morton}}, \bibinfo {author} {\bibfnamefont {D.~W.}\ \bibnamefont
  {Silverstein}}, \ and\ \bibinfo {author} {\bibfnamefont {L.}~\bibnamefont
  {Jensen}},\ }\bibfield  {title} {\emph {\enquote {\bibinfo {title}
  {Theoretical studies of plasmonics using electronic structure methods},}\
  }}\href {\doibase 10.1021/cr100265f} {\bibfield  {journal} {\bibinfo
  {journal} {Chem. Rev.}\ }\textbf {\bibinfo {volume} {111}},\ \bibinfo {pages}
  {3962} (\bibinfo {year} {2011})}\BibitemShut {NoStop}%
\bibitem [{\citenamefont {Brack}(1993)}]{brack93}%
  \BibitemOpen
  \bibfield  {author} {\bibinfo {author} {\bibfnamefont {M.}~\bibnamefont
  {Brack}},\ }\bibfield  {title} {\emph {\enquote {\bibinfo {title} {The
  physics of simple metal clusters: self-consistent jellium model and
  semiclassical approaches},}\ }}\href@noop {} {\bibfield  {journal} {\bibinfo
  {journal} {Rev. Mod. Phys.}\ }\textbf {\bibinfo {volume} {65}},\ \bibinfo
  {pages} {677} (\bibinfo {year} {1993})}\BibitemShut {NoStop}%
\bibitem [{\citenamefont {Ekardt}(1985)}]{ekardt85}%
  \BibitemOpen
  \bibfield  {author} {\bibinfo {author} {\bibfnamefont {W.}~\bibnamefont
  {Ekardt}},\ }\bibfield  {title} {\emph {\enquote {\bibinfo {title}
  {Size-dependent photoabsorption and photoemission of small metal
  particles},}\ }}\href@noop {} {\bibfield  {journal} {\bibinfo  {journal}
  {Phys. Rev. B}\ }\textbf {\bibinfo {volume} {31}},\ \bibinfo {pages} {6360}
  (\bibinfo {year} {1985})}\BibitemShut {NoStop}%
\bibitem [{\citenamefont {Teperik}\ \emph {et~al.}(2013)\citenamefont
  {Teperik}, \citenamefont {Nordlander}, \citenamefont {Aizpurua},\ and\
  \citenamefont {Borisov}}]{Teperik:2013dd}%
  \BibitemOpen
  \bibfield  {author} {\bibinfo {author} {\bibfnamefont {T.~V.}\ \bibnamefont
  {Teperik}}, \bibinfo {author} {\bibfnamefont {P.}~\bibnamefont {Nordlander}},
  \bibinfo {author} {\bibfnamefont {J.}~\bibnamefont {Aizpurua}}, \ and\
  \bibinfo {author} {\bibfnamefont {A.~G.}\ \bibnamefont {Borisov}},\
  }\bibfield  {title} {\emph {\enquote {\bibinfo {title} {{Robust subnanometric
  plasmon ruler by rescaling of the nonlocal optical response}},}\ }}\href@noop
  {} {\bibfield  {journal} {\bibinfo  {journal} {Phys. Rev. Lett.}\ }\textbf
  {\bibinfo {volume} {110}},\ \bibinfo {pages} {263901} (\bibinfo {year}
  {2013})}\BibitemShut {NoStop}%
\bibitem [{\citenamefont {Yan}\ \emph {et~al.}(2015)\citenamefont {Yan},
  \citenamefont {Wubs},\ and\ \citenamefont {Asger~Mortensen}}]{Yan:2015gx}%
  \BibitemOpen
  \bibfield  {author} {\bibinfo {author} {\bibfnamefont {W.}~\bibnamefont
  {Yan}}, \bibinfo {author} {\bibfnamefont {M.}~\bibnamefont {Wubs}}, \ and\
  \bibinfo {author} {\bibfnamefont {N.}~\bibnamefont {Asger~Mortensen}},\
  }\bibfield  {title} {\emph {\enquote {\bibinfo {title} {{Projected Dipole
  Model for Quantum Plasmonics}},}\ }}\href@noop {} {\bibfield  {journal}
  {\bibinfo  {journal} {Phys. Rev. Lett.}\ }\textbf {\bibinfo {volume} {115}},\
  \bibinfo {pages} {137403} (\bibinfo {year} {2015})}\BibitemShut {NoStop}%
\bibitem [{\citenamefont {Esteban}\ \emph {et~al.}(2012)\citenamefont
  {Esteban}, \citenamefont {Borisov},\ and\ \citenamefont
  {Nordlander}}]{Esteban:2012}%
  \BibitemOpen
  \bibfield  {author} {\bibinfo {author} {\bibfnamefont {R.}~\bibnamefont
  {Esteban}}, \bibinfo {author} {\bibfnamefont {A.~G.}\ \bibnamefont
  {Borisov}}, \ and\ \bibinfo {author} {\bibfnamefont {P.}~\bibnamefont
  {Nordlander}},\ }\bibfield  {title} {\emph {\enquote {\bibinfo {title}
  {{Bridging quantum and classical plasmonics with a quantum-corrected
  model}},}\ }}\href@noop {} {\bibfield  {journal} {\bibinfo  {journal} {Nature
  Commun.}\ }\textbf {\bibinfo {volume} {3}},\ \bibinfo {pages} {825} (\bibinfo
  {year} {2012})}\BibitemShut {NoStop}%
\bibitem [{\citenamefont {Barbry}\ \emph
  {et~al.}(2015{\natexlab{a}})\citenamefont {Barbry}, \citenamefont {Koval},
  \citenamefont {Marchesin}, \citenamefont {Esteban}, \citenamefont {Borisov},
  \citenamefont {Aizpurua},\ and\ \citenamefont {Sanchez-Portal}}]{barbry2015}%
  \BibitemOpen
  \bibfield  {author} {\bibinfo {author} {\bibfnamefont {M.}~\bibnamefont
  {Barbry}}, \bibinfo {author} {\bibfnamefont {P.}~\bibnamefont {Koval}},
  \bibinfo {author} {\bibfnamefont {F.}~\bibnamefont {Marchesin}}, \bibinfo
  {author} {\bibfnamefont {R.}~\bibnamefont {Esteban}}, \bibinfo {author}
  {\bibfnamefont {A.~G.}\ \bibnamefont {Borisov}}, \bibinfo {author}
  {\bibfnamefont {J.}~\bibnamefont {Aizpurua}}, \ and\ \bibinfo {author}
  {\bibfnamefont {D.}~\bibnamefont {Sanchez-Portal}},\ }\bibfield  {title}
  {\emph {\enquote {\bibinfo {title} {Atomistic near-field nanoplasmonics:
  Reaching atomic-scale resolution in nanooptics},}\ }}\href@noop {} {\bibfield
   {journal} {\bibinfo  {journal} {Nano Lett.}\ }\textbf {\bibinfo {volume}
  {15}},\ \bibinfo {pages} {3410} (\bibinfo {year}
  {2015}{\natexlab{a}})}\BibitemShut {NoStop}%
\bibitem [{\citenamefont {Zhang}\ \emph {et~al.}(2014)\citenamefont {Zhang},
  \citenamefont {Feist}, \citenamefont {Rubio}, \citenamefont
  {Garc\'{i}a-Gonz\'alez},\ and\ \citenamefont
  {Garc\'{i}a-Vidal}}]{Zhang:2014rubio}%
  \BibitemOpen
  \bibfield  {author} {\bibinfo {author} {\bibfnamefont {P.}~\bibnamefont
  {Zhang}}, \bibinfo {author} {\bibfnamefont {J.}~\bibnamefont {Feist}},
  \bibinfo {author} {\bibfnamefont {A.}~\bibnamefont {Rubio}}, \bibinfo
  {author} {\bibfnamefont {P.}~\bibnamefont {Garc\'{i}a-Gonz\'alez}}, \ and\
  \bibinfo {author} {\bibfnamefont {F.~J.}\ \bibnamefont {Garc\'{i}a-Vidal}},\
  }\bibfield  {title} {\emph {\enquote {\bibinfo {title} {\textit{Ab initio}
  nanoplasmonics: The impact of atomic structure},}\ }}\href@noop {} {\bibfield
   {journal} {\bibinfo  {journal} {Phys. Rev. B}\ }\textbf {\bibinfo {volume}
  {90}},\ \bibinfo {pages} {161407} (\bibinfo {year} {2014})}\BibitemShut
  {NoStop}%
\bibitem [{\citenamefont {Li}\ \emph {et~al.}(2013)\citenamefont {Li},
  \citenamefont {Hayashi},\ and\ \citenamefont {Guo}}]{li2013}%
  \BibitemOpen
  \bibfield  {author} {\bibinfo {author} {\bibfnamefont {J.-H.}\ \bibnamefont
  {Li}}, \bibinfo {author} {\bibfnamefont {M.}~\bibnamefont {Hayashi}}, \ and\
  \bibinfo {author} {\bibfnamefont {G.-Y.}\ \bibnamefont {Guo}},\ }\bibfield
  {title} {\emph {\enquote {\bibinfo {title} {Plasmonic excitations in
  quantum-sized sodium nanoparticles studied by time-dependent density
  functional calculations},}\ }}\href@noop {} {\bibfield  {journal} {\bibinfo
  {journal} {Phys. Rev. B}\ }\textbf {\bibinfo {volume} {88}},\ \bibinfo
  {pages} {155437} (\bibinfo {year} {2013})}\BibitemShut {NoStop}%
\bibitem [{\citenamefont {Iida}\ \emph {et~al.}(2014)\citenamefont {Iida},
  \citenamefont {Noda}, \citenamefont {Ishimura},\ and\ \citenamefont
  {Nobusada}}]{iida2014}%
  \BibitemOpen
  \bibfield  {author} {\bibinfo {author} {\bibfnamefont {K.}~\bibnamefont
  {Iida}}, \bibinfo {author} {\bibfnamefont {M.}~\bibnamefont {Noda}}, \bibinfo
  {author} {\bibfnamefont {K.}~\bibnamefont {Ishimura}}, \ and\ \bibinfo
  {author} {\bibfnamefont {K.}~\bibnamefont {Nobusada}},\ }\bibfield  {title}
  {\emph {\enquote {\bibinfo {title} {First-principles computational
  visualization of localized surface plasmon resonance in gold nanoclusters},}\
  }}\href {\doibase 10.1021/jp5088042} {\bibfield  {journal} {\bibinfo
  {journal} {J. Phys. Chem. A}\ }\textbf {\bibinfo {volume} {118}},\ \bibinfo
  {pages} {11317} (\bibinfo {year} {2014})}\BibitemShut {NoStop}%
\bibitem [{\citenamefont {Pitarke}\ \emph {et~al.}(2007)\citenamefont
  {Pitarke}, \citenamefont {Silkin}, \citenamefont {Chulkov},\ and\
  \citenamefont {Echenique}}]{pitarke07}%
  \BibitemOpen
  \bibfield  {author} {\bibinfo {author} {\bibfnamefont {J.~M.}\ \bibnamefont
  {Pitarke}}, \bibinfo {author} {\bibfnamefont {V.~M.}\ \bibnamefont {Silkin}},
  \bibinfo {author} {\bibfnamefont {E.~V.}\ \bibnamefont {Chulkov}}, \ and\
  \bibinfo {author} {\bibfnamefont {P.~M.}\ \bibnamefont {Echenique}},\
  }\bibfield  {title} {\emph {\enquote {\bibinfo {title} {Theory of surface
  plasmons and surface-plasmon polaritons},}\ }}\href@noop {} {\bibfield
  {journal} {\bibinfo  {journal} {Rep. Prog. Phys.}\ }\textbf {\bibinfo
  {volume} {70}},\ \bibinfo {pages} {1} (\bibinfo {year} {2007})}\BibitemShut
  {NoStop}%
\bibitem [{\citenamefont {Raza}\ \emph {et~al.}(2011)\citenamefont {Raza},
  \citenamefont {Toscano}, \citenamefont {Jauho}, \citenamefont {Wubs},\ and\
  \citenamefont {Mortensen}}]{Raza:2011io}%
  \BibitemOpen
  \bibfield  {author} {\bibinfo {author} {\bibfnamefont {S.}~\bibnamefont
  {Raza}}, \bibinfo {author} {\bibfnamefont {G.}~\bibnamefont {Toscano}},
  \bibinfo {author} {\bibfnamefont {A.~P.}\ \bibnamefont {Jauho}}, \bibinfo
  {author} {\bibfnamefont {M.}~\bibnamefont {Wubs}}, \ and\ \bibinfo {author}
  {\bibfnamefont {N.~A.}\ \bibnamefont {Mortensen}},\ }\bibfield  {title}
  {\emph {\enquote {\bibinfo {title} {{Unusual resonances in nanoplasmonic
  structures due to nonlocal response}},}\ }}\href@noop {} {\bibfield
  {journal} {\bibinfo  {journal} {Phys. Rev. B}\ }\textbf {\bibinfo {volume}
  {84}},\ \bibinfo {pages} {121412} (\bibinfo {year} {2011})}\BibitemShut
  {NoStop}%
\bibitem [{\citenamefont {Cirac{\`\i}}\ \emph
  {et~al.}(2013{\natexlab{a}})\citenamefont {Cirac{\`\i}}, \citenamefont
  {Pendry},\ and\ \citenamefont {Smith}}]{Ciraci:2013dz}%
  \BibitemOpen
  \bibfield  {author} {\bibinfo {author} {\bibfnamefont {C.}~\bibnamefont
  {Cirac{\`\i}}}, \bibinfo {author} {\bibfnamefont {J.~B.}\ \bibnamefont
  {Pendry}}, \ and\ \bibinfo {author} {\bibfnamefont {D.~R.}\ \bibnamefont
  {Smith}},\ }\bibfield  {title} {\emph {\enquote {\bibinfo {title}
  {{Hydrodynamic Model for Plasmonics: A Macroscopic Approach to a Microscopic
  Problem}},}\ }}\href@noop {} {\bibfield  {journal} {\bibinfo  {journal}
  {ChemPhysChem}\ }\textbf {\bibinfo {volume} {14}},\ \bibinfo {pages} {1109}
  (\bibinfo {year} {2013}{\natexlab{a}})}\BibitemShut {NoStop}%
\bibitem [{\citenamefont {Raza}\ \emph {et~al.}(2015)\citenamefont {Raza},
  \citenamefont {Bozhevolnyi}, \citenamefont {Wubs},\ and\ \citenamefont
  {Mortensen}}]{Raza:2015ef}%
  \BibitemOpen
  \bibfield  {author} {\bibinfo {author} {\bibfnamefont {S.}~\bibnamefont
  {Raza}}, \bibinfo {author} {\bibfnamefont {S.~I.}\ \bibnamefont
  {Bozhevolnyi}}, \bibinfo {author} {\bibfnamefont {M.}~\bibnamefont {Wubs}}, \
  and\ \bibinfo {author} {\bibfnamefont {N.~A.}\ \bibnamefont {Mortensen}},\
  }\bibfield  {title} {\emph {\enquote {\bibinfo {title} {{Nonlocal optical
  response in metallic nanostructures}},}\ }}\href@noop {} {\bibfield
  {journal} {\bibinfo  {journal} {J. Phys. Condens. Matter}\ }\textbf {\bibinfo
  {volume} {27}},\ \bibinfo {pages} {183204} (\bibinfo {year}
  {2015})}\BibitemShut {NoStop}%
\bibitem [{\citenamefont {Parr}\ and\ \citenamefont {Yang}(1989)}]{parrbook}%
  \BibitemOpen
  \bibinfo {editor} {\bibfnamefont {R.~G.}\ \bibnamefont {Parr}}\ and\ \bibinfo
  {editor} {\bibfnamefont {W.}~\bibnamefont {Yang}},\ eds.,\ \href@noop {}
  {\emph {\bibinfo {title} {Density Functional Theory of Atoms and
  Molecules}}}\ (\bibinfo  {publisher} {Oxford University Press},\ \bibinfo
  {year} {1989})\BibitemShut {NoStop}%
\bibitem [{\citenamefont {Heinrichs}(1973)}]{Heinrichs:1973hn}%
  \BibitemOpen
  \bibfield  {author} {\bibinfo {author} {\bibfnamefont {J.}~\bibnamefont
  {Heinrichs}},\ }\bibfield  {title} {\emph {\enquote {\bibinfo {title}
  {{Hydrodynamic Theory of Surface-Plasmon Dispersion}},}\ }}\href@noop {}
  {\bibfield  {journal} {\bibinfo  {journal} {Phys. Rev. B}\ }\textbf {\bibinfo
  {volume} {7}},\ \bibinfo {pages} {3487} (\bibinfo {year} {1973})}\BibitemShut
  {NoStop}%
\bibitem [{\citenamefont {Eguiluz}\ \emph {et~al.}(1975)\citenamefont
  {Eguiluz}, \citenamefont {Ying},\ and\ \citenamefont
  {Quinn}}]{Eguiluz:1975be}%
  \BibitemOpen
  \bibfield  {author} {\bibinfo {author} {\bibfnamefont {A.}~\bibnamefont
  {Eguiluz}}, \bibinfo {author} {\bibfnamefont {S.}~\bibnamefont {Ying}}, \
  and\ \bibinfo {author} {\bibfnamefont {J.}~\bibnamefont {Quinn}},\ }\bibfield
   {title} {\emph {\enquote {\bibinfo {title} {{Influence of the electron
  density profile on surface plasmons in a hydrodynamic model}},}\ }}\href@noop
  {} {\bibfield  {journal} {\bibinfo  {journal} {Phys. Rev. B}\ }\textbf
  {\bibinfo {volume} {11}},\ \bibinfo {pages} {2118} (\bibinfo {year}
  {1975})}\BibitemShut {NoStop}%
\bibitem [{\citenamefont {Eguiluz}\ and\ \citenamefont
  {Quinn}(1976)}]{Eguiluz:1976wk}%
  \BibitemOpen
  \bibfield  {author} {\bibinfo {author} {\bibfnamefont {A.}~\bibnamefont
  {Eguiluz}}\ and\ \bibinfo {author} {\bibfnamefont {J.}~\bibnamefont
  {Quinn}},\ }\bibfield  {title} {\emph {\enquote {\bibinfo {title}
  {{Hydrodynamic model for surface plasmons in metals and degenerate
  semiconductors}},}\ }}\href@noop {} {\bibfield  {journal} {\bibinfo
  {journal} {Phys. Rev. B}\ }\textbf {\bibinfo {volume} {14}},\ \bibinfo
  {pages} {1347} (\bibinfo {year} {1976})}\BibitemShut {NoStop}%
\bibitem [{\citenamefont {Ruppin}(1975)}]{ruppin75}%
  \BibitemOpen
  \bibfield  {author} {\bibinfo {author} {\bibfnamefont {R.}~\bibnamefont
  {Ruppin}},\ }\bibfield  {title} {\emph {\enquote {\bibinfo {title} {Optical
  properties of small metal spheres},}\ }}\href@noop {} {\bibfield  {journal}
  {\bibinfo  {journal} {Phys. Rev. B}\ }\textbf {\bibinfo {volume} {11}},\
  \bibinfo {pages} {2871} (\bibinfo {year} {1975})}\BibitemShut {NoStop}%
\bibitem [{\citenamefont {Ruppin}(1978)}]{ruppin78}%
  \BibitemOpen
  \bibfield  {author} {\bibinfo {author} {\bibfnamefont {R.}~\bibnamefont
  {Ruppin}},\ }\bibfield  {title} {\emph {\enquote {\bibinfo {title} {Plasmon
  frequencies of small metal spheres},}\ }}\href@noop {} {\bibfield  {journal}
  {\bibinfo  {journal} {J. Phys. Chem. Solids}\ }\textbf {\bibinfo {volume}
  {39}},\ \bibinfo {pages} {233 } (\bibinfo {year} {1978})}\BibitemShut
  {NoStop}%
\bibitem [{\citenamefont {Dasgupta}\ and\ \citenamefont
  {Fuchs}(1981)}]{fuchs81}%
  \BibitemOpen
  \bibfield  {author} {\bibinfo {author} {\bibfnamefont {B.~B.}\ \bibnamefont
  {Dasgupta}}\ and\ \bibinfo {author} {\bibfnamefont {R.}~\bibnamefont
  {Fuchs}},\ }\bibfield  {title} {\emph {\enquote {\bibinfo {title}
  {Polarizability of a small sphere including nonlocal effects},}\ }}\href
  {\doibase 10.1103/PhysRevB.24.554} {\bibfield  {journal} {\bibinfo  {journal}
  {Phys. Rev. B}\ }\textbf {\bibinfo {volume} {24}},\ \bibinfo {pages} {554}
  (\bibinfo {year} {1981})}\BibitemShut {NoStop}%
\bibitem [{\citenamefont {Mock}\ \emph {et~al.}(2008)\citenamefont {Mock},
  \citenamefont {Hill}, \citenamefont {Degiron}, \citenamefont {Zauscher},
  \citenamefont {Chilkoti},\ and\ \citenamefont {Smith}}]{Anonymous:Vp6g18Xg}%
  \BibitemOpen
  \bibfield  {author} {\bibinfo {author} {\bibfnamefont {J.~J.}\ \bibnamefont
  {Mock}}, \bibinfo {author} {\bibfnamefont {R.}~\bibnamefont {Hill}}, \bibinfo
  {author} {\bibfnamefont {A.}~\bibnamefont {Degiron}}, \bibinfo {author}
  {\bibfnamefont {S.}~\bibnamefont {Zauscher}}, \bibinfo {author}
  {\bibfnamefont {A.}~\bibnamefont {Chilkoti}}, \ and\ \bibinfo {author}
  {\bibfnamefont {D.~R.}\ \bibnamefont {Smith}},\ }\bibfield  {title} {\emph
  {\enquote {\bibinfo {title} {{Distance-dependent plasmon resonant coupling
  between a gold nanoparticle and gold film}},}\ }}\href@noop {} {\bibfield
  {journal} {\bibinfo  {journal} {Nano Lett.}\ }\textbf {\bibinfo {volume}
  {8}},\ \bibinfo {pages} {2245} (\bibinfo {year} {2008})}\BibitemShut
  {NoStop}%
\bibitem [{\citenamefont {Hill}\ \emph {et~al.}(2010)\citenamefont {Hill},
  \citenamefont {Mock}, \citenamefont {Urzhumov},\ and\ \citenamefont
  {Sebba}}]{Hill:2010ez}%
  \BibitemOpen
  \bibfield  {author} {\bibinfo {author} {\bibfnamefont {R.}~\bibnamefont
  {Hill}}, \bibinfo {author} {\bibfnamefont {J.~J.}\ \bibnamefont {Mock}},
  \bibinfo {author} {\bibfnamefont {Y.~A.}\ \bibnamefont {Urzhumov}}, \ and\
  \bibinfo {author} {\bibfnamefont {D.~S.}\ \bibnamefont {Sebba}},\ }\bibfield
  {title} {\emph {\enquote {\bibinfo {title} {{Leveraging nanoscale plasmonic
  modes to achieve reproducible enhancement of light}},}\ }}\href@noop {}
  {\bibfield  {journal} {\bibinfo  {journal} {Nano Lett.}\ }\textbf {\bibinfo
  {volume} {10}},\ \bibinfo {pages} {4150} (\bibinfo {year}
  {2010})}\BibitemShut {NoStop}%
\bibitem [{\citenamefont {Mortensen}\ \emph {et~al.}(2014)\citenamefont
  {Mortensen}, \citenamefont {Raza}, \citenamefont {Wubs}, \citenamefont
  {S{\o}ndergaard},\ and\ \citenamefont {Bozhevolnyi}}]{Mortensen:2014kc}%
  \BibitemOpen
  \bibfield  {author} {\bibinfo {author} {\bibfnamefont {N.~A.}\ \bibnamefont
  {Mortensen}}, \bibinfo {author} {\bibfnamefont {S.}~\bibnamefont {Raza}},
  \bibinfo {author} {\bibfnamefont {M.}~\bibnamefont {Wubs}}, \bibinfo {author}
  {\bibfnamefont {T.}~\bibnamefont {S{\o}ndergaard}}, \ and\ \bibinfo {author}
  {\bibfnamefont {S.~I.}\ \bibnamefont {Bozhevolnyi}},\ }\bibfield  {title}
  {\emph {\enquote {\bibinfo {title} {{A generalized non-local optical response
  theory for plasmonic nanostructures}},}\ }}\href@noop {} {\bibfield
  {journal} {\bibinfo  {journal} {Nat. Commun.}\ }\textbf {\bibinfo {volume}
  {5}},\ \bibinfo {pages} {3809} (\bibinfo {year} {2014})}\BibitemShut
  {NoStop}%
\bibitem [{\citenamefont {Toscano}\ \emph {et~al.}(2013)\citenamefont
  {Toscano}, \citenamefont {Raza}, \citenamefont {Yan}, \citenamefont
  {Jeppesen},\ and\ \citenamefont {Xiao}}]{Toscano:2013hr}%
  \BibitemOpen
  \bibfield  {author} {\bibinfo {author} {\bibfnamefont {G.}~\bibnamefont
  {Toscano}}, \bibinfo {author} {\bibfnamefont {S.}~\bibnamefont {Raza}},
  \bibinfo {author} {\bibfnamefont {W.}~\bibnamefont {Yan}}, \bibinfo {author}
  {\bibfnamefont {C.}~\bibnamefont {Jeppesen}}, \ and\ \bibinfo {author}
  {\bibfnamefont {S.}~\bibnamefont {Xiao}},\ }\bibfield  {title} {\emph
  {\enquote {\bibinfo {title} {{Nonlocal response in plasmonic waveguiding with
  extreme light confinement}},}\ }}\href@noop {} {\bibfield  {journal}
  {\bibinfo  {journal} {Nanophotonics}\ }\textbf {\bibinfo {volume} {2}},\
  \bibinfo {pages} {161} (\bibinfo {year} {2013})}\BibitemShut {NoStop}%
\bibitem [{\citenamefont {Filter}\ \emph {et~al.}(2014)\citenamefont {Filter},
  \citenamefont {B{\"o}sel}, \citenamefont {Toscano}, \citenamefont {Lederer},\
  and\ \citenamefont {Rockstuhl}}]{Filter:2014}%
  \BibitemOpen
  \bibfield  {author} {\bibinfo {author} {\bibfnamefont {R.}~\bibnamefont
  {Filter}}, \bibinfo {author} {\bibfnamefont {C.}~\bibnamefont {B{\"o}sel}},
  \bibinfo {author} {\bibfnamefont {G.}~\bibnamefont {Toscano}}, \bibinfo
  {author} {\bibfnamefont {F.}~\bibnamefont {Lederer}}, \ and\ \bibinfo
  {author} {\bibfnamefont {C.}~\bibnamefont {Rockstuhl}},\ }\bibfield  {title}
  {\emph {\enquote {\bibinfo {title} {{Nonlocal effects: relevance for the
  spontaneous emission rates of quantum emitters coupled to plasmonic
  structures }},}\ }}\href@noop {} {\bibfield  {journal} {\bibinfo  {journal}
  {Opt. Lett.}\ }\textbf {\bibinfo {volume} {39}},\ \bibinfo {pages} {6118}
  (\bibinfo {year} {2014})}\BibitemShut {NoStop}%
\bibitem [{\citenamefont {Cirac{\`\i}}\ \emph
  {et~al.}(2013{\natexlab{b}})\citenamefont {Cirac{\`\i}}, \citenamefont
  {Urzhumov},\ and\ \citenamefont {Smith}}]{Ciraci:2013jt}%
  \BibitemOpen
  \bibfield  {author} {\bibinfo {author} {\bibfnamefont {C.}~\bibnamefont
  {Cirac{\`\i}}}, \bibinfo {author} {\bibfnamefont {Y.~A.}\ \bibnamefont
  {Urzhumov}}, \ and\ \bibinfo {author} {\bibfnamefont {D.~R.}\ \bibnamefont
  {Smith}},\ }\bibfield  {title} {\emph {\enquote {\bibinfo {title} {{Effects
  of classical nonlocality on the optical response of three-dimensional
  plasmonic nanodimers}},}\ }}\href@noop {} {\bibfield  {journal} {\bibinfo
  {journal} {J. Opt. Soc. Am. B}\ }\textbf {\bibinfo {volume} {30}},\ \bibinfo
  {pages} {2731} (\bibinfo {year} {2013}{\natexlab{b}})}\BibitemShut {NoStop}%
\bibitem [{\citenamefont {Fern{\'a}ndez-Dom{\'\i}nguez}\ \emph
  {et~al.}(2012{\natexlab{a}})\citenamefont {Fern{\'a}ndez-Dom{\'\i}nguez},
  \citenamefont {Zhang}, \citenamefont {Luo}, \citenamefont {Maier},
  \citenamefont {Garc{\'\i}a-Vidal},\ and\ \citenamefont
  {Pendry}}]{FernandezDominguez:2012eg}%
  \BibitemOpen
  \bibfield  {author} {\bibinfo {author} {\bibfnamefont {A.~I.}\ \bibnamefont
  {Fern{\'a}ndez-Dom{\'\i}nguez}}, \bibinfo {author} {\bibfnamefont
  {P.}~\bibnamefont {Zhang}}, \bibinfo {author} {\bibfnamefont
  {Y.}~\bibnamefont {Luo}}, \bibinfo {author} {\bibfnamefont {S.~A.}\
  \bibnamefont {Maier}}, \bibinfo {author} {\bibfnamefont {F.~J.}\ \bibnamefont
  {Garc{\'\i}a-Vidal}}, \ and\ \bibinfo {author} {\bibfnamefont {J.~B.}\
  \bibnamefont {Pendry}},\ }\bibfield  {title} {\emph {\enquote {\bibinfo
  {title} {{Transformation-optics insight into nonlocal effects in separated
  nanowires}},}\ }}\href@noop {} {\bibfield  {journal} {\bibinfo  {journal}
  {Phys. Rev. B}\ }\textbf {\bibinfo {volume} {86}},\ \bibinfo {pages} {241110}
  (\bibinfo {year} {2012}{\natexlab{a}})}\BibitemShut {NoStop}%
\bibitem [{\citenamefont {Christensen}\ \emph {et~al.}(2014)\citenamefont
  {Christensen}, \citenamefont {Yan}, \citenamefont {Raza}, \citenamefont
  {Jauho}, \citenamefont {Mortensen}, \citenamefont {{Morte}},\ and\
  \citenamefont {Wubs}}]{Christensen:2014tm}%
  \BibitemOpen
  \bibfield  {author} {\bibinfo {author} {\bibfnamefont {T.}~\bibnamefont
  {Christensen}}, \bibinfo {author} {\bibfnamefont {W.}~\bibnamefont {Yan}},
  \bibinfo {author} {\bibfnamefont {S.}~\bibnamefont {Raza}}, \bibinfo {author}
  {\bibfnamefont {A.-P.}\ \bibnamefont {Jauho}}, \bibinfo {author}
  {\bibfnamefont {N.~A.}\ \bibnamefont {Mortensen}}, \bibinfo {author}
  {\bibnamefont {{Morte}}}, \ and\ \bibinfo {author} {\bibfnamefont
  {M.}~\bibnamefont {Wubs}},\ }\bibfield  {title} {\emph {\enquote {\bibinfo
  {title} {{Nonlocal Response of Metallic Nanospheres Probed by Light,
  Electrons, and Atoms}},}\ }}\href@noop {} {\bibfield  {journal} {\bibinfo
  {journal} {ACS Nano}\ }\textbf {\bibinfo {volume} {8}},\ \bibinfo {pages}
  {1745} (\bibinfo {year} {2014})}\BibitemShut {NoStop}%
\bibitem [{\citenamefont {Luo}\ \emph {et~al.}(2013)\citenamefont {Luo},
  \citenamefont {Fernandez-Dominguez}, \citenamefont {Wiener},\ and\
  \citenamefont {Maier}}]{Luo:2013jx}%
  \BibitemOpen
  \bibfield  {author} {\bibinfo {author} {\bibfnamefont {Y.}~\bibnamefont
  {Luo}}, \bibinfo {author} {\bibfnamefont {A.~I.}\ \bibnamefont
  {Fernandez-Dominguez}}, \bibinfo {author} {\bibfnamefont {A.}~\bibnamefont
  {Wiener}}, \ and\ \bibinfo {author} {\bibfnamefont {S.~A.}\ \bibnamefont
  {Maier}},\ }\bibfield  {title} {\emph {\enquote {\bibinfo {title} {{Surface
  plasmons and nonlocality: A simple model}},}\ }}\href@noop {} {\bibfield
  {journal} {\bibinfo  {journal} {Phys. Rev. Lett.}\ }\textbf {\bibinfo
  {volume} {111}},\ \bibinfo {pages} {093901} (\bibinfo {year}
  {2013})}\BibitemShut {NoStop}%
\bibitem [{\citenamefont {Wiener}\ \emph {et~al.}(2013)\citenamefont {Wiener},
  \citenamefont {Fern{\'a}ndez-Dom{\'\i}nguez}, \citenamefont {Pendry},
  \citenamefont {Horsfield},\ and\ \citenamefont {Maier}}]{Wiener:2013gl}%
  \BibitemOpen
  \bibfield  {author} {\bibinfo {author} {\bibfnamefont {A.}~\bibnamefont
  {Wiener}}, \bibinfo {author} {\bibfnamefont {A.~I.}\ \bibnamefont
  {Fern{\'a}ndez-Dom{\'\i}nguez}}, \bibinfo {author} {\bibfnamefont {J.~B.}\
  \bibnamefont {Pendry}}, \bibinfo {author} {\bibfnamefont {A.~P.}\
  \bibnamefont {Horsfield}}, \ and\ \bibinfo {author} {\bibfnamefont {S.~A.}\
  \bibnamefont {Maier}},\ }\bibfield  {title} {\emph {\enquote {\bibinfo
  {title} {{Nonlocal propagation and tunnelling of surface plasmons in metallic
  hourglass waveguides}},}\ }}\href@noop {} {\bibfield  {journal} {\bibinfo
  {journal} {Opt. Express}\ }\textbf {\bibinfo {volume} {21}},\ \bibinfo
  {pages} {27509} (\bibinfo {year} {2013})}\BibitemShut {NoStop}%
\bibitem [{\citenamefont {Wiener}\ \emph {et~al.}(2012)\citenamefont {Wiener},
  \citenamefont {Fernandez-Dominguez},\ and\ \citenamefont
  {Horsfield}}]{Wiener:2012fd}%
  \BibitemOpen
  \bibfield  {author} {\bibinfo {author} {\bibfnamefont {A.}~\bibnamefont
  {Wiener}}, \bibinfo {author} {\bibfnamefont {A.~I.}\ \bibnamefont
  {Fernandez-Dominguez}}, \ and\ \bibinfo {author} {\bibfnamefont {A.~P.}\
  \bibnamefont {Horsfield}},\ }\bibfield  {title} {\emph {\enquote {\bibinfo
  {title} {{Nonlocal effects in the nanofocusing performance of plasmonic
  tips}},}\ }}\href@noop {} {\bibfield  {journal} {\bibinfo  {journal} {Nano
  Lett.}\ }\textbf {\bibinfo {volume} {12}},\ \bibinfo {pages} {3308} (\bibinfo
  {year} {2012})}\BibitemShut {NoStop}%
\bibitem [{\citenamefont {Stella}\ \emph {et~al.}(2013)\citenamefont {Stella},
  \citenamefont {Zhang},\ and\ \citenamefont
  {Garc{\'\i}a-Vidal}}]{Stella:2013by}%
  \BibitemOpen
  \bibfield  {author} {\bibinfo {author} {\bibfnamefont {L.}~\bibnamefont
  {Stella}}, \bibinfo {author} {\bibfnamefont {P.}~\bibnamefont {Zhang}}, \
  and\ \bibinfo {author} {\bibfnamefont {F.~J.}\ \bibnamefont
  {Garc{\'\i}a-Vidal}},\ }\bibfield  {title} {\emph {\enquote {\bibinfo {title}
  {{Performance of nonlocal optics when applied to plasmonic
  nanostructures}},}\ }}\href@noop {} {\bibfield  {journal} {\bibinfo
  {journal} {J. Phys. Chem. C}\ }\textbf {\bibinfo {volume} {117}},\ \bibinfo
  {pages} {8941} (\bibinfo {year} {2013})}\BibitemShut {NoStop}%
\bibitem [{\citenamefont {Lerm\'e}\ \emph {et~al.}(1999)\citenamefont
  {Lerm\'e}, \citenamefont {Palpant}, \citenamefont {Cottancin}, \citenamefont
  {Pellarin}, \citenamefont {Pr\'evel}, \citenamefont {Vialle},\ and\
  \citenamefont {Broyer}}]{lerme99}%
  \BibitemOpen
  \bibfield  {author} {\bibinfo {author} {\bibfnamefont {J.}~\bibnamefont
  {Lerm\'e}}, \bibinfo {author} {\bibfnamefont {B.}~\bibnamefont {Palpant}},
  \bibinfo {author} {\bibfnamefont {E.}~\bibnamefont {Cottancin}}, \bibinfo
  {author} {\bibfnamefont {M.}~\bibnamefont {Pellarin}}, \bibinfo {author}
  {\bibfnamefont {B.}~\bibnamefont {Pr\'evel}}, \bibinfo {author}
  {\bibfnamefont {J.~L.}\ \bibnamefont {Vialle}}, \ and\ \bibinfo {author}
  {\bibfnamefont {M.}~\bibnamefont {Broyer}},\ }\bibfield  {title} {\emph
  {\enquote {\bibinfo {title} {Quantum extension of mie's theory in the dipolar
  approximation},}\ }}\href {\doibase 10.1103/PhysRevB.60.16151} {\bibfield
  {journal} {\bibinfo  {journal} {Phys. Rev. B}\ }\textbf {\bibinfo {volume}
  {60}},\ \bibinfo {pages} {16151} (\bibinfo {year} {1999})}\BibitemShut
  {NoStop}%
\bibitem [{\citenamefont {Zapata}\ \emph {et~al.}(2015)\citenamefont {Zapata},
  \citenamefont {Camacho~Beltr{\'a}n}, \citenamefont {Borisov},\ and\
  \citenamefont {Aizpurua}}]{Zapata:2015fq}%
  \BibitemOpen
  \bibfield  {author} {\bibinfo {author} {\bibfnamefont {M.}~\bibnamefont
  {Zapata}}, \bibinfo {author} {\bibfnamefont {{\'A}.~S.}\ \bibnamefont
  {Camacho~Beltr{\'a}n}}, \bibinfo {author} {\bibfnamefont {A.~G.}\
  \bibnamefont {Borisov}}, \ and\ \bibinfo {author} {\bibfnamefont
  {J.}~\bibnamefont {Aizpurua}},\ }\bibfield  {title} {\emph {\enquote
  {\bibinfo {title} {{Quantum effects in the optical response of extended
  plasmonic gaps: validation of the quantum corrected model in core-shell
  nanomatryushkas}},}\ }}\href@noop {} {\bibfield  {journal} {\bibinfo
  {journal} {Opt. Express}\ }\textbf {\bibinfo {volume} {23}},\ \bibinfo
  {pages} {8134} (\bibinfo {year} {2015})}\BibitemShut {NoStop}%
\bibitem [{\citenamefont {Domps}\ \emph {et~al.}(1998)\citenamefont {Domps},
  \citenamefont {Reinhard},\ and\ \citenamefont {Suraud}}]{domps98}%
  \BibitemOpen
  \bibfield  {author} {\bibinfo {author} {\bibfnamefont {A.}~\bibnamefont
  {Domps}}, \bibinfo {author} {\bibfnamefont {P.-G.}\ \bibnamefont {Reinhard}},
  \ and\ \bibinfo {author} {\bibfnamefont {E.}~\bibnamefont {Suraud}},\
  }\bibfield  {title} {\emph {\enquote {\bibinfo {title} {Time-dependent
  thomas-fermi approach for electron dynamics in metal clusters},}\
  }}\href@noop {} {\bibfield  {journal} {\bibinfo  {journal} {Phys. Rev.
  Lett.}\ }\textbf {\bibinfo {volume} {80}},\ \bibinfo {pages} {5520} (\bibinfo
  {year} {1998})}\BibitemShut {NoStop}%
\bibitem [{\citenamefont {Xiang}\ \emph {et~al.}(2014)\citenamefont {Xiang},
  \citenamefont {Zhang}, \citenamefont {Neuhauser},\ and\ \citenamefont
  {Lu}}]{xiang2014}%
  \BibitemOpen
  \bibfield  {author} {\bibinfo {author} {\bibfnamefont {H.}~\bibnamefont
  {Xiang}}, \bibinfo {author} {\bibfnamefont {X.}~\bibnamefont {Zhang}},
  \bibinfo {author} {\bibfnamefont {D.}~\bibnamefont {Neuhauser}}, \ and\
  \bibinfo {author} {\bibfnamefont {G.}~\bibnamefont {Lu}},\ }\bibfield
  {title} {\emph {\enquote {\bibinfo {title} {{Size-Dependent Plasmonic
  Resonances from Large-Scale Quantum Simulations}},}\ }}\href@noop {}
  {\bibfield  {journal} {\bibinfo  {journal} {J. Phys. Chem. Lett.}\ }\textbf
  {\bibinfo {volume} {5}},\ \bibinfo {pages} {1163} (\bibinfo {year}
  {2014})}\BibitemShut {NoStop}%
\bibitem [{\citenamefont {Ball}\ \emph {et~al.}(1973)\citenamefont {Ball},
  \citenamefont {Wheeler},\ and\ \citenamefont {Firemen}}]{ball73}%
  \BibitemOpen
  \bibfield  {author} {\bibinfo {author} {\bibfnamefont {J.~A.}\ \bibnamefont
  {Ball}}, \bibinfo {author} {\bibfnamefont {J.~A.}\ \bibnamefont {Wheeler}}, \
  and\ \bibinfo {author} {\bibfnamefont {E.~L.}\ \bibnamefont {Firemen}},\
  }\bibfield  {title} {\emph {\enquote {\bibinfo {title} {Photoabsorption and
  charge oscillation of the thomas-fermi atom},}\ }}\href@noop {} {\bibfield
  {journal} {\bibinfo  {journal} {Rev. Mod. Phys.}\ }\textbf {\bibinfo {volume}
  {45}},\ \bibinfo {pages} {333} (\bibinfo {year} {1973})}\BibitemShut
  {NoStop}%
\bibitem [{\citenamefont {Walecka}(1976)}]{walecka76}%
  \BibitemOpen
  \bibfield  {author} {\bibinfo {author} {\bibfnamefont {J.}~\bibnamefont
  {Walecka}},\ }\bibfield  {title} {\emph {\enquote {\bibinfo {title}
  {Collective excitations in atoms},}\ }}\href@noop {} {\bibfield  {journal}
  {\bibinfo  {journal} {Phys. Lett. A}\ }\textbf {\bibinfo {volume} {58}},\
  \bibinfo {pages} {83 } (\bibinfo {year} {1976})}\BibitemShut {NoStop}%
\bibitem [{\citenamefont {Monaghan}(1974)}]{mona74}%
  \BibitemOpen
  \bibfield  {author} {\bibinfo {author} {\bibfnamefont {J.}~\bibnamefont
  {Monaghan}},\ }\bibfield  {title} {\emph {\enquote {\bibinfo {title}
  {Collective oscillations in many electron atoms. {III}. {P}hotoabsorption},}\
  }}\href@noop {} {\bibfield  {journal} {\bibinfo  {journal} {Aus. J. Phys.}\
  }\textbf {\bibinfo {volume} {27}},\ \bibinfo {pages} {667} (\bibinfo {year}
  {1974})}\BibitemShut {NoStop}%
\bibitem [{\citenamefont {Bennett}(1970)}]{bennet70}%
  \BibitemOpen
  \bibfield  {author} {\bibinfo {author} {\bibfnamefont {A.~J.}\ \bibnamefont
  {Bennett}},\ }\bibfield  {title} {\emph {\enquote {\bibinfo {title}
  {Influence of the electron charge distribution on surface-plasmon
  dispersion},}\ }}\href {\doibase 10.1103/PhysRevB.1.203} {\bibfield
  {journal} {\bibinfo  {journal} {Phys. Rev. B}\ }\textbf {\bibinfo {volume}
  {1}},\ \bibinfo {pages} {203} (\bibinfo {year} {1970})}\BibitemShut {NoStop}%
\bibitem [{\citenamefont {Schwartz}\ and\ \citenamefont
  {Schaich}(1982)}]{Schwartz82}%
  \BibitemOpen
  \bibfield  {author} {\bibinfo {author} {\bibfnamefont {C.}~\bibnamefont
  {Schwartz}}\ and\ \bibinfo {author} {\bibfnamefont {W.~L.}\ \bibnamefont
  {Schaich}},\ }\bibfield  {title} {\emph {\enquote {\bibinfo {title}
  {Hydrodynamic models of surface plasmons},}\ }}\href {\doibase
  10.1103/PhysRevB.26.7008} {\bibfield  {journal} {\bibinfo  {journal} {Phys.
  Rev. B}\ }\textbf {\bibinfo {volume} {26}},\ \bibinfo {pages} {7008}
  (\bibinfo {year} {1982})}\BibitemShut {NoStop}%
\bibitem [{\citenamefont {David}\ and\ \citenamefont {Garc{\'\i}a~de
  Abajo}(2014)}]{David:2014iw}%
  \BibitemOpen
  \bibfield  {author} {\bibinfo {author} {\bibfnamefont {C.}~\bibnamefont
  {David}}\ and\ \bibinfo {author} {\bibfnamefont {F.~J.}\ \bibnamefont
  {Garc{\'\i}a~de Abajo}},\ }\bibfield  {title} {\emph {\enquote {\bibinfo
  {title} {{Surface Plasmon Dependence on the Electron Density Profile at Metal
  Surfaces}},}\ }}\href@noop {} {\bibfield  {journal} {\bibinfo  {journal} {ACS
  Nano}\ }\textbf {\bibinfo {volume} {8}},\ \bibinfo {pages} {9558} (\bibinfo
  {year} {2014})}\BibitemShut {NoStop}%
\bibitem [{\citenamefont {Malzacher}\ and\ \citenamefont
  {Dreizler}(1982)}]{dreiz82}%
  \BibitemOpen
  \bibfield  {author} {\bibinfo {author} {\bibfnamefont {P.}~\bibnamefont
  {Malzacher}}\ and\ \bibinfo {author} {\bibfnamefont {R.~M.}\ \bibnamefont
  {Dreizler}},\ }\bibfield  {title} {\emph {\enquote {\bibinfo {title} {{Charge
  oscillations and photoabsorption of the
  {T}homas-{F}ermi-{D}irac-{W}eizs{\"a}cker atom}},}\ }}\href@noop {}
  {\bibfield  {journal} {\bibinfo  {journal} {Z. Phys. A}\ }\textbf {\bibinfo
  {volume} {307}},\ \bibinfo {pages} {211} (\bibinfo {year}
  {1982})}\BibitemShut {NoStop}%
\bibitem [{\citenamefont {Banerjee}\ and\ \citenamefont
  {Harbola}(2000)}]{harbola00}%
  \BibitemOpen
  \bibfield  {author} {\bibinfo {author} {\bibfnamefont {A.}~\bibnamefont
  {Banerjee}}\ and\ \bibinfo {author} {\bibfnamefont {M.~K.}\ \bibnamefont
  {Harbola}},\ }\bibfield  {title} {\emph {\enquote {\bibinfo {title}
  {Hydrodynamic approach to time-dependent density functional theory; response
  properties of metal clusters},}\ }}\href@noop {} {\bibfield  {journal}
  {\bibinfo  {journal} {J. Chem. Phys.}\ }\textbf {\bibinfo {volume} {113}},\
  \bibinfo {pages} {5614} (\bibinfo {year} {2000})}\BibitemShut {NoStop}%
\bibitem [{\citenamefont {Banerjee}\ and\ \citenamefont
  {Harbola}(2008)}]{harbola2008}%
  \BibitemOpen
  \bibfield  {author} {\bibinfo {author} {\bibfnamefont {A.}~\bibnamefont
  {Banerjee}}\ and\ \bibinfo {author} {\bibfnamefont {M.~K.}\ \bibnamefont
  {Harbola}},\ }\bibfield  {title} {\emph {\enquote {\bibinfo {title}
  {Hydrodynamical approach to collective oscillations in metal clusters},}\
  }}\href@noop {} {\bibfield  {journal} {\bibinfo  {journal} {Phys. Lett. A}\
  }\textbf {\bibinfo {volume} {372}},\ \bibinfo {pages} {2881 } (\bibinfo
  {year} {2008})}\BibitemShut {NoStop}%
\bibitem [{\citenamefont {Bonitz}\ \emph {et~al.}(2014)\citenamefont {Bonitz},
  \citenamefont {Lopez}, \citenamefont {Becker},\ and\ \citenamefont
  {Thomsen}}]{plasmabook}%
  \BibitemOpen
  \bibinfo {editor} {\bibfnamefont {M.}~\bibnamefont {Bonitz}}, \bibinfo
  {editor} {\bibfnamefont {J.}~\bibnamefont {Lopez}}, \bibinfo {editor}
  {\bibfnamefont {K.}~\bibnamefont {Becker}}, \ and\ \bibinfo {editor}
  {\bibfnamefont {H.}~\bibnamefont {Thomsen}},\ eds.,\ \href@noop {} {\emph
  {\bibinfo {title} {Complex Plasmas: Scientific Challenges and Technological
  Opportunities}}}\ (\bibinfo  {publisher} {Springer},\ \bibinfo {year}
  {2014})\BibitemShut {NoStop}%
\bibitem [{\citenamefont {Manfredi}(2005)}]{manfredi05}%
  \BibitemOpen
  \bibfield  {author} {\bibinfo {author} {\bibfnamefont {G.}~\bibnamefont
  {Manfredi}},\ }\bibfield  {title} {\emph {\enquote {\bibinfo {title} {How to
  model quantum plasma},}\ }}\href@noop {} {\bibfield  {journal} {\bibinfo
  {journal} {Fields Inst. Comm.}\ }\textbf {\bibinfo {volume} {46}},\ \bibinfo
  {pages} {263} (\bibinfo {year} {2005})}\BibitemShut {NoStop}%
\bibitem [{\citenamefont {Shukla}\ and\ \citenamefont
  {Eliasson}(2012)}]{shukla12}%
  \BibitemOpen
  \bibfield  {author} {\bibinfo {author} {\bibfnamefont {P.~K.}\ \bibnamefont
  {Shukla}}\ and\ \bibinfo {author} {\bibfnamefont {B.}~\bibnamefont
  {Eliasson}},\ }\bibfield  {title} {\emph {\enquote {\bibinfo {title} {Novel
  attractive force between ions in quantum plasmas},}\ }}\href@noop {}
  {\bibfield  {journal} {\bibinfo  {journal} {Phys. Rev. Lett.}\ }\textbf
  {\bibinfo {volume} {108}},\ \bibinfo {pages} {165007} (\bibinfo {year}
  {2012})}\BibitemShut {NoStop}%
\bibitem [{\citenamefont {Akbari-Moghanjoughi}(2015)}]{akbari15}%
  \BibitemOpen
  \bibfield  {author} {\bibinfo {author} {\bibfnamefont {M.}~\bibnamefont
  {Akbari-Moghanjoughi}},\ }\bibfield  {title} {\emph {\enquote {\bibinfo
  {title} {Hydrodynamic limit of wigner-poisson kinetic theory: Revisited},}\
  }}\href@noop {} {\bibfield  {journal} {\bibinfo  {journal} {Phys. Plasmas}\
  }\textbf {\bibinfo {volume} {22}},\ \bibinfo {pages} {022103} (\bibinfo
  {year} {2015})}\BibitemShut {NoStop}%
\bibitem [{\citenamefont {Zaremba}\ and\ \citenamefont
  {Tso}(1994)}]{zaremba94}%
  \BibitemOpen
  \bibfield  {author} {\bibinfo {author} {\bibfnamefont {E.}~\bibnamefont
  {Zaremba}}\ and\ \bibinfo {author} {\bibfnamefont {H.~C.}\ \bibnamefont
  {Tso}},\ }\bibfield  {title} {\emph {\enquote {\bibinfo {title}
  {{T}homas-{F}ermi-{D}irac-von {W}eizs\"acker hydrodynamics in parabolic
  wells},}\ }}\href {\doibase 10.1103/PhysRevB.49.8147} {\bibfield  {journal}
  {\bibinfo  {journal} {Phys. Rev. B}\ }\textbf {\bibinfo {volume} {49}},\
  \bibinfo {pages} {8147} (\bibinfo {year} {1994})}\BibitemShut {NoStop}%
\bibitem [{\citenamefont {van Zyl}\ and\ \citenamefont
  {Zaremba}(1999)}]{zaremba99}%
  \BibitemOpen
  \bibfield  {author} {\bibinfo {author} {\bibfnamefont {B.~P.}\ \bibnamefont
  {van Zyl}}\ and\ \bibinfo {author} {\bibfnamefont {E.}~\bibnamefont
  {Zaremba}},\ }\bibfield  {title} {\emph {\enquote {\bibinfo {title}
  {{T}homas-{F}ermi-{D}irac-von {W}eizs\"acker hydrodynamics in laterally
  modulated electronic systems},}\ }}\href@noop {} {\bibfield  {journal}
  {\bibinfo  {journal} {Phys. Rev. B}\ }\textbf {\bibinfo {volume} {59}},\
  \bibinfo {pages} {2079} (\bibinfo {year} {1999})}\BibitemShut {NoStop}%
\bibitem [{\citenamefont {Yan}(2015)}]{Yan:2015ff}%
  \BibitemOpen
  \bibfield  {author} {\bibinfo {author} {\bibfnamefont {W.}~\bibnamefont
  {Yan}},\ }\bibfield  {title} {\emph {\enquote {\bibinfo {title}
  {{Hydrodynamic theory for quantum plasmonics: Linear-response dynamics of the
  inhomogeneous electron gas}},}\ }}\href@noop {} {\bibfield  {journal}
  {\bibinfo  {journal} {Phys. Rev. B}\ }\textbf {\bibinfo {volume} {91}},\
  \bibinfo {pages} {115416} (\bibinfo {year} {2015})}\BibitemShut {NoStop}%
\bibitem [{\citenamefont {Toscano}\ \emph {et~al.}(2015)\citenamefont
  {Toscano}, \citenamefont {Straubel}, \citenamefont {Kwiatkowski},
  \citenamefont {Rockstuhl}, \citenamefont {Evers}, \citenamefont {Xu},
  \citenamefont {Mortensen},\ and\ \citenamefont {Wubs}}]{Toscano:2015iw}%
  \BibitemOpen
  \bibfield  {author} {\bibinfo {author} {\bibfnamefont {G.}~\bibnamefont
  {Toscano}}, \bibinfo {author} {\bibfnamefont {J.}~\bibnamefont {Straubel}},
  \bibinfo {author} {\bibfnamefont {A.}~\bibnamefont {Kwiatkowski}}, \bibinfo
  {author} {\bibfnamefont {C.}~\bibnamefont {Rockstuhl}}, \bibinfo {author}
  {\bibfnamefont {F.}~\bibnamefont {Evers}}, \bibinfo {author} {\bibfnamefont
  {H.}~\bibnamefont {Xu}}, \bibinfo {author} {\bibfnamefont {N.~A.}\
  \bibnamefont {Mortensen}}, \ and\ \bibinfo {author} {\bibfnamefont
  {M.}~\bibnamefont {Wubs}},\ }\bibfield  {title} {\emph {\enquote {\bibinfo
  {title} {{Resonance shifts and spill-out effects in self-consistent
  hydrodynamic nanoplasmonics}},}\ }}\href@noop {} {\bibfield  {journal}
  {\bibinfo  {journal} {Nature Commun.}\ }\textbf {\bibinfo {volume} {6}},\
  \bibinfo {pages} {7132} (\bibinfo {year} {2015})}\BibitemShut {NoStop}%
\bibitem [{\citenamefont {Li}\ \emph {et~al.}(2015)\citenamefont {Li},
  \citenamefont {Fang}, \citenamefont {Weng}, \citenamefont {Zhang},
  \citenamefont {Dou}, \citenamefont {Yang},\ and\ \citenamefont
  {Yuan}}]{Li:2015io}%
  \BibitemOpen
  \bibfield  {author} {\bibinfo {author} {\bibfnamefont {X.}~\bibnamefont
  {Li}}, \bibinfo {author} {\bibfnamefont {H.}~\bibnamefont {Fang}}, \bibinfo
  {author} {\bibfnamefont {X.}~\bibnamefont {Weng}}, \bibinfo {author}
  {\bibfnamefont {L.}~\bibnamefont {Zhang}}, \bibinfo {author} {\bibfnamefont
  {X.}~\bibnamefont {Dou}}, \bibinfo {author} {\bibfnamefont {A.}~\bibnamefont
  {Yang}}, \ and\ \bibinfo {author} {\bibfnamefont {X.}~\bibnamefont {Yuan}},\
  }\bibfield  {title} {\emph {\enquote {\bibinfo {title} {{Electronic spill-out
  induced spectral broadening in quantum hydrodynamic nanoplasmonics}},}\
  }}\href@noop {} {\bibfield  {journal} {\bibinfo  {journal} {Opt. Express}\
  }\textbf {\bibinfo {volume} {23}},\ \bibinfo {pages} {29738} (\bibinfo {year}
  {2015})}\BibitemShut {NoStop}%
\bibitem [{\citenamefont {Tokatly}\ and\ \citenamefont
  {Pankratov}(2000)}]{Tokatly:2000fy}%
  \BibitemOpen
  \bibfield  {author} {\bibinfo {author} {\bibfnamefont {I.}~\bibnamefont
  {Tokatly}}\ and\ \bibinfo {author} {\bibfnamefont {O.}~\bibnamefont
  {Pankratov}},\ }\bibfield  {title} {\emph {\enquote {\bibinfo {title}
  {{Hydrodynamics beyond local equilibrium: Application to electron gas}},}\
  }}\href@noop {} {\bibfield  {journal} {\bibinfo  {journal} {Phys. Rev. B}\
  }\textbf {\bibinfo {volume} {62}},\ \bibinfo {pages} {2759} (\bibinfo {year}
  {2000})}\BibitemShut {NoStop}%
\bibitem [{\citenamefont {Tokatly}\ and\ \citenamefont
  {Pankratov}(1999)}]{Tokatly:1999id}%
  \BibitemOpen
  \bibfield  {author} {\bibinfo {author} {\bibfnamefont {I.}~\bibnamefont
  {Tokatly}}\ and\ \bibinfo {author} {\bibfnamefont {O.}~\bibnamefont
  {Pankratov}},\ }\bibfield  {title} {\emph {\enquote {\bibinfo {title}
  {{Hydrodynamic theory of an electron gas}},}\ }}\href@noop {} {\bibfield
  {journal} {\bibinfo  {journal} {Phys. Rev. B}\ }\textbf {\bibinfo {volume}
  {60}},\ \bibinfo {pages} {15550} (\bibinfo {year} {1999})}\BibitemShut
  {NoStop}%
\bibitem [{\citenamefont {Boardman}(1982)}]{boardmanbook}%
  \BibitemOpen
  \bibinfo {editor} {\bibfnamefont {A.}~\bibnamefont {Boardman}},\ ed.,\
  \href@noop {} {\emph {\bibinfo {title} {Electromagnetic Surface Modes
  Hydrodynamic Theory of Plasmon-Polaritonson Plane Surfaces}}}\ (\bibinfo
  {publisher} {Wiley},\ \bibinfo {year} {1982})\BibitemShut {NoStop}%
\bibitem [{\citenamefont {Dreizler}\ and\ \citenamefont
  {Gross}(1990)}]{dreibook}%
  \BibitemOpen
  \bibfield  {author} {\bibinfo {author} {\bibfnamefont {R.~M.}\ \bibnamefont
  {Dreizler}}\ and\ \bibinfo {author} {\bibfnamefont {E.~K.~U.}\ \bibnamefont
  {Gross}},\ }\href@noop {} {\emph {\bibinfo {title} {Density functional theory
  -- {A}n approach to the quantum many-body problem}}}\ (\bibinfo  {publisher}
  {Springer},\ \bibinfo {address} {Berlin},\ \bibinfo {year}
  {1990})\BibitemShut {NoStop}%
\bibitem [{\citenamefont {Ho}\ \emph {et~al.}(2008)\citenamefont {Ho},
  \citenamefont {Lign{\`e}res},\ and\ \citenamefont {Carter}}]{Ho:2008jq}%
  \BibitemOpen
  \bibfield  {author} {\bibinfo {author} {\bibfnamefont {G.~S.}\ \bibnamefont
  {Ho}}, \bibinfo {author} {\bibfnamefont {V.~L.}\ \bibnamefont
  {Lign{\`e}res}}, \ and\ \bibinfo {author} {\bibfnamefont {E.~A.}\
  \bibnamefont {Carter}},\ }\bibfield  {title} {\emph {\enquote {\bibinfo
  {title} {{Introducing PROFESS: A new program for orbital-free density
  functional theory calculations}},}\ }}\href@noop {} {\bibfield  {journal}
  {\bibinfo  {journal} {Comput. Phys. Commun.}\ }\textbf {\bibinfo {volume}
  {179}},\ \bibinfo {pages} {839} (\bibinfo {year} {2008})}\BibitemShut
  {NoStop}%
\bibitem [{\citenamefont {Perdew}\ and\ \citenamefont
  {Zunger}(1981)}]{Perdew:1981dv}%
  \BibitemOpen
  \bibfield  {author} {\bibinfo {author} {\bibfnamefont {J.~P.}\ \bibnamefont
  {Perdew}}\ and\ \bibinfo {author} {\bibfnamefont {A.}~\bibnamefont
  {Zunger}},\ }\bibfield  {title} {\emph {\enquote {\bibinfo {title}
  {{Self-interaction correction to density-functional approximations for
  many-electron systems}},}\ }}\href@noop {} {\bibfield  {journal} {\bibinfo
  {journal} {Phys. Rev. B}\ }\textbf {\bibinfo {volume} {23}},\ \bibinfo
  {pages} {5048} (\bibinfo {year} {1981})}\BibitemShut {NoStop}%
\bibitem [{\citenamefont {Wang}\ and\ \citenamefont
  {Carter}(2000)}]{wangcarterof}%
  \BibitemOpen
  \bibfield  {author} {\bibinfo {author} {\bibfnamefont {Y.}~\bibnamefont
  {Wang}}\ and\ \bibinfo {author} {\bibfnamefont {E.~A.}\ \bibnamefont
  {Carter}},\ }in\ \href@noop {} {\emph {\bibinfo {booktitle} {Progress in
  Theoretical Chemistry and Physics}}},\ \bibinfo {editor} {edited by\ \bibinfo
  {editor} {\bibfnamefont {S.}~\bibnamefont {Schwartz}}}\ (\bibinfo
  {publisher} {Kluwer},\ \bibinfo {address} {Dordrecht},\ \bibinfo {year}
  {2000})\ p.\ \bibinfo {pages} {117}\BibitemShut {NoStop}%
\bibitem [{\citenamefont {Yariv}(1988)}]{Yariv:1988vc}%
  \BibitemOpen
  \bibfield  {author} {\bibinfo {author} {\bibfnamefont {A.}~\bibnamefont
  {Yariv}},\ }\href@noop {} {\emph {\bibinfo {title} {{Quantum electronics; 3rd
  ed.}}}}\ (\bibinfo  {publisher} {Wiley},\ \bibinfo {address} {New York, NY},\
  \bibinfo {year} {1988})\BibitemShut {NoStop}%
\bibitem [{\citenamefont {\textsc{Comsol} Multiphysics}()}]{comsol}%
  \BibitemOpen
  \bibfield  {author} {\bibinfo {author} {\bibnamefont {\textsc{Comsol}
  Multiphysics}},\ }\href@noop {} {}\bibinfo {note}
  {\url{http://www.comsol.com}}\BibitemShut {NoStop}%
\bibitem [{\citenamefont {Cirac{\`\i}}\ \emph
  {et~al.}(2013{\natexlab{c}})\citenamefont {Cirac{\`\i}}, \citenamefont
  {Urzhumov},\ and\ \citenamefont {Smith}}]{Ciraci:2013wi}%
  \BibitemOpen
  \bibfield  {author} {\bibinfo {author} {\bibfnamefont {C.}~\bibnamefont
  {Cirac{\`\i}}}, \bibinfo {author} {\bibfnamefont {Y.~A.}\ \bibnamefont
  {Urzhumov}}, \ and\ \bibinfo {author} {\bibfnamefont {D.~R.}\ \bibnamefont
  {Smith}},\ }\bibfield  {title} {\emph {\enquote {\bibinfo {title} {{Far-field
  analysis of axially symmetric three-dimensional directional cloaks}},}\
  }}\href@noop {} {\bibfield  {journal} {\bibinfo  {journal} {Opt. Express}\
  }\textbf {\bibinfo {volume} {21}},\ \bibinfo {pages} {9397} (\bibinfo {year}
  {2013}{\natexlab{c}})}\BibitemShut {NoStop}%
\bibitem [{\citenamefont {Ekardt}(1997)}]{ekardt97}%
  \BibitemOpen
  \bibfield  {author} {\bibinfo {author} {\bibfnamefont {W.}~\bibnamefont
  {Ekardt}},\ }\bibfield  {title} {\emph {\enquote {\bibinfo {title} {The
  super-atom model: link between the metal atom and the infinite metal},}\
  }}\href {\doibase 10.1007/s002570050378} {\bibfield  {journal} {\bibinfo
  {journal} {Zeitschrift f{\"u}r Physik B Condensed Matter}\ }\textbf {\bibinfo
  {volume} {103}},\ \bibinfo {pages} {305} (\bibinfo {year}
  {1997})}\BibitemShut {NoStop}%
\bibitem [{\citenamefont {Rubio}\ \emph {et~al.}(1991)\citenamefont {Rubio},
  \citenamefont {Balbas},\ and\ \citenamefont {Alonso}}]{rubio91}%
  \BibitemOpen
  \bibfield  {author} {\bibinfo {author} {\bibfnamefont {A.}~\bibnamefont
  {Rubio}}, \bibinfo {author} {\bibfnamefont {L.}~\bibnamefont {Balbas}}, \
  and\ \bibinfo {author} {\bibfnamefont {J.}~\bibnamefont {Alonso}},\
  }\bibfield  {title} {\emph {\enquote {\bibinfo {title} {Response properties
  of sodium clusters within a jellium-like model with finite surface
  thickness},}\ }}\href@noop {} {\bibfield  {journal} {\bibinfo  {journal} {Z.
  Phys. D}\ }\textbf {\bibinfo {volume} {19}},\ \bibinfo {pages} {93} (\bibinfo
  {year} {1991})}\BibitemShut {NoStop}%
\bibitem [{\citenamefont {Bonatsos}\ \emph {et~al.}(2000)\citenamefont
  {Bonatsos}, \citenamefont {Karoussos}, \citenamefont {Lenis}, \citenamefont
  {Raychev}, \citenamefont {Roussev},\ and\ \citenamefont
  {Terziev}}]{occup2000}%
  \BibitemOpen
  \bibfield  {author} {\bibinfo {author} {\bibfnamefont {D.}~\bibnamefont
  {Bonatsos}}, \bibinfo {author} {\bibfnamefont {N.}~\bibnamefont {Karoussos}},
  \bibinfo {author} {\bibfnamefont {D.}~\bibnamefont {Lenis}}, \bibinfo
  {author} {\bibfnamefont {P.~P.}\ \bibnamefont {Raychev}}, \bibinfo {author}
  {\bibfnamefont {R.~P.}\ \bibnamefont {Roussev}}, \ and\ \bibinfo {author}
  {\bibfnamefont {P.~A.}\ \bibnamefont {Terziev}},\ }\bibfield  {title} {\emph
  {\enquote {\bibinfo {title} {Unified description of magic numbers of metal
  clusters in terms of the three-dimensional \textbf{\textit{q}}-deformed
  harmonic oscillator},}\ }}\href@noop {} {\bibfield  {journal} {\bibinfo
  {journal} {Phys. Rev. A}\ }\textbf {\bibinfo {volume} {62}},\ \bibinfo
  {pages} {013203} (\bibinfo {year} {2000})}\BibitemShut {NoStop}%
\bibitem [{\citenamefont {de~Heer}(1993)}]{vander93}%
  \BibitemOpen
  \bibfield  {author} {\bibinfo {author} {\bibfnamefont {W.~A.}\ \bibnamefont
  {de~Heer}},\ }\bibfield  {title} {\emph {\enquote {\bibinfo {title} {The
  physics of simple metal clusters: experimental aspects and simple models},}\
  }}\href {\doibase 10.1103/RevModPhys.65.611} {\bibfield  {journal} {\bibinfo
  {journal} {Rev. Mod. Phys.}\ }\textbf {\bibinfo {volume} {65}},\ \bibinfo
  {pages} {611} (\bibinfo {year} {1993})}\BibitemShut {NoStop}%
\bibitem [{\citenamefont {Ekardt}(1984)}]{ekardt84}%
  \BibitemOpen
  \bibfield  {author} {\bibinfo {author} {\bibfnamefont {W.}~\bibnamefont
  {Ekardt}},\ }\bibfield  {title} {\emph {\enquote {\bibinfo {title} {Work
  function of small metal particles: Self-consistent spherical
  jellium-background model},}\ }}\href@noop {} {\bibfield  {journal} {\bibinfo
  {journal} {Phys. Rev. B}\ }\textbf {\bibinfo {volume} {29}},\ \bibinfo
  {pages} {1558} (\bibinfo {year} {1984})}\BibitemShut {NoStop}%
\bibitem [{\citenamefont {King}\ \emph {et~al.}(2015)\citenamefont {King},
  \citenamefont {Liu}, \citenamefont {Yang}, \citenamefont {Cerjan},
  \citenamefont {Everitt}, \citenamefont {Nordlander},\ and\ \citenamefont
  {Halas}}]{King:2015gr}%
  \BibitemOpen
  \bibfield  {author} {\bibinfo {author} {\bibfnamefont {N.~S.}\ \bibnamefont
  {King}}, \bibinfo {author} {\bibfnamefont {L.}~\bibnamefont {Liu}}, \bibinfo
  {author} {\bibfnamefont {X.}~\bibnamefont {Yang}}, \bibinfo {author}
  {\bibfnamefont {B.}~\bibnamefont {Cerjan}}, \bibinfo {author} {\bibfnamefont
  {H.~O.}\ \bibnamefont {Everitt}}, \bibinfo {author} {\bibfnamefont
  {P.}~\bibnamefont {Nordlander}}, \ and\ \bibinfo {author} {\bibfnamefont
  {N.~J.}\ \bibnamefont {Halas}},\ }\bibfield  {title} {\emph {\enquote
  {\bibinfo {title} {{Fano Resonant Aluminum Nanoclusters for Plasmonic
  Colorimetric Sensing}},}\ }}\href@noop {} {\bibfield  {journal} {\bibinfo
  {journal} {ACS Nano}\ }\textbf {\bibinfo {volume} {9}},\ \bibinfo {pages}
  {10628} (\bibinfo {year} {2015})}\BibitemShut {NoStop}%
\bibitem [{\citenamefont {Cirac{\`\i}}\ \emph {et~al.}(2015)\citenamefont
  {Cirac{\`\i}}, \citenamefont {Scalora},\ and\ \citenamefont
  {Smith}}]{Ciraci:2015tp}%
  \BibitemOpen
  \bibfield  {author} {\bibinfo {author} {\bibfnamefont {C.}~\bibnamefont
  {Cirac{\`\i}}}, \bibinfo {author} {\bibfnamefont {M.}~\bibnamefont
  {Scalora}}, \ and\ \bibinfo {author} {\bibfnamefont {D.~R.}\ \bibnamefont
  {Smith}},\ }\bibfield  {title} {\emph {\enquote {\bibinfo {title}
  {{Third-harmonic generation in the presence of classical nonlocal effects in
  gap-plasmon nanostructures}},}\ }}\href@noop {} {\bibfield  {journal}
  {\bibinfo  {journal} {Phys. Rev. B}\ }\textbf {\bibinfo {volume} {91}},\
  \bibinfo {pages} {205403} (\bibinfo {year} {2015})}\BibitemShut {NoStop}%
\bibitem [{\citenamefont {Argyropoulos}\ \emph {et~al.}(2014)\citenamefont
  {Argyropoulos}, \citenamefont {Cirac{\`\i}},\ and\ \citenamefont
  {Smith}}]{Argyropoulos:2014bw}%
  \BibitemOpen
  \bibfield  {author} {\bibinfo {author} {\bibfnamefont {C.}~\bibnamefont
  {Argyropoulos}}, \bibinfo {author} {\bibfnamefont {C.}~\bibnamefont
  {Cirac{\`\i}}}, \ and\ \bibinfo {author} {\bibfnamefont {D.~R.}\ \bibnamefont
  {Smith}},\ }\bibfield  {title} {\emph {\enquote {\bibinfo {title} {Enhanced
  optical bistability with film-coupled plasmonic nanocubes},}\ }}\href@noop {}
  {\bibfield  {journal} {\bibinfo  {journal} {Appl. Phys. Lett.}\ }\textbf
  {\bibinfo {volume} {104}},\ \bibinfo {eid} {063108} (\bibinfo {year}
  {2014})}\BibitemShut {NoStop}%
\bibitem [{\citenamefont {Levy}\ \emph {et~al.}(1984)\citenamefont {Levy},
  \citenamefont {Perdew},\ and\ \citenamefont {Sahni}}]{levy84}%
  \BibitemOpen
  \bibfield  {author} {\bibinfo {author} {\bibfnamefont {M.}~\bibnamefont
  {Levy}}, \bibinfo {author} {\bibfnamefont {J.~P.}\ \bibnamefont {Perdew}}, \
  and\ \bibinfo {author} {\bibfnamefont {V.}~\bibnamefont {Sahni}},\ }\bibfield
   {title} {\emph {\enquote {\bibinfo {title} {Exact differential equation for
  the density and ionization energy of a many-particle system},}\ }}\href@noop
  {} {\bibfield  {journal} {\bibinfo  {journal} {Phys. Rev. A}\ }\textbf
  {\bibinfo {volume} {30}},\ \bibinfo {pages} {2745} (\bibinfo {year}
  {1984})}\BibitemShut {NoStop}%
\bibitem [{\citenamefont {Snider}\ and\ \citenamefont
  {Sorbello}(1983)}]{snidersorbello83}%
  \BibitemOpen
  \bibfield  {author} {\bibinfo {author} {\bibfnamefont {D.~R.}\ \bibnamefont
  {Snider}}\ and\ \bibinfo {author} {\bibfnamefont {R.~S.}\ \bibnamefont
  {Sorbello}},\ }\bibfield  {title} {\emph {\enquote {\bibinfo {title}
  {Density-functional calculation of the static electronic polarizability of a
  small metal sphere},}\ }}\href@noop {} {\bibfield  {journal} {\bibinfo
  {journal} {Phys. Rev. B}\ }\textbf {\bibinfo {volume} {28}},\ \bibinfo
  {pages} {5702} (\bibinfo {year} {1983})}\BibitemShut {NoStop}%
\bibitem [{\citenamefont {Brack}(1989)}]{brack89}%
  \BibitemOpen
  \bibfield  {author} {\bibinfo {author} {\bibfnamefont {M.}~\bibnamefont
  {Brack}},\ }\bibfield  {title} {\emph {\enquote {\bibinfo {title} {Multipole
  vibrations of small alkali-metal spheres in a semiclassical description},}\
  }}\href@noop {} {\bibfield  {journal} {\bibinfo  {journal} {Phys. Rev. B}\
  }\textbf {\bibinfo {volume} {39}},\ \bibinfo {pages} {3533} (\bibinfo {year}
  {1989})}\BibitemShut {NoStop}%
\bibitem [{\citenamefont {Perdew}\ \emph {et~al.}(1982)\citenamefont {Perdew},
  \citenamefont {Parr}, \citenamefont {Levy},\ and\ \citenamefont
  {Balduz}}]{perdewHOMO}%
  \BibitemOpen
  \bibfield  {author} {\bibinfo {author} {\bibfnamefont {J.~P.}\ \bibnamefont
  {Perdew}}, \bibinfo {author} {\bibfnamefont {R.~G.}\ \bibnamefont {Parr}},
  \bibinfo {author} {\bibfnamefont {M.}~\bibnamefont {Levy}}, \ and\ \bibinfo
  {author} {\bibfnamefont {J.~L.}\ \bibnamefont {Balduz}},\ }\bibfield  {title}
  {\emph {\enquote {\bibinfo {title} {Density-functional theory for fractional
  particle number: Derivative discontinuities of the energy},}\ }}\href
  {\doibase 10.1103/PhysRevLett.49.1691} {\bibfield  {journal} {\bibinfo
  {journal} {Phys. Rev. Lett.}\ }\textbf {\bibinfo {volume} {49}},\ \bibinfo
  {pages} {1691} (\bibinfo {year} {1982})}\BibitemShut {NoStop}%
\bibitem [{\citenamefont {Della~Sala}\ \emph {et~al.}(2015)\citenamefont
  {Della~Sala}, \citenamefont {Fabiano},\ and\ \citenamefont
  {Constantin}}]{dskato15}%
  \BibitemOpen
  \bibfield  {author} {\bibinfo {author} {\bibfnamefont {F.}~\bibnamefont
  {Della~Sala}}, \bibinfo {author} {\bibfnamefont {E.}~\bibnamefont {Fabiano}},
  \ and\ \bibinfo {author} {\bibfnamefont {L.~A.}\ \bibnamefont {Constantin}},\
  }\bibfield  {title} {\emph {\enquote {\bibinfo {title} {Kohn-sham kinetic
  energy density in the nuclear and asymptotic regions: Deviations from the von
  weizs\"acker behavior and applications to density functionals},}\
  }}\href@noop {} {\bibfield  {journal} {\bibinfo  {journal} {Phys. Rev. B}\
  }\textbf {\bibinfo {volume} {91}},\ \bibinfo {pages} {035126} (\bibinfo
  {year} {2015})}\BibitemShut {NoStop}%
\bibitem [{\citenamefont {Casida}\ \emph {et~al.}(1998)\citenamefont {Casida},
  \citenamefont {Jamorski}, \citenamefont {Casida},\ and\ \citenamefont
  {Salahub}}]{casida98}%
  \BibitemOpen
  \bibfield  {author} {\bibinfo {author} {\bibfnamefont {M.~E.}\ \bibnamefont
  {Casida}}, \bibinfo {author} {\bibfnamefont {C.}~\bibnamefont {Jamorski}},
  \bibinfo {author} {\bibfnamefont {K.~C.}\ \bibnamefont {Casida}}, \ and\
  \bibinfo {author} {\bibfnamefont {D.~R.}\ \bibnamefont {Salahub}},\
  }\bibfield  {title} {\emph {\enquote {\bibinfo {title} {Molecular excitation
  energies to high-lying bound states from time-dependent density-functional
  response theory: Characterization and correction of the time-dependent local
  density approximation ionization threshold},}\ }}\href@noop {} {\bibfield
  {journal} {\bibinfo  {journal} {J. Chem. Phys.}\ }\textbf {\bibinfo {volume}
  {108}},\ \bibinfo {pages} {4439} (\bibinfo {year} {1998})}\BibitemShut
  {NoStop}%
\bibitem [{\citenamefont {Engquist}\ and\ \citenamefont {Majda}(1977)}]{abc}%
  \BibitemOpen
  \bibfield  {author} {\bibinfo {author} {\bibfnamefont {B.}~\bibnamefont
  {Engquist}}\ and\ \bibinfo {author} {\bibfnamefont {A.}~\bibnamefont
  {Majda}},\ }\bibfield  {title} {\emph {\enquote {\bibinfo {title} {Absorbing
  boundary conditions for numerical simulation of waves},}\ }}\href@noop {}
  {\bibfield  {journal} {\bibinfo  {journal} {PNAS}\ }\textbf {\bibinfo
  {volume} {74}},\ \bibinfo {pages} {1765} (\bibinfo {year}
  {1977})}\BibitemShut {NoStop}%
\bibitem [{\citenamefont {Zangwill}\ and\ \citenamefont
  {Soven}(1980)}]{zangsoven80}%
  \BibitemOpen
  \bibfield  {author} {\bibinfo {author} {\bibfnamefont {A.}~\bibnamefont
  {Zangwill}}\ and\ \bibinfo {author} {\bibfnamefont {P.}~\bibnamefont
  {Soven}},\ }\bibfield  {title} {\emph {\enquote {\bibinfo {title}
  {Density-functional approach to local-field effects in finite systems:
  Photoabsorption in the rare gases},}\ }}\href@noop {} {\bibfield  {journal}
  {\bibinfo  {journal} {Phys. Rev. A}\ }\textbf {\bibinfo {volume} {21}},\
  \bibinfo {pages} {1561} (\bibinfo {year} {1980})}\BibitemShut {NoStop}%
\bibitem [{\citenamefont {Bertsch}(1990)}]{bertsch90}%
  \BibitemOpen
  \bibfield  {author} {\bibinfo {author} {\bibfnamefont {G.}~\bibnamefont
  {Bertsch}},\ }\bibfield  {title} {\emph {\enquote {\bibinfo {title} {An {RPA}
  program for jellium spheres},}\ }}\href@noop {} {\bibfield  {journal}
  {\bibinfo  {journal} {Comput. Phys. Commun.}\ }\textbf {\bibinfo {volume}
  {60}},\ \bibinfo {pages} {247 } (\bibinfo {year} {1990})}\BibitemShut
  {NoStop}%
\bibitem [{\citenamefont {Prodan}\ and\ \citenamefont
  {Nordlander}(2002)}]{prodan2002}%
  \BibitemOpen
  \bibfield  {author} {\bibinfo {author} {\bibfnamefont {E.}~\bibnamefont
  {Prodan}}\ and\ \bibinfo {author} {\bibfnamefont {P.}~\bibnamefont
  {Nordlander}},\ }\bibfield  {title} {\emph {\enquote {\bibinfo {title}
  {Electronic structure and polarizability of metallic nanoshells},}\
  }}\href@noop {} {\bibfield  {journal} {\bibinfo  {journal} {Chem. Phys.
  Lett.}\ }\textbf {\bibinfo {volume} {352}},\ \bibinfo {pages} {140} (\bibinfo
  {year} {2002})}\BibitemShut {NoStop}%
\bibitem [{\citenamefont {Yannouleas}\ \emph {et~al.}(1993)\citenamefont
  {Yannouleas}, \citenamefont {Vigezzi},\ and\ \citenamefont
  {Broglia}}]{yanno93}%
  \BibitemOpen
  \bibfield  {author} {\bibinfo {author} {\bibfnamefont {C.}~\bibnamefont
  {Yannouleas}}, \bibinfo {author} {\bibfnamefont {E.}~\bibnamefont {Vigezzi}},
  \ and\ \bibinfo {author} {\bibfnamefont {R.~A.}\ \bibnamefont {Broglia}},\
  }\bibfield  {title} {\emph {\enquote {\bibinfo {title} {Evolution of the
  optical properties of alkali-metal microclusters towards the bulk: The matrix
  random-phase-approximation description},}\ }}\href@noop {} {\bibfield
  {journal} {\bibinfo  {journal} {Phys. Rev. B}\ }\textbf {\bibinfo {volume}
  {47}},\ \bibinfo {pages} {9849} (\bibinfo {year} {1993})}\BibitemShut
  {NoStop}%
\bibitem [{\citenamefont {Reiners}\ \emph {et~al.}(1995)\citenamefont
  {Reiners}, \citenamefont {Ellert}, \citenamefont {Schmidt},\ and\
  \citenamefont {Haberland}}]{Reiners:1995gq}%
  \BibitemOpen
  \bibfield  {author} {\bibinfo {author} {\bibfnamefont {T.}~\bibnamefont
  {Reiners}}, \bibinfo {author} {\bibfnamefont {C.}~\bibnamefont {Ellert}},
  \bibinfo {author} {\bibfnamefont {M.}~\bibnamefont {Schmidt}}, \ and\
  \bibinfo {author} {\bibfnamefont {H.}~\bibnamefont {Haberland}},\ }\bibfield
  {title} {\emph {\enquote {\bibinfo {title} {{Size Dependence of the
  Optical-Response of Spherical Sodium Clusters}},}\ }}\href@noop {} {\bibfield
   {journal} {\bibinfo  {journal} {Phys. Rev. Lett.}\ }\textbf {\bibinfo
  {volume} {74}},\ \bibinfo {pages} {1558} (\bibinfo {year}
  {1995})}\BibitemShut {NoStop}%
\bibitem [{\citenamefont {Parks}\ and\ \citenamefont
  {Mcdonald}(1989)}]{Parks:1989ui}%
  \BibitemOpen
  \bibfield  {author} {\bibinfo {author} {\bibfnamefont {J.~H.}\ \bibnamefont
  {Parks}}\ and\ \bibinfo {author} {\bibfnamefont {S.~A.}\ \bibnamefont
  {Mcdonald}},\ }\bibfield  {title} {\emph {\enquote {\bibinfo {title}
  {{Evolution of the Collective-Mode Resonance in Small Adsorbed Sodium
  Clusters}},}\ }}\href@noop {} {\bibfield  {journal} {\bibinfo  {journal}
  {Phys. Rev. Lett.}\ }\textbf {\bibinfo {volume} {62}},\ \bibinfo {pages}
  {2301} (\bibinfo {year} {1989})}\BibitemShut {NoStop}%
\bibitem [{\citenamefont {Scholl}\ \emph {et~al.}(2012)\citenamefont {Scholl},
  \citenamefont {Koh},\ and\ \citenamefont {Dionne}}]{Scholl:2012dsa}%
  \BibitemOpen
  \bibfield  {author} {\bibinfo {author} {\bibfnamefont {J.~A.}\ \bibnamefont
  {Scholl}}, \bibinfo {author} {\bibfnamefont {A.~L.}\ \bibnamefont {Koh}}, \
  and\ \bibinfo {author} {\bibfnamefont {J.~A.}\ \bibnamefont {Dionne}},\
  }\bibfield  {title} {\emph {\enquote {\bibinfo {title} {{Quantum plasmon
  resonances of individual metallic nanoparticles}},}\ }}\href@noop {}
  {\bibfield  {journal} {\bibinfo  {journal} {Nature}\ }\textbf {\bibinfo
  {volume} {483}},\ \bibinfo {pages} {421} (\bibinfo {year}
  {2012})}\BibitemShut {NoStop}%
\bibitem [{\citenamefont {Toscano}\ \emph {et~al.}(2012)\citenamefont
  {Toscano}, \citenamefont {Raza}, \citenamefont {Jauho}, \citenamefont
  {Mortensen},\ and\ \citenamefont {Wubs}}]{Toscano:2012fh}%
  \BibitemOpen
  \bibfield  {author} {\bibinfo {author} {\bibfnamefont {G.}~\bibnamefont
  {Toscano}}, \bibinfo {author} {\bibfnamefont {S.}~\bibnamefont {Raza}},
  \bibinfo {author} {\bibfnamefont {A.-P.}\ \bibnamefont {Jauho}}, \bibinfo
  {author} {\bibfnamefont {N.~A.}\ \bibnamefont {Mortensen}}, \ and\ \bibinfo
  {author} {\bibfnamefont {M.}~\bibnamefont {Wubs}},\ }\bibfield  {title}
  {\emph {\enquote {\bibinfo {title} {{Modified field enhancement and
  extinction by plasmonic nanowire dimers due to nonlocal response}},}\
  }}\href@noop {} {\bibfield  {journal} {\bibinfo  {journal} {Opt. Express}\
  }\textbf {\bibinfo {volume} {20}},\ \bibinfo {pages} {4176} (\bibinfo {year}
  {2012})}\BibitemShut {NoStop}%
\bibitem [{\citenamefont {Raza}\ \emph {et~al.}(2013)\citenamefont {Raza},
  \citenamefont {Raza}, \citenamefont {Stenger}, \citenamefont {Stenger},
  \citenamefont {Kadkhodazadeh}, \citenamefont {Kadkhodazadeh}, \citenamefont
  {Fischer}, \citenamefont {Fischer}, \citenamefont {Kostesha}, \citenamefont
  {Jauho}, \citenamefont {Burrows}, \citenamefont {Wubs},\ and\ \citenamefont
  {Mortensen}}]{Raza:2013gn}%
  \BibitemOpen
  \bibfield  {author} {\bibinfo {author} {\bibfnamefont {S.}~\bibnamefont
  {Raza}}, \bibinfo {author} {\bibfnamefont {S.}~\bibnamefont {Raza}}, \bibinfo
  {author} {\bibfnamefont {N.}~\bibnamefont {Stenger}}, \bibinfo {author}
  {\bibfnamefont {N.}~\bibnamefont {Stenger}}, \bibinfo {author} {\bibfnamefont
  {S.}~\bibnamefont {Kadkhodazadeh}}, \bibinfo {author} {\bibfnamefont
  {S.}~\bibnamefont {Kadkhodazadeh}}, \bibinfo {author} {\bibfnamefont {S.~V.}\
  \bibnamefont {Fischer}}, \bibinfo {author} {\bibfnamefont {S.~V.}\
  \bibnamefont {Fischer}}, \bibinfo {author} {\bibfnamefont {N.}~\bibnamefont
  {Kostesha}}, \bibinfo {author} {\bibfnamefont {A.-P.}\ \bibnamefont {Jauho}},
  \bibinfo {author} {\bibfnamefont {A.}~\bibnamefont {Burrows}}, \bibinfo
  {author} {\bibfnamefont {M.}~\bibnamefont {Wubs}}, \ and\ \bibinfo {author}
  {\bibfnamefont {N.~A.}\ \bibnamefont {Mortensen}},\ }\bibfield  {title}
  {\emph {\enquote {\bibinfo {title} {{Blueshift of the surface plasmon
  resonance in silver nanoparticles studied with EELS}},}\ }}\href@noop {}
  {\bibfield  {journal} {\bibinfo  {journal} {Nanophotonics}\ }\textbf
  {\bibinfo {volume} {2}},\ \bibinfo {pages} {131} (\bibinfo {year}
  {2013})}\BibitemShut {NoStop}%
\bibitem [{\citenamefont {Borensztein}\ \emph {et~al.}(1986)\citenamefont
  {Borensztein}, \citenamefont {De~Andr\`es}, \citenamefont {Monreal},
  \citenamefont {Lopez-Rios},\ and\ \citenamefont {Flores}}]{bore86}%
  \BibitemOpen
  \bibfield  {author} {\bibinfo {author} {\bibfnamefont {Y.}~\bibnamefont
  {Borensztein}}, \bibinfo {author} {\bibfnamefont {P.}~\bibnamefont
  {De~Andr\`es}}, \bibinfo {author} {\bibfnamefont {R.}~\bibnamefont
  {Monreal}}, \bibinfo {author} {\bibfnamefont {T.}~\bibnamefont {Lopez-Rios}},
  \ and\ \bibinfo {author} {\bibfnamefont {F.}~\bibnamefont {Flores}},\
  }\bibfield  {title} {\emph {\enquote {\bibinfo {title} {Blue shift of the
  dipolar plasma resonance in small silver particles on an alumina surface},}\
  }}\href@noop {} {\bibfield  {journal} {\bibinfo  {journal} {Phys. Rev. B}\
  }\textbf {\bibinfo {volume} {33}},\ \bibinfo {pages} {2828} (\bibinfo {year}
  {1986})}\BibitemShut {NoStop}%
\bibitem [{\citenamefont {Lerm\'e}\ \emph {et~al.}(1998)\citenamefont
  {Lerm\'e}, \citenamefont {Palpant}, \citenamefont {Pr\'evel}, \citenamefont
  {Pellarin}, \citenamefont {Treilleux}, \citenamefont {Vialle}, \citenamefont
  {Perez},\ and\ \citenamefont {Broyer}}]{lerme98}%
  \BibitemOpen
  \bibfield  {author} {\bibinfo {author} {\bibfnamefont {J.}~\bibnamefont
  {Lerm\'e}}, \bibinfo {author} {\bibfnamefont {B.}~\bibnamefont {Palpant}},
  \bibinfo {author} {\bibfnamefont {B.}~\bibnamefont {Pr\'evel}}, \bibinfo
  {author} {\bibfnamefont {M.}~\bibnamefont {Pellarin}}, \bibinfo {author}
  {\bibfnamefont {M.}~\bibnamefont {Treilleux}}, \bibinfo {author}
  {\bibfnamefont {J.~L.}\ \bibnamefont {Vialle}}, \bibinfo {author}
  {\bibfnamefont {A.}~\bibnamefont {Perez}}, \ and\ \bibinfo {author}
  {\bibfnamefont {M.}~\bibnamefont {Broyer}},\ }\bibfield  {title} {\emph
  {\enquote {\bibinfo {title} {Quenching of the size effects in free and
  matrix-embedded silver clusters},}\ }}\href@noop {} {\bibfield  {journal}
  {\bibinfo  {journal} {Phys. Rev. Lett.}\ }\textbf {\bibinfo {volume} {80}},\
  \bibinfo {pages} {5105} (\bibinfo {year} {1998})}\BibitemShut {NoStop}%
\bibitem [{\citenamefont {Liebsch}(1993)}]{lieb93}%
  \BibitemOpen
  \bibfield  {author} {\bibinfo {author} {\bibfnamefont {A.}~\bibnamefont
  {Liebsch}},\ }\bibfield  {title} {\emph {\enquote {\bibinfo {title}
  {Surface-plasmon dispersion and size dependence of mie resonance: Silver
  versus simple metals},}\ }}\href@noop {} {\bibfield  {journal} {\bibinfo
  {journal} {Phys. Rev. B}\ }\textbf {\bibinfo {volume} {48}},\ \bibinfo
  {pages} {11317} (\bibinfo {year} {1993})}\BibitemShut {NoStop}%
\bibitem [{\citenamefont {Monreal}\ \emph {et~al.}(2013)\citenamefont
  {Monreal}, \citenamefont {Antosiewicz},\ and\ \citenamefont
  {Apell}}]{apell2013}%
  \BibitemOpen
  \bibfield  {author} {\bibinfo {author} {\bibfnamefont {R.~C.}\ \bibnamefont
  {Monreal}}, \bibinfo {author} {\bibfnamefont {T.~J.}\ \bibnamefont
  {Antosiewicz}}, \ and\ \bibinfo {author} {\bibfnamefont {S.~P.}\ \bibnamefont
  {Apell}},\ }\bibfield  {title} {\emph {\enquote {\bibinfo {title}
  {Competition between surface screening and size quantization for surface
  plasmons in nanoparticles},}\ }}\href@noop {} {\bibfield  {journal} {\bibinfo
   {journal} {New J. Phys.}\ }\textbf {\bibinfo {volume} {15}},\ \bibinfo
  {pages} {083044} (\bibinfo {year} {2013})}\BibitemShut {NoStop}%
\bibitem [{\citenamefont {Hohenberg}\ and\ \citenamefont
  {Kohn}(1964)}]{PhysRev.136.B864}%
  \BibitemOpen
  \bibfield  {author} {\bibinfo {author} {\bibfnamefont {P.}~\bibnamefont
  {Hohenberg}}\ and\ \bibinfo {author} {\bibfnamefont {W.}~\bibnamefont
  {Kohn}},\ }\bibfield  {title} {\emph {\enquote {\bibinfo {title}
  {Inhomogeneous electron gas},}\ }}\href {\doibase 10.1103/PhysRev.136.B864}
  {\bibfield  {journal} {\bibinfo  {journal} {Phys. Rev.}\ }\textbf {\bibinfo
  {volume} {136}},\ \bibinfo {pages} {B864} (\bibinfo {year}
  {1964})}\BibitemShut {NoStop}%
\bibitem [{\citenamefont {Romero}\ \emph {et~al.}(2006)\citenamefont {Romero},
  \citenamefont {Aizpurua}, \citenamefont {Bryant},\ and\ \citenamefont
  {Garc{\'\i}a~de Abajo}}]{Romero:2006ip}%
  \BibitemOpen
  \bibfield  {author} {\bibinfo {author} {\bibfnamefont {I.}~\bibnamefont
  {Romero}}, \bibinfo {author} {\bibfnamefont {J.}~\bibnamefont {Aizpurua}},
  \bibinfo {author} {\bibfnamefont {G.~W.}\ \bibnamefont {Bryant}}, \ and\
  \bibinfo {author} {\bibfnamefont {F.~J.}\ \bibnamefont {Garc{\'\i}a~de
  Abajo}},\ }\bibfield  {title} {\emph {\enquote {\bibinfo {title} {{Plasmons
  in nearly touching metallic nanoparticles: singular response in the limit of
  touching dimers}},}\ }}\href@noop {} {\bibfield  {journal} {\bibinfo
  {journal} {Opt. Express}\ }\textbf {\bibinfo {volume} {14}},\ \bibinfo
  {pages} {9988} (\bibinfo {year} {2006})}\BibitemShut {NoStop}%
\bibitem [{\citenamefont {Fern{\'a}ndez-Dom{\'\i}nguez}\ \emph
  {et~al.}(2012{\natexlab{b}})\citenamefont {Fern{\'a}ndez-Dom{\'\i}nguez},
  \citenamefont {Maier},\ and\ \citenamefont
  {Pendry}}]{FernandezDominguez:2012ih}%
  \BibitemOpen
  \bibfield  {author} {\bibinfo {author} {\bibfnamefont {A.~I.}\ \bibnamefont
  {Fern{\'a}ndez-Dom{\'\i}nguez}}, \bibinfo {author} {\bibfnamefont {S.~A.}\
  \bibnamefont {Maier}}, \ and\ \bibinfo {author} {\bibfnamefont {J.~B.}\
  \bibnamefont {Pendry}},\ }\bibfield  {title} {\emph {\enquote {\bibinfo
  {title} {{Transformation optics description of touching metal
  nanospheres}},}\ }}\href@noop {} {\bibfield  {journal} {\bibinfo  {journal}
  {Phys. Rev. B}\ }\textbf {\bibinfo {volume} {85}},\ \bibinfo {pages} {165148}
  (\bibinfo {year} {2012}{\natexlab{b}})}\BibitemShut {NoStop}%
\bibitem [{\citenamefont {Fern{\'a}ndez-Dom{\'\i}nguez}\ \emph
  {et~al.}(2010)\citenamefont {Fern{\'a}ndez-Dom{\'\i}nguez}, \citenamefont
  {Maier},\ and\ \citenamefont {Pendry}}]{FernandezDominguez:2010dq}%
  \BibitemOpen
  \bibfield  {author} {\bibinfo {author} {\bibfnamefont {A.~I.}\ \bibnamefont
  {Fern{\'a}ndez-Dom{\'\i}nguez}}, \bibinfo {author} {\bibfnamefont {S.~A.}\
  \bibnamefont {Maier}}, \ and\ \bibinfo {author} {\bibfnamefont {J.~B.}\
  \bibnamefont {Pendry}},\ }\bibfield  {title} {\emph {\enquote {\bibinfo
  {title} {{Collection and Concentration of Light by Touching Spheres: A
  Transformation Optics Approach}},}\ }}\href@noop {} {\bibfield  {journal}
  {\bibinfo  {journal} {Phys. Rev. Lett.}\ }\textbf {\bibinfo {volume} {105}},\
  \bibinfo {pages} {266807} (\bibinfo {year} {2010})}\BibitemShut {NoStop}%
\bibitem [{\citenamefont {Laricchia}\ \emph {et~al.}(2014)\citenamefont
  {Laricchia}, \citenamefont {Constantin}, \citenamefont {Fabiano},\ and\
  \citenamefont {Sala}}]{laricchia14}%
  \BibitemOpen
  \bibfield  {author} {\bibinfo {author} {\bibfnamefont {S.}~\bibnamefont
  {Laricchia}}, \bibinfo {author} {\bibfnamefont {L.~A.}\ \bibnamefont
  {Constantin}}, \bibinfo {author} {\bibfnamefont {E.}~\bibnamefont {Fabiano}},
  \ and\ \bibinfo {author} {\bibfnamefont {F.~D.}\ \bibnamefont {Sala}},\
  }\bibfield  {title} {\emph {\enquote {\bibinfo {title} {Laplacian-level
  kinetic energy approximations based on the fourth-order gradient expansion:
  Global assessment and application to the subsystem formulation of density
  functional theory},}\ }}\href@noop {} {\bibfield  {journal} {\bibinfo
  {journal} {J. Chem. Theory Comput.}\ }\textbf {\bibinfo {volume} {10}},\
  \bibinfo {pages} {164} (\bibinfo {year} {2014})}\BibitemShut {NoStop}%
\end{thebibliography}
\end{document}